*Title*: A strong test of the Maximum Entropy Theory of Ecology

*Article type*: Note

*Author affiliation*: Xiao Xiao[1, 2, *], Daniel J. McGlinn[1, 2], and Ethan P. White[1, 2]

[1]Department of Biology, Utah State University, 5305 Old Main Hill, Logan, UT 84322-5305;

[2]Ecology Center, Utah State University, 5205 Old Main Hill, Logan, UT 84322-5205

[*]Corresponding author

*Email addresses*: xiao@weecology.org; ethan@weecology.org; daniel.mcglinn@usu.edu

*Keywords*: biodiversity, body size distributions, macroecology, maximum entropy, species abundance distribution, unified theory

*Online supplementary material*: Appendices A to E, Figure B1, Figure D1

*Figures to print in color*: Figure 1, Figure 2

**Abstract**

The Maximum Entropy Theory of Ecology (METE) is a unified theory of biodiversity that predicts a large number of macroecological patterns using only information on the species richness, total abundance, and total metabolic rate of the community. We evaluated four major predictions of METE simultaneously at an unprecedented scale using data from 60 globally distributed forest communities including over 300,000 individuals and nearly 2000 species. METE successfully captured 96% and 89% of the variation in the species abundance distribution and the individual size distribution, but performed poorly when characterizing the size-density relationship and intraspecific distribution of individual size. Specifically, METE predicted a negative correlation between size and species abundance, which is weak in natural communities. By evaluating multiple predictions with large quantities of data, our study not only identifies a mismatch between abundance and body size in METE, but also demonstrates the importance of conducting strong tests of ecological theories.

## Introduction

The structure of ecological communities can be quantified using a variety of relationships, including many of the most well-studied patterns in ecology such as the distribution of individuals among species (the species abundance distribution or SAD), the increase of species richness with area (the species area relationship or SAR), and the distributions of energy consumption and body size (Brown 1995; Rosenzweig 1995; McGill et al. 2007; White et al. 2007). With the increasing consensus that these patterns are not fully independent, a growing number of unified theories have been proposed to identify links between the patterns and unite them under a single framework (e.g., Hanski and Gyllenberg 1997; Hubbell 2001; Harte 2011; see McGill 2010 for a review). Among these unified theories there are generally two different approaches, one based on processes and the other based on constraints. With the process-based approach, characteristics of the community are captured by explicitly modeling a few key ecological processes (e.g., Hanski and Gyllenberg 1997; Hubbell 2001). While this approach has the potential to directly establish connection between patterns and processes, it has been found that the same empirical patterns can result from different processes (Cohen 1968; Pielou 1975), and process-specific parameters are often hard to obtain (Hubbell 2001; Jones and Muller-Landau 2008). Alternatively, the constraint-based approach suggests that many macroecological patterns are emergent statistical properties arising from general constraints on the system, while processes are only indirectly incorporated through their effect on the constraints (e.g., Harte 2011; Locey and White 2013). This approach attempts to provide a general explanation of the observed patterns that does not rely on specific processes, which allows predictions to be made with little detailed information about the system.

One of the newest and most parsimonious constraint-based approaches is the Maximum Entropy Theory of Ecology (METE; Harte et al. 2008; Harte et al. 2009; Harte 2011). METE adopts the Maximum Entropy Principle from information theory, which identifies the most likely (least biased) state of a system given a set of constraints (Jaynes 2003). Assuming that the allocation of individuals and energy consumption within a community is constrained by three state variables (total species richness, total number of individuals, and total energy consumption), METE makes predictions for the SAD as well as multiple patterns related to energy use. Spatial patterns such as the SAR and the endemics area relationship can also be predicted with an additional constraint on the area sampled (Harte et al. 2008; Harte et al. 2009; Harte 2011). METE is one of the growing number of theoretical approaches that attempt to synthesize traditionally distinct areas of macroecology dealing with the distributions of individuals and the distributions of energy and biomass (Dewar and Porté 2008; Morlon et al. 2009; O'Dwyer et al. 2009), and thus provides a very general characterization of the structure of ecological systems. With no tunable parameters and no specific assumptions about biological processes, it can potentially be applied to any community where the values of the state variables can be obtained.

Previous studies have evaluated the performance of METE with separate datasets for different patterns and have shown that METE generally provides good characterizations of these patterns across geographical locations and taxonomic groups (Harte et al. 2008; Harte et al. 2009; Harte 2011; White et al. 2012a; McGlinn et al. 2013). However, these tests are relatively weak as they focus on one pattern at a time (McGill 2003). As a unified theory with multiple predictions, METE allows stronger tests to be made by testing the ability of the theory to characterize multiple patterns simultaneously for the same data (McGill 2003; McGill et al.

2006). In this study, we conduct a strong test of the non-spatial predictions of METE using data from 60 globally distributed forest communities to simultaneously evaluate four predictions of the theory (Fig. 1) including the SAD (the distribution of individuals among species) and energetic analogs of the individual size distribution (ISD; the distribution of body size among individuals regardless of their species identity) (Enquist and Niklas 2001; Muller-Landau et al. 2006), the size-density relationship (SDR; the correlation between species abundance and average individual size within species) (Cotgreave 1993), and the intraspecific individual size distribution (iISD; the distribution of body size among individuals within a species) (Gouws et al. 2011). Our analysis shows mixed support for METE across its four predictions, with METE successfully capturing the variation in two patterns while failing for the other two. We discuss the ecological implications of our findings, as well as the importance of conducting strong multi-pattern tests in the evaluation of ecological theories.

## Methods

1. Predicted patterns of METE

METE assumes that allocation of individuals and energy consumption within a community is constrained by three state variables: species richness ($S_0$), total number of individuals ($N_0$), and total metabolic rate summed over all individuals in the community ($E_0$) (Harte et al. 2008; Harte et al. 2009; Harte 2011). Define $R(n, \varepsilon)$ as the joint probability that a species randomly picked from the community has abundance $n$ and an individual randomly picked from such a species has metabolic rate between $(\varepsilon, \varepsilon + \Delta\varepsilon)$, two constraints are then established on the ratio between the state variables:

$$\sum_{n=1}^{N_0} \int_{\varepsilon=1}^{E_0} d\varepsilon \cdot nR(n, \varepsilon) = \frac{N_0}{S_0} \qquad (1)$$

which represents the average abundance per species, and

$$\sum_{n=1}^{N_0} \int_{\varepsilon=1}^{E_0} d\varepsilon \cdot n\varepsilon R(n, \varepsilon) = \frac{E_0}{S_0} \qquad (2)$$

which represents the average total metabolic rate per species. Note that the lower limit of individual metabolic rate is set to be 1, and all measures of metabolic rate are rescaled accordingly.

The forms of the four macroecological patterns that METE predicts can then be derived from $R(n, \varepsilon)$ (see Harte 2011 and **Appendix A** for detailed derivation) and are given by the following four equations. SAD takes the form

$$\Phi(n) \approx \frac{1}{Cn} e^{-(\lambda_1+\lambda_2)n} \qquad (3)$$

which is an upper-truncated Fisher's log-series distribution. Here $\lambda_1$ and $\lambda_2$ are Lagrange multipliers obtained by applying the Maximum Entropy Principle with respect to the constraints, and $C$ is the proper normalization constant. The Individual-level Energy Distribution (which is the energetic equivalent of the ISD) takes the form

$$\Psi(\varepsilon) = \frac{S_0}{N_0 Z} \cdot \frac{e^{-\gamma}}{(1-e^{-\gamma})^2} \cdot \left(1 - (N_0+1)e^{-\gamma N_0} + N_0 e^{-\gamma(N_0+1)}\right) \qquad (4)$$

where $\gamma = \lambda_1 + \lambda_2 \cdot \varepsilon$. Conditioned on abundance $n$, the Species-level Energy Distribution (which is the energetic equivalent of the iISD) is given by

$$\Theta(\varepsilon|n) = \frac{n\lambda_2 e^{-\lambda_2 n\varepsilon}}{e^{-\lambda_2 n} - e^{-\lambda_2 n E_0}} \qquad (5)$$

which is an exponential distribution with parameter $\lambda_2 n$. The expected value of the iISD $\Theta(\varepsilon|n)$ then gives the Average Species Energy Distribution (which is the energetic equivalent of the SDR), i.e., the expected average metabolic rate (size) for individuals within a species with abundance $n$:

$$\bar{\varepsilon}(n) = \frac{1}{n\lambda_2(e^{-\lambda_2 n} - e^{-\lambda_2 n E_0})} \cdot \left[e^{-\lambda_2 n}(\lambda_2 n + 1) - e^{-\lambda_2 n E_0}(\lambda_2 n E_0 + 1)\right] \qquad (6)$$

## 2. Data

METE predicts the iISD to be an exponential distribution (Eqn 5; also see Fig. 1D) where the smallest size class is the most abundant, regardless of species identity or abundance. However, most animal species exhibit interior modes of adult body size (e.g., Koons et al. 2009; Gouws et al. 2011; but see Harte 2011) and large variation in minimum (and maximum) body size among species associated with these modal values (Gouws et al. 2011). In other words, the body sizes of conspecifics are clustered around some intermediate value, while individuals that are much larger or smaller are rare. Consequently, assembling all individuals across species in such communities often yields multimodal ISD (Thibault et al. 2011), as opposed to monotonically decreasing predicted by METE (Eqn 4; also see Fig. 1B). As such animal communities are expected a priori to violate two of the predictions of METE. Therefore, to ensure that the performance of METE was not trivially rejected because of the life history trait of determinate growth, in our analysis we focused exclusively on trees, which are known to have iISDs (Condit et al. 1998) and ISDs (Enquist and Niklas 2001; Muller-Landau et al. 2006) that are well characterized by monotonically declining distributions and which arguably have the greatest prevalence of high quality individual level size data among indeterminately growing taxonomic groups.

We compiled forest plot data from previous publications, publicly available databases, and data obtained through personal communication (Table 1). All plots have been fully surveyed with size measurement for all individuals above plot-specific minimum thresholds. For those plots where surveys have been conducted multiple times, we adopted data from the most recent one unless otherwise specified (see Table 1). Individuals that were dead, not identified to species/morphospecies, and/or missing size measurements were excluded. Individuals with size measurements below or equal to the designated minimum thresholds were excluded as well, because it is unclear whether these size classes were thoroughly surveyed. Overall our analysis encompassed 60 plots that were at least 1 ha in size and had a richness of at least 14 (Table 1), with 1943 species/morphospecies and 379022 individuals in total.

3. Analyses

The scaling relationship between diameter and metabolic rate can be described with good approximation by metabolic theory as $B \propto D^2 \cdot e^{-E/kT}$, where $B$ is metabolic rate, $D$ is diameter, $T$ is temperature, $E$ is the activation energy, and $k$ is the Boltzmann's constant (West et al. 1999; Gillooly et al. 2001). Assuming that $E$ is constant across species and $T$ is constant within a community, the temperature-dependent term $e^{-E/kT}$ is constant within a community, and can be dropped when the metabolic rate of individuals are rescaled. We thus used $(D/D_{min})^2$ as the surrogate for individual metabolic rate, where $D_{min}$ is the diameter of the smallest individual in the community, which sets the minimal individual metabolic rate to be 1 following METE's assumption (see Eqn 2). Applying alternative models that more accurately capture nonlinearities between diameter, mass and metabolic rate did not have any qualitative effect on our results (**Appendix B**). For individuals with multiple stems, we adopted the pipe model to combine the records, i.e., $D = \sqrt{\sum d_i^2}$, where $d_i$'s were diameter of individual stems (Ernest et al. 2009). Since metabolic rate scales as $D^2$, the pipe model preserves the total area as well as the total metabolic rate for all stems combined.

We obtained the Lagrange multipliers $\lambda_1$ and $\lambda_2$ in each community with inputs $S_0$, $N_0$, and $E_0$ (i.e., the sum over the rescaled individual metabolic rates) (see **Appendix A**). Predictions for the four ecological patterns were obtained from Eqns 3-6 and further transformed to facilitate comparison with observations. For the SAD and the ISD, we converted the predicted probability distributions (Eqns 3 & 4) to rank distributions of abundance (i.e., abundance at each rank from the most abundant species to the least abundant species) and size (i.e., scaled metabolic rate at each rank from the largest individual to the smallest individual across all species) (Harte et al. 2008; Harte 2011; White et al. 2012a), which were compared with the empirical rank distributions of abundance and size. For the SDR, predicted average metabolic rate was obtained from Eqn 6 for species with abundance *n*, which was compared to the observed average metabolic rate for that species. For the iISD, we converted the predicted exponential distribution (Eqn 5) into a rank distribution of individual size for each species, and compared the scaled metabolic rate predicted at each rank to the observed value. Alternative analyses for the two continuous distributions, the ISD and the iISD, did not change our results (**Appendix C**).

The explanatory power of METE for each pattern was quantified using the coefficient of determination $R^2$, which was calculated as

$$R^2 = 1 - \sum_i [\log_{10}(obs_i) - \log_{10}(pred_i)]^2 \,/\, \sum_i [\log_{10}(obs_i) - \overline{\log_{10}(obs_i)}]^2 \qquad (7)$$

where $obs_i$ and $pred_i$ were the *i*th observed value and METE's prediction, respectively. Both observed and predicted values were log-transformed for homoscedasticity. Note that $R^2$ measures the proportion of variation in the observation explained by the prediction; it is based on the 1:1 line when the observed values are plotted against the predicted values, not the regression line. Thus it is possible for $R^2$ to be negative, which is an indication that the prediction is worse than taking the average of the observation.

## Results

The results for all forest plots combined are summarized in Fig.2, with observations plotted against predictions for each macroecological pattern. METE provides excellent predictions for the SAD ($R^2$ = 0.96) and the ISD ($R^2$ = 0.89), though the largest size classes deviate slightly but consistently in the ISD. However, the SDR ($R^2$ = -2.24) and the iISD ($R^2$ = 0.15) are not well characterized by the theory.

Further examination of the four macroecological patterns within each community (**Appendix D**, Fig. A4; also see insets in Fig. 2) confirms METE's ability to consistently characterize the SAD (all $R^2$ values > 0.60, 59/60 $R^2$ values > 0.8) and the ISD (all $R^2$ values > 0.48, 49/60 $R^2$ values > 0.8), as well as its inadequacy in characterizing the SDR (all $R^2$ values below zero) and the iISD (maximal $R^2$ = 0.30, 49/60 $R^2$ values < 0).

**Discussion**

Macroecological theories increasingly attempt to make predictions across numerous ecological patterns (McGill 2010), by either directly modeling ecological processes or imposing constraints on the system. Among the constraint-based theories, METE is unique in that it makes simultaneous predictions for two distinct sets of ecological patterns, synthesizing traditionally separate areas of macroecology dealing with distributions of individuals and distributions related to body size and energy use (see also Dewar and Porté 2008; Morlon et al. 2009; O'Dwyer et al. 2009). Using only information on the species richness, total abundance, and total energy use as inputs, METE attempts to characterize various aspects of community structure without tunable parameters or additional assumptions, making it one of the most parsimonious of the current unified theories.

Our analysis shows that METE accurately captures the general form of the SAD (allocation of individuals among species) and ISD (allocation of energy/biomass among individuals) within and among 60 forest communities (Fig. 2A, B; Fig. A4). The SAD and the ISD are among the most well-studied patterns in ecology, and numerous models exist for both patterns. For instance, with metabolic theory and demographic equilibrium models, Muller-Landau *et al.* (2006) identified four possible predictions for the ISD under different assumptions of growth and mortality rates. For the SAD more than twenty models have been proposed (Marquet et al. 2003; McGill et al. 2007), ranging from purely statistical to mechanistic.

Our study demonstrates METE's high predictive power for these two patterns, but it does not imply that it is the best model when each pattern is considered independently. Indeed, our results reveal a slight but consistent departure of individuals in the largest size class from the ISD predicted by METE, which may result from mortality unrelated to energy use (Muller-Landau et al. 2006). Moreover, while METE does generally outperform the most common model of the species abundance distribution (White et al. 2012a), model comparisons for the ISD using AIC suggest that the maximum likelihood Weibull distribution (one of the distributions for tree diameter in Muller-Landau et al. 2006) almost always outperforms METE (though METE's performance is comparable to that of the other two distributions, the exponential and the Pareto; see **Appendix E**). Quantitatively comparing theories that make multiple predictions is challenging and there is no general approach for properly comparing models that make different numbers of predictions. When comparing general theories to single prediction models with tunable parameters it is not surprising that theories such as METE fail to provide the best quantitative fit (White et al. 2012b). However, as a constraint-based unified theory, METE's strength lies in its ability to link together ecological phenomena that were previously considered distinct, and to make predictions based on first principles with minimal inputs. The agreement between METE's predictions and the observed SAD and ISD supports the notion that the majority of variation in these macroecological patterns can be characterized by variation in the state variables $S_0$, $N_0$, and $E_0$ alone (Harte 2011; Supp et al. 2012; White et al. 2012a).

While METE performs well in characterizing the SAD and ISD, it performs poorly when predicting the distribution of energy at the species level (Fig. 2C, D; Fig. A4). These deviations from the predictions reveal a mismatch between the predicted metabolic rate of individuals and their species' abundances. METE predicts a monotonically decreasing relationship between species abundance and average intraspecific metabolic rate, i.e., species with higher abundance are also smaller in size on average and are more likely to contain smaller individuals (Eqns 5, 6, Fig. 1C). Evaluating the total (instead of average) intraspecific metabolic rate, this relationship translates roughly into Damuth's energetic equivalence rule (Damuth 1981), where the total energy consumption within a species does not depend on species identity or abundance (Harte et al. 2008; Harte 2011). While Damuth's rule has been argued to apply at global scales (Damuth 1981; White et al. 2007), our results indicate that it does not hold locally, in concordance with a number of previous studies (Brown and Maurer 1987; Blackburn and Gaston 1997; White et al. 2007).

The consistency of our results across 60 forest communities (as well as confirmative evidence from a concurrent study of a single herbaceous plant community; Newman *et al.* in review) provides strong evidence for METE's mixed performance among the four macroecological patterns. However, several limitations of the study are worth noting. First, we only analyzed a single taxonomic group (trees). This was in part because individual level size data collected in standardized ways is available for a large number of tree communities, and in part based on a prior knowledge that the form of the ISD and the iISD (Condit et al. 1998; Enquist and Niklas 2001; Muller-Landau et al. 2006) had a reasonable chance of being well characterized by the theory (see **Methods**). While we know that the SAD predictions of the theory perform well in general (White et al. 2012a), further tests are necessary to determine if the simultaneous good fit of the ISD predictions is supported in other taxonomic groups. There is some evidence that this result holds in invertebrate communities (Harte 2011). Second, we estimated the metabolic rate of individuals based on predictions of metabolic theory rather than direct measurement. While our results were not sensitive to the use of other equations used for estimating metabolic rate (**Appendix B**), it is possible that directly measured metabolic rates could result in different fits to the theory (but see Newman *et al.* in review, which adopts a different method to obtain metabolic rate yet reaches similar conclusions).

Models and theories can be evaluated at multiple levels which yield different strengths of inference (McGill 2003; McGill et al. 2006), progressing from matching theory to empirical observations on a single pattern, to testing against a null hypothesis, to evaluating multiple a priori predictions, to eventually comparing between multiple competing models. With quantitative predictions on various ecological patterns, METE and other unified theories allow for simultaneous examination of multiple predictions, which provides a much stronger test compared to curve-fitting for a single pattern and can often reveal important insight into theories that are otherwise overlooked by single pattern tests (e.g., Adler 2004). As a comprehensive analysis on the performance of METE in predicting abundance and energy distributions in the same datasets, our study demonstrates the importance of moving towards stronger tests in ecology, especially when multiple intercorrelated predictions are available; while previous studies have shown that METE does an impressive job characterizing a single pattern (White et al. 2012a; McGlinn et al. 2013), concurrently evaluating all predictions of the theory identifies a mismatch between species' abundance and individual size that consistently deviates from empirical patterns.

The fact that METE fails to provide good characterization of all four patterns of community structure and performs more poorly than alternative models in some cases can be interpreted in two ways. First, the aspects of community structure that are poorly characterized by the theory may be more adequately characterized by explicitly modeling ecological processes. For example, O'Dwyer *et al.* (2009) has developed a model that incorporates individual demographic rates of birth, death, and growth, which likewise yields predictions of abundance and body size distributions. It is worth noting, however, that the process-based approach and the constraint-based approach do not have to be mutually exclusive. While O'Dwyer *et al.* (2009) suggested that size-related patterns may reflect ecological processes, the agreement between their model and METE in the predicted SAD (both log-series), as well as METE's excellent performance for the ISD, support the idea that information in the underlying processes can be summarized in constraints alone for some macroecological patterns. Alternatively, the constraint-based approach may be sufficient in characterizing patterns of abundance and of body size, but the current form of METE may be incorrect. Specifically, its limitations revealed in our analyses may be remedied by either relaxing the current constraints to remove the association between species level body size and abundance from the theory, or by adding additional constraints to the system so that energetic equivalence among species no longer holds (J. Harte, pers. comm.). While the success of METE in characterizing the SAD and the ISD adds to the growing support for the constraint-based approach for studying macroecological patterns, further work is clearly needed to develop unified theories for community structure whether they are based on specific biological processes or emergent statistical properties.

## Acknowledgements

We thank John Harte, Erica Newman, the rest of the Harte Lab, and members of the Weecology Lab for extensive feedback on this research, general insights into MaxEnt, and for being incredibly supportive of our efforts to evaluate METE. Nathan G. Swenson provided data for wood density in Luquillo forest plot and gave insightful comments. Robert K. Peet provided data for the North Carolina forest plots. The Serimbu (provided by T. Kohyama), Lahei (provided by T. B. Nishimura), and Shirakami (provided by T. Nakashizuka) datasets were obtained from the PlotNet Forest Database. The ACA Amazon (provided by N. Pitman) and DeWalt Bolivia (provided by S. DeWalt) datasets where obtained from SALVIAS. The BCI forest dynamics research project was made possible by National Science Foundation grants to Stephen P. Hubbell: DEB-0640386, DEB-0425651, DEB-0346488, DEB-0129874, DEB-00753102, DEB-9909347, DEB-9615226, DEB-9615226, DEB-9405933, DEB-9221033, DEB-9100058, DEB-8906869, DEB-8605042, DEB-8206992, DEB-7922197, support from the Center for Tropical Forest Science, the Smithsonian Tropical Research Institute, the John D. and Catherine T. MacArthur Foundation, the Mellon Foundation, the Small World Institute Fund, and numerous private individuals, and through the hard work of over 100 people from 10 countries over the past two decades. The UCSC Forest Ecology Research Plot was made possible by National Science Foundation grants to Gregory S. Gilbert (DEB-0515520 and DEB-084259), by the Pepper-Giberson Chair Fund, the University of California, and the hard work of dozens of UCSC students. These two projects are part the Center for Tropical Forest Science, a global network of large-scale demographic tree plots. The Luquillo Experimental Forest Long-Term Ecological Research Program was supported by grants BSR-8811902, DEB 9411973, DEB 0080538, DEB 0218039, DEB 0620910 and DEB 0963447 from NSF to the Institute for Tropical Ecosystem Studies, University of Puerto Rico, and to the International Institute of Tropical Forestry USDA

Forest Service, as part of the Luquillo Long-Term Ecological Research Program. The U.S. Forest Service (Dept. of Agriculture) and the University of Puerto Rico gave additional support. This research was supported by a CAREER award from the U.S. National Science Foundation to E. P. White (DEB-0953694).

## Figure Legends

**Figure 1.** An illustration of the four patterns with data from Barro Colorado Island: A) Rank-abundance distribution; B) Individual size distribution (ISD); C) Size-density relationship (SDR); D) Intraspecific individual size distribution (iISD) of the most abundant species, *Hybanthus prunifolius*. Grey dots or bars in each panel represent empirical observations and magenta curve represents METE's prediction.

**Figure 2.** METE's predictions are plotted against empirical observations across 60 communities for A) SAD (each data point is the abundance of a species at a single rank in one community), B) ISD (each data point is the metabolic rate of an individual at a single rank in one community), C) SDR (each data point is the average metabolic rate within one species in one community), and D) iISD (each data point is the metabolic rate of an individual at a single rank belonging to a specific species in one community). The diagonal black line in each panel is the 1:1 line. The points are color-coded to reflect the density of neighbouring points, with warm (red) colors representing higher densities and cold (blue) colors representing lower densities. The inset reflects the distribution of $R^2$ among 60 communities from negative (left) to 1 (right).

**Figure 1.**

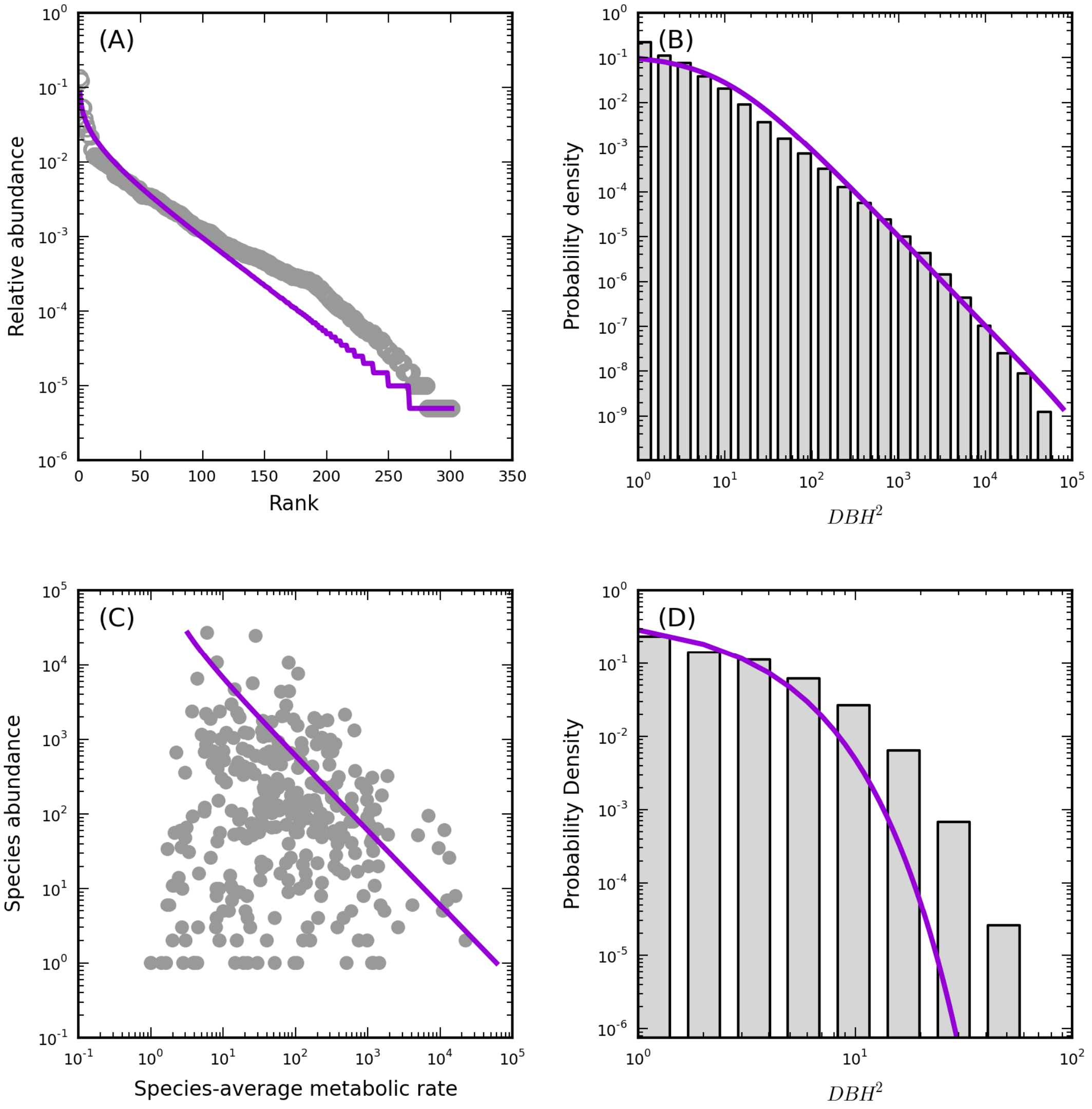

490 **Figure 2.**

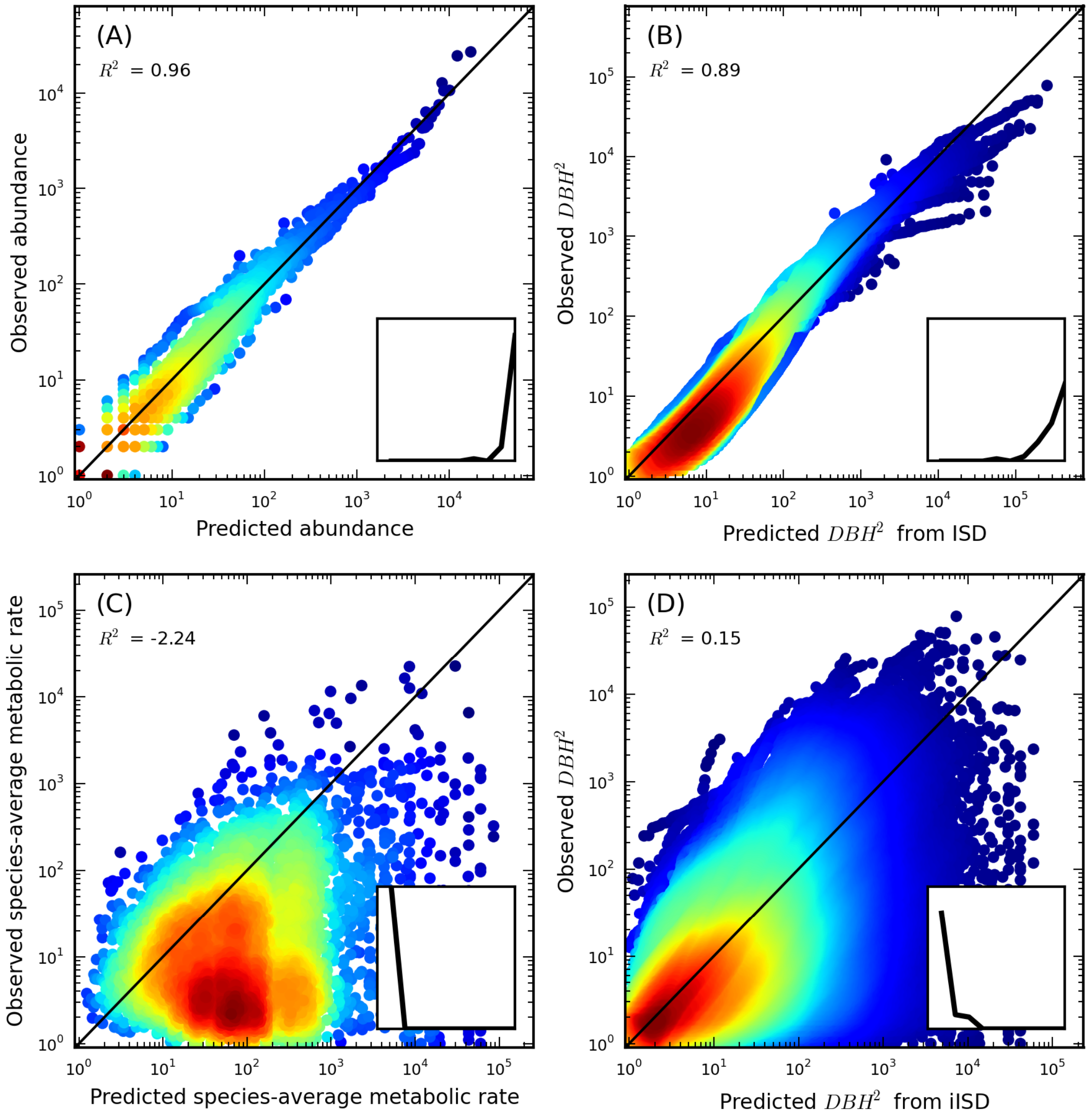

491

**Table 1.** Summary of datasets.

| Dataset | Description | Area of Individual Plots (ha) | Number of Plots | Survey Year | References |
|---|---|---|---|---|---|
| Serimbu | Tropical rainforest | 1 | 2 | 1995[*] | 1, 2, 3 |
| La Selva | Tropical wet forest | 2.24 | 5 | 2009 | 4, 5 |
| ACA Amazon Forest Inventories | Tropical moist forest | 1 | 1 | 2000-2001 | 6 |
| BCI | Tropical moist forest | 50 | 1 | 2010 | 7, 8, 9 |
| DeWalt Bolivia forest plots | Tropical moist forest | 1 | 2 | N/A | 10 |
| Lahei | Tropical moist forest | 1 | 3 | 1998 | 3, 11, 12 |
| Luquillo | Tropical moist forest | 16 | 1 | 1994-1996[†] | 13, 14 |
| Sherman | Tropical moist forest | 5.96 | 1 | 1999 | 15, 16, 17 |
| Cocoli | Tropical moist forest | 4 | 1 | 1998 | 15, 16, 17 |
| Western Ghats | Wet evergreen / moist / dry deciduous forests | 1 | 34 | 1996-1997 | 18 |
| UCSC FERP | Mediterranean mixed evergreen forest | 6 | 1 | 2007 | 19 |
| Shirakami | Beech forest | 1 | 2 | 2006 | 3, 20 |
| Oosting | Hardwood forest | 6.55 | 1 | 1989 | 21, 22 |
| North Carolina forest plots | Mixed hardwoods / pine forest | 1.3 – 5.65 | 5 | 1990-1993[‡] | 23, 24, 25 |

[1]Kohyama et al. (2001) [2]Kohyama et al. (2003) [3]PlotNet (2007) [4]Baribault et al. (2011) [5]Baribault et al. (2012) [6]Pitman et al. (2005) [7]Condit (1998a) [8]Hubbell et al. (2005) [9]Hubbell et al. (1999) [10]DeWalt et al. (1999) [11]Nishimura et al. (2006) [12]Nishimura and Suzuki (2001) [13]Zimmerman et al. (1994) [14]Thompson et al. (2002) [15]Condit (1998b) [16]Condit et al. (2004) [17]Pyke et al. (2001) [18]Ramesh et al. (2010) [19]Gilbert et al. (2010) [20]Nakashizuka et al. (2003) [21]Reed et al. (1993) [22]Palmer et al. (2007) [23]McDonald et al. (2002) [24]Peet and Christensen (1987) [25]Xi et al. (2008)

---

[*] One plot has a more recent survey in 1998, however it lacks species ID.

[†] We chose Census 2 because information for multiple stems is not available in Census 3, and the unit of diameter is unclear in Census 4.

[‡] We chose survey individually for each plot based on expert opinion to minimize the effect of hurricane disturbance.

## Appendix A. Derivation for the Equations

The equations we adopted in our analysis (see **Methods:** 1. Predicted patterns of METE) are largely identical to those in Harte (2011), except for a few minor modifications. Below we briefly summarize the derivations, and derive those that are slightly different. See Harte (2011) for the step-by-step procedure.

**Table A1**. List of equations in our analysis and the location of their counterparts in Harte (2011).

| Equation in this study | Equation in Harte 2011 |
|---|---|
| Eqn 1 | Eqn 7.2 |
| Eqn 2 | Eqn 7.3 |
| Eqn 3 | N/A |
| Eqn 4 | N/A |
| Eqn 5 | Eqn 7.25 |
| Eqn 6 | N/A |

The distribution of central significance on which all other predictions are based is $R(n, \varepsilon)$, the joint probability that a species randomly picked from the community has abundance $n$ and an individual randomly picked from such a species has metabolic rate between ($\varepsilon$, $\varepsilon + \Delta\varepsilon$). By maximizing information entropy $I = -\sum_{n=1}^{N_0} \int_{\varepsilon=1}^{E_0} d\varepsilon \cdot R(n,\varepsilon) \log(R(n,\varepsilon))$ with respect to the constraint on average abundance per species

$$\sum_{n=1}^{N_0} \int_{\varepsilon=1}^{E_0} d\varepsilon \cdot nR(n,\varepsilon) = \frac{N_0}{S_0}$$ (Eqn 1 in the main text; Eqn 7.2 in Harte 2011)

and the constraint on total metabolic rate per species

$$\sum_{n=1}^{N_0} \int_{\varepsilon=1}^{E_0} d\varepsilon \cdot n\varepsilon R(n,\varepsilon) = \frac{E_0}{S_0}$$ (Eqn 2 in the main text; Eqn 7.3 in Harte 2011)

as well as the normalization condition $\sum_{n=1}^{N_0}\int_{\varepsilon=1}^{E_0} d\varepsilon \cdot R(n,\varepsilon) = 1$ (Eqn 7.1 in Harte 2011), $R(n, \varepsilon)$ can be obtained as

$$R(n,\varepsilon) = \frac{1}{Z} e^{-\lambda_1 n} e^{-\lambda_2 n\varepsilon} \quad \text{(Eqn 7.13 in Harte 2011)}$$

where the normalization constant $Z$ is given by

$$Z = \sum_{n=1}^{N_0}\int_{\varepsilon=1}^{E_0} d\varepsilon \cdot e^{-\lambda_1 n} e^{-\lambda_2 n\varepsilon} \quad \text{(Eqn 7.14 in Harte 2011)}$$

With reasonable approximations, the Lagrange multipliers $\lambda_1$ and $\lambda_2$ are given by

$$\sum_{n=1}^{N_0} e^{-(\lambda_1+\lambda_2)\cdot n} \Big/ \sum_{n=1}^{N_0} \frac{e^{-(\lambda_1+\lambda_2)n}}{n} \approx \frac{N_0}{S_0} \quad \text{(Eqn 7.26 in Harte 2011)}$$

$$\lambda_2 \approx \frac{S_0}{E_0 - N_0} \quad \text{(Eqn 7.27 in Harte 2011)}$$

**Derivation for equations not found in Harte (2011):**

1. Species-abundance distribution (SAD; Eqn 3 in main text)

From Eqn 7.23 in Harte (2011):

$$\Phi(n) = \int_{\varepsilon=1}^{E_0} d\varepsilon \cdot R(n,\varepsilon) = \frac{e^{-(\lambda_1+\lambda_2)n} - e^{-(\lambda_1+E_0\lambda_2)n}}{\lambda_2 Z n} \quad \text{(Eqn A1)}$$

Note that this distribution is properly normalized, i.e., $\sum_{n=1}^{N_0} \Phi(n) = 1$.

Given that $E_0$ is large, the second term in the numerator, $e^{-(\lambda_1+E_0\lambda_2)n}$ , is much smaller than the first term $e^{-(\lambda_1+\lambda_2)n}$. Dropping the second term,

$$\Phi(n) \approx \frac{e^{-(\lambda_1+\lambda_2)n}}{\lambda_2 Z n} \quad \text{(Eqn A2)}$$

This approximation leads to the familiar Fisher's log-series distribution, upper-truncated at $N_0$. However, the form in Eqn A2 is not properly normalized, which can cause problems when the SAD is converted to the RAD (rank-abundance distribution). To ensure the proper normalization of $\Phi(n)$, we replace the constant term in the Eqn A2, $\lambda_2 Z$, with constant $C$, where

$$C = \sum_{n=1}^{N_0} \frac{e^{-(\lambda_1+\lambda_2)n}}{n} \quad \text{(Eqn A3)}$$

2. The energetic analog of the individual size distribution (ISD; Eqn 4 in main text)

From Eqn 7.6 in Harte (2011):

$$\Psi(\varepsilon) = \frac{S_0}{N_0}\sum_{n=1}^{N_0} n \cdot R(n,\varepsilon)$$

$$= \frac{S_0}{N_0 Z}\sum_{n=1}^{N_0} n \cdot e^{-\lambda_1 n}\, e^{-\lambda_2 n\varepsilon}$$

$$= \frac{S_0}{N_0 Z}\sum_{n=1}^{N_0} n \cdot e^{-(\lambda_1+\lambda_2\varepsilon)n}$$

$$= \frac{S_0}{N_0 Z} \cdot e^{-(\lambda_1+\lambda_2\varepsilon)} \cdot \frac{1-(N_0+1)e^{-N_0(\lambda_1+\lambda_2\varepsilon)} + N_0 e^{-(N_0+1)(\lambda_1+\lambda_2\varepsilon)}}{(1-e^{-(\lambda_1+\lambda_2\varepsilon)})^2}$$

$$= \frac{S_0}{N_0 Z} \cdot \frac{e^{-\gamma}}{(1-e^{-\gamma})^2} \cdot \left(1-(N_0+1)e^{-\gamma N_0} + N_0 e^{-\gamma(N_0+1)}\right) \qquad \text{(Eqn A4)}$$

where $\gamma = \lambda_1 + \lambda_2 \cdot \varepsilon$. Note that Eqn A4 is not identical to Eqn 7.24 in Harte (2011), which contains a minor error (J. Harte, pers. comm.). However, the trivial difference is unlikely to invalidate or significantly change any published results.

3. The energetic analog of the size-density relationship (Eqn 6 in main text)

From Eqn 7.25 in Harte (2011):

$$\Theta(\varepsilon|n) = \frac{n\lambda_2 e^{-\lambda_2 n\varepsilon}}{e^{-\lambda_2 n} - e^{-\lambda_2 n E_0}} \qquad \text{(Eqn A5)}$$

Then

$$\bar{\varepsilon}(n) = \int_{\varepsilon=1}^{E_0} d\varepsilon \cdot \varepsilon \cdot \Theta(\varepsilon|n)$$

$$= \int_{\varepsilon=1}^{E_0} d\varepsilon \cdot \varepsilon \cdot \frac{n\lambda_2 e^{-\lambda_2 n\varepsilon}}{e^{-\lambda_2 n} - e^{-\lambda_2 n E_0}}$$

$$= \frac{n\lambda_2}{e^{-\lambda_2 n} - e^{-\lambda_2 n E_0}} \int_{\varepsilon=1}^{E_0} d\varepsilon \cdot \varepsilon \cdot e^{-\lambda_2 n\varepsilon}$$

$$= \frac{1}{n\lambda_2(e^{-\lambda_2 n} - e^{-\lambda_2 n E_0})} \cdot \left[e^{-\lambda_2 n}(\lambda_2 n + 1) - e^{-\lambda_2 n E_0}(\lambda_2 n E_0 + 1)\right] \qquad \text{(Eqn A6)}$$

## Appendix B. Alternative Scaling Relationship between Diameter and Metabolic Rate

While we converted diameter ($D$) to metabolic rate ($B$) with $B \propto D^2$ in our analyses, alternative relationships between diameter and metabolic rate have been proposed. Specifically, it has been suggested that the aboveground biomass of tropical trees is a function of diameter, wood density, and forest type (Chave et al. 2005), while the relationship between aboveground biomass and metabolic rate is a biphasic, mixed-power function (Mori et al. 2010). Here we demonstrate that adopting this alternative scaling relationship does not quantitatively change our results.

We compiled species-specific wood density (wood specific gravity; WSG) from previous publications (Reyes et al. 1992; Chave et al. 2009; Zanne et al. 2009; Wright et al. 2010; Swenson et al. 2012). Since WSG information is not available for every species, we included only communities of tropical forest where no less than 70% of individuals belonged to species with known WSG to ensure the accuracy of our analysis. This criterion was met by five communities (BCI, Cocoli, Plots 4 and 5 in LaSelva, and Luquillo) out of all 60 that we examined. Individuals in these communities for which WSG information were not available were assigned average WSG value across all species in the WSG compilation.

We obtained metabolic rate of each individual using the alternative scaling relationships specified in Chave et al. (2005) and Mori et al. (2010). METE was then applied to each community following the steps described in **Methods** in the main text, and its predictions were compared to the observed values for the ISD, the SDR, and the iISD (Fig. B1). Though the patterns differ slightly in shape with metabolic rates obtained from the alternative method, the explanatory power of METE for each pattern does not change qualitatively, i.e., METE characterizes the ISD with high accuracy but is unable to explain much variation in the SDR or the iISD, regardless of the method used to calculate metabolic rate (compare Fig. B1 with corresponding communities in Fig. D1).

**Figure B1.** METE's predictions are plotted against observed values for A) SAD (which remains unchanged), B) ISD, C) SDR, and D) iISD for each of the five communities individually. Here the metabolic rate was obtained with alternative scaling method, which slightly changes the shape of the ISD, the size-density relationship, and the iISD, without significant impact on the explanatory power of METE. (See Pages 42 – 46)

## Appendix C. Alternative Analyses for the ISD and the iISD

In our analyses in the main text, we converted all three probability distributions (SAD, ISD, and iISD) into distributions of rank, and compared the predicted values at each rank against the observed values. While this approach has been widely adopted (Harte et al. 2008; Harte 2011; White et al. 2012), it may not be entirely adequate for continuous distributions such as the ISD and the iISD, where empirical measurements are usually rounded off to decimals and thus may not be directly comparable to the truly continuous values obtained from the predicted distributions of rank. Here we conduct additional analyses for the ISD and the iISD with alternative approaches directly applied on the probability distributions without converting them to distributions of rank to demonstrate the robustness of our results.

For the ISD, we grouped the scaled individual metabolic rates into log(1.7) bins (i.e., 1-1.7, 1.7-2.89, 2.89-4.913, etc.), which resulted in 10 to 21 bins for each forest community. The predicted frequency for each bin was then calculated from the cumulative distribution of $\Psi(\varepsilon)$ (Eqn 4 in the main text) and compared with the observed frequency. The predictive power of METE for the ISD does not change qualitatively when the ISD is analyzed as frequencies ($R^2$ = 0.93; Fig. C1) instead of as ranked metabolic rates ($R^2$ = 0.89; Fig. 2B in the main text).

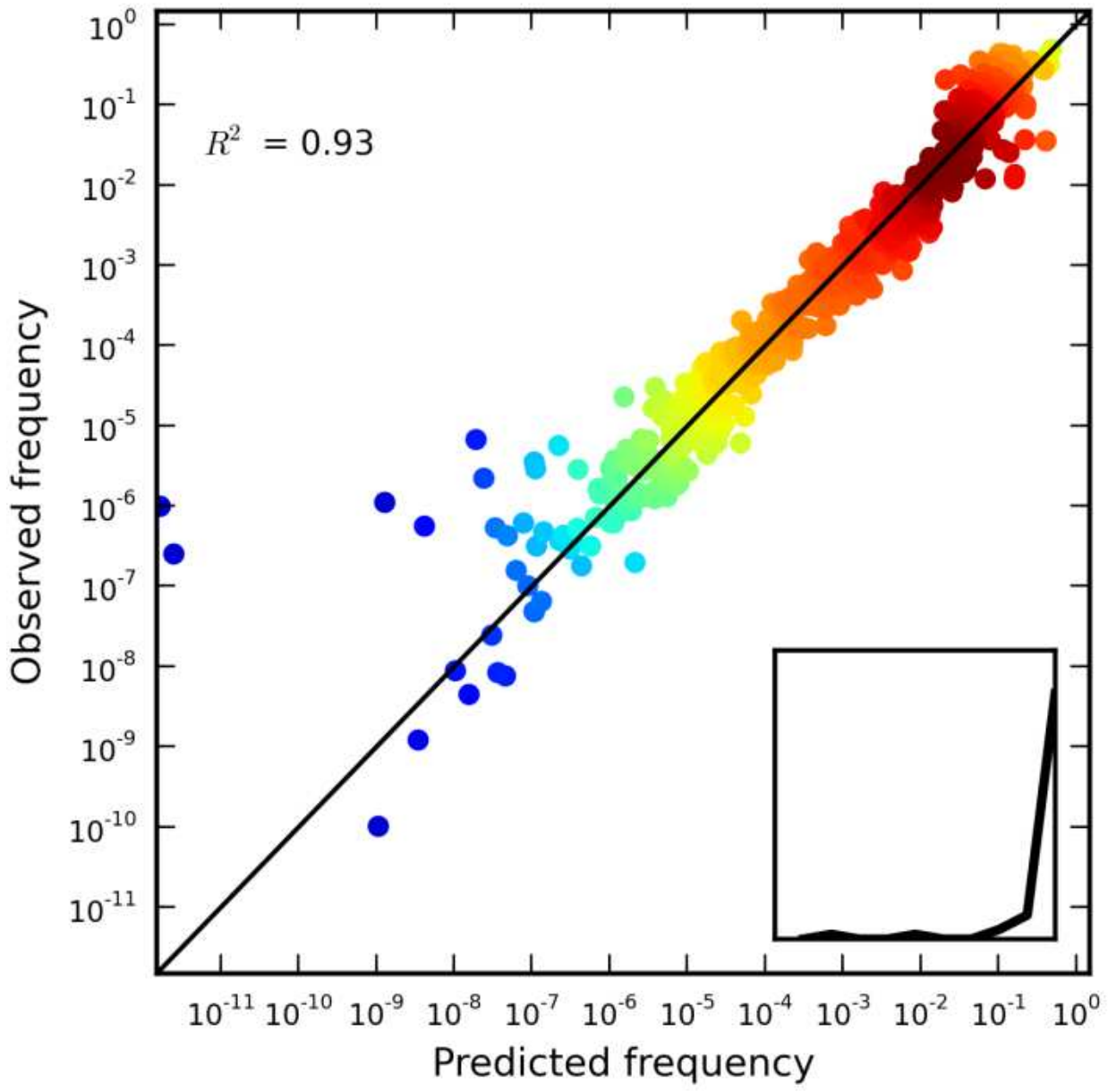

**Figure C1.** Plot METE's predictions against empirical observations across 60 communities for the ISD, which is analyzed as binned frequencies. The diagonal black line is the 1:1 line. The points are color-coded to reflect the density of neighbouring points, with warm (red) colors representing higher densities and cold (blue) colors representing lower densities. The inset in the lower right corner shows the distribution of $R^2$ among individual communities from below zero (left) to 1 (right).

The iISDs for most species contain too few individuals for the above analysis with binned frequencies. Instead, we directly looked at the shape of the distribution. METE predicts that the iISD for each species within a community follows an exponential distribution left-truncated at 1, with the parameter of the distribution proportional to the abundance of the species (see Eqn 5 in main text). Deviation from METE's prediction can occur in one or both of two ways: 1. The observed iISDs are not well characterized by exponential distributions; 2. Assuming that the iISDs can be characterized by exponential distributions (which may or may not be true), the parameter of the distributions that best capture the observed iISDs differ from those predicted by METE (Eqn 5 in main text). Here we show that METE's prediction for iISD fails in both aspects, which is consistent with our results in the main text (Fig. 2D).

1. Characterizing iISDs with exponential distributions

In each community, we fitted an exponential distribution left-truncated at 1 (the minimal rescaled metabolic rate within each community) to rescaled individual metabolic rates for each species with at least 5 individuals, and obtained the maximum likelihood (MLE) parameter of the distribution. For each species, 5,000 independent samples were drawn from a left-truncated exponential distribution with the MLE parameter, where the sample size was equal to the abundance of the species. The two-sample Kolmogorov-Smirnov test was then applied to evaluate if the empirical iISD differ significantly from each sample drawn from the left-truncated exponential distribution. If the proportion of tests (among all 5,000) where the empirical iISD and the randomly generated sample differ in distribution is higher than the significance level ($\alpha$) of the tests, the empirical iISD for the focal species does not conform to a left-truncated exponential distribution.

Fig. C2 shows a histogram of proportions of Kolmogorov-Smirnov tests that are significant at $\alpha = 0.05$ among species (with abundance >= 5) across all 60 communities. Overall the iISDs for more than half of the species are deemed to be significantly different from the left-truncated exponential distribution, which implies that the form of iISD predicted by METE does not hold.

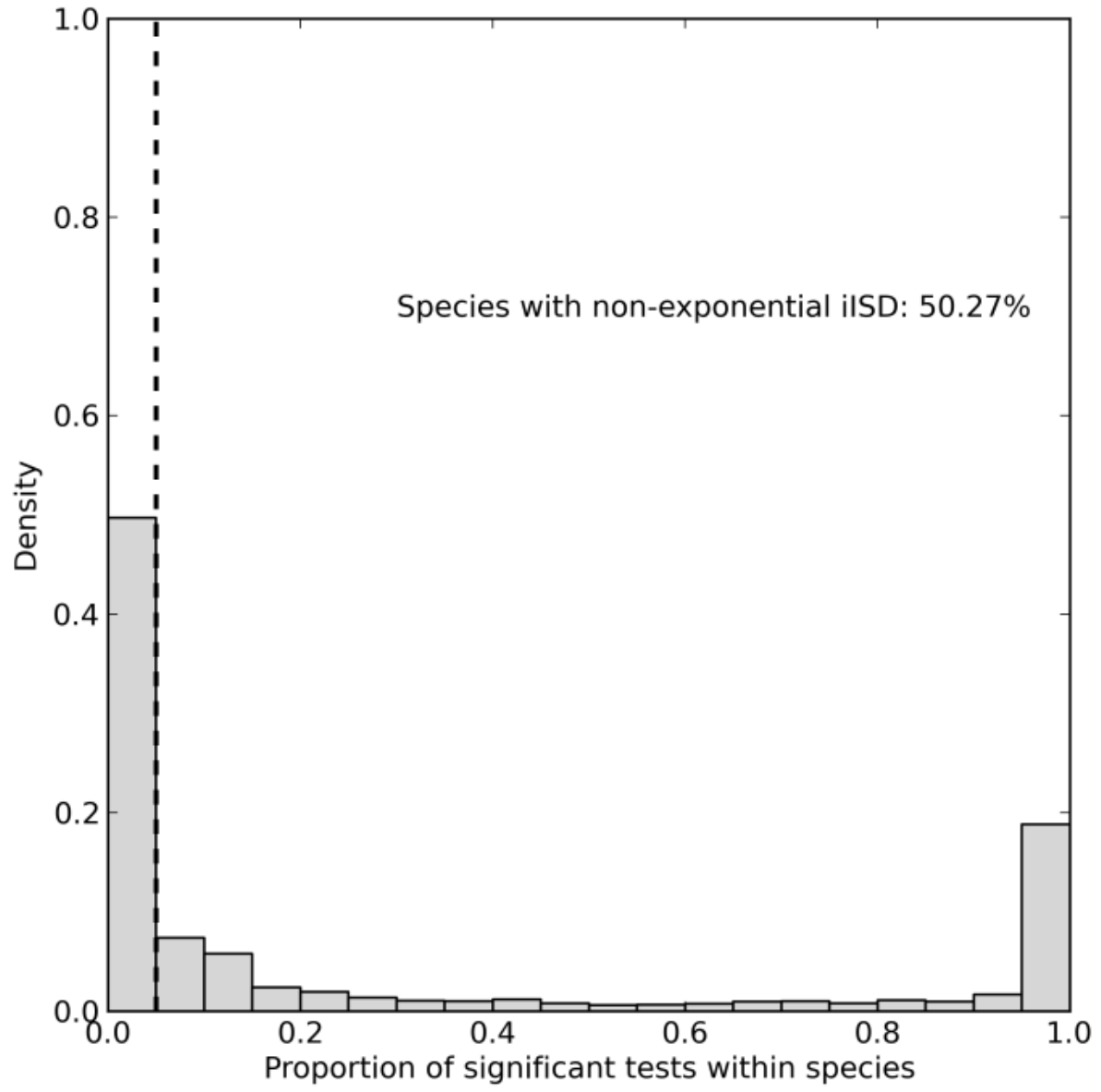

**Figure C2.** Histogram of the proportion of Kolmogorov-Smirnov tests that are significant for each species. The dashed vertical line represents the significance level of the tests $\alpha$ =0.05. Species for which the proportion of tests (among 5,000) with significant results is higher than 0.05 have iISDs that differ significantly from the left-truncated exponential distribution.

2. Comparing MLE parameter with METE's predicted parameter

We further compared the MLE parameter of left-truncated exponential distribution for each species to the parameter predicted by METE ($\lambda_2 n_i$ see Eqn 5 in main text) **(**Fig. C3). Note that this analysis is biased in favor of METE, as we have already shown that left-truncated exponential distribution does not provide a good characterization of empirical iISD for most species (Fig. C2). The fact that the $R^2$ for the iISD is below zero even when METE is evaluated with this biased analysis further strengthens our conclusion that METE is unable to meaningfully capture any variation in the iISD.

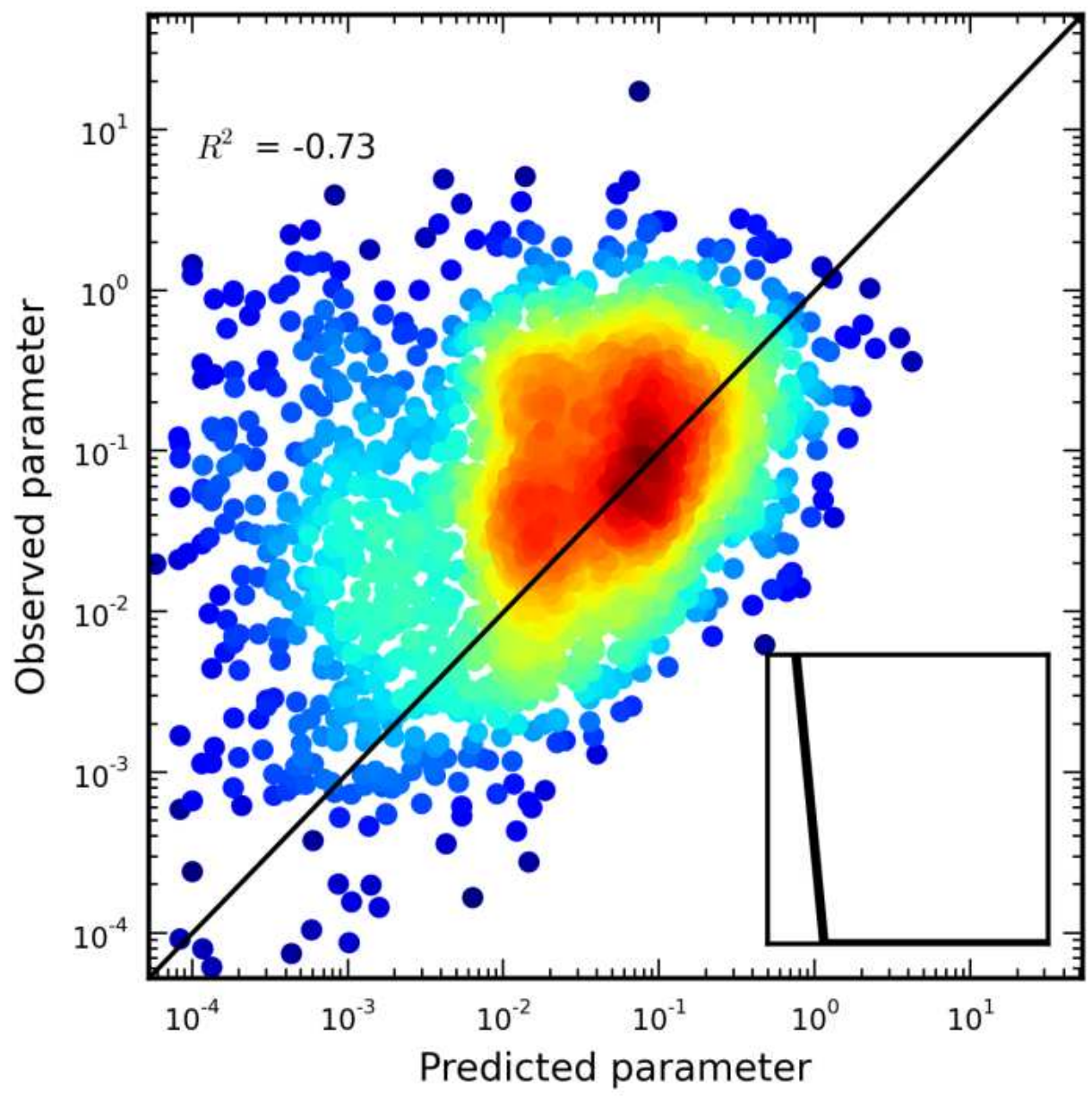

**Figure C3.** The iISD parameter predicted by METE is plotted against the MLE parameter for the empirical distribution for each species (with no fewer than 5 individuals) in each of the 60 communities. The diagonal black line is the 1:1 line. The points are color-coded to reflect the density of neighbouring points, with warm (red) colors representing higher densities and cold (blue) colors representing lower densities. The inset reflects the distribution of $R^2$ among 60 communities from negative (left) to 1 (right).

## Appendix D. Evaluation of METE within Communities

**Figure D1.** METE's predictions for A) SAD, B) ISD, C) SDR, and D) iISD are plotted against observed patterns in each of the 60 forest communities. In A), the grey circles represent observed abundance of species in each community at each rank from the most abundant to the least abundant. In B), the grey bars represent the proportion of individuals within each size bin in each community. In C), each grey dot represents one species with a specific abundance and species-average metabolic rate in the community. The magenta curves in subplots A), B), and C) represent the relationships predicted by METE. In D), the size for each individual within a species predicted by METE is plotted against its observed size, while the diagonal line is the 1:1 line. (See Pages 47 – 106)

## Appendix E. Model Comparison for ISD

Muller-Landau et al. (2006) proposed four possible distributions (exponential, Pareto, Weibull, and quasi-Weibull) for diameter in old-growth forests, under different assumptions of growth and mortality. Here we compare the fit of three of the four distributions (exponential, Pareto, and Weibull) to the fit of the ISD predicted by METE (Eqn 8) using data from the 60 forest communities. The quasi-Weibull distribution, which has been shown to provide the best fit for the majority of communities (Muller-Landau et al. 2006), is not evaluated due to the difficulty in obtaining its maximum likelihood parameters when it is left-truncated.

All distributions are left-truncated to account for the fact that individuals below the minimal threshold in each community where excluded from the datasets. With the minimal size rescaled as 1 across communities (see **Methods)**, the left-truncated exponential distribution takes the form

$$f(D) = \lambda e^{-\lambda(D-1)} \qquad \text{(Eqn E1)}$$

the left-truncated Pareto distribution takes the form

$$f(D) = \frac{\alpha}{D^{\alpha+1}} \qquad \text{(Eqn E2)}$$

the left-truncated Weibull distribution takes the form

$$f(D) = \frac{k}{\lambda}\left(\frac{D}{\lambda}\right)^{k-1} e^{-\left(D/\lambda\right)^k} / e^{-\left(1/\lambda\right)^k} \qquad \text{(Eqn E3)}$$

where the diameter $D$ >= 1 for all three distributions.

Parameters in Eqns E1, E2 and E3 were obtained with maximum likelihood method (MLE) for each community. While analytical solutions exist for parameters in Eqn E1 and Eqn E2, MLE solutions for parameters in Eqn E3 can only be obtained numerically. The three distributions of $D$ were then transformed into distributions of $D^2$ (surrogate for metabolic rate; see Methods) to be consistent with METE's prediction (Eqn 8) as:

$$g(D^2) = \frac{1}{2D} f(D) \qquad \text{(Eqn E4)}$$

where f($D$) is the left-truncated exponential, Pareto, or Weibull distribution in Eqns E1, E2 or E3.

The fit of the ISD predicted by METE and the other three distributions was evaluated with Akaike's Information Criterion (AIC; Burnham and Anderson 2002). $AIC_c$, a second-order variant of AIC which corrects for finite sample size, was computed for each distribution as

$$AIC_c = 2k - 2\ln(L) + \frac{2k(k+1)}{n-k-1} \qquad \text{(Eqn E5)}$$

where $k$ is the number of parameters in the corresponding distribution, $n$ is the number of individuals in the community, and $L$ is the likelihood of the distribution across all individuals (Burnham and Anderson 2002). Within a community, the distribution with a lower $AIC_c$ value provides a better fit.

Our results show that overall the Weibull distribution provides the best fit for the ISD, which outperforms the other three distributions (i.e., has the smallest $AIC_c$ value) in 50 out of 60 communities. While METE is exceeded by the Weibull distribution in all except 3 communities, its performance is comparable to that of the other two distributions, with METE outperforming the exponential distribution in 24 communities and the Pareto distribution in 33 (Table E1).

**Table E1.** The $AIC_c$ value of the four distributions of ISD across communities. The distribution with the best fit (lowest $AIC_c$ value) for each community is in red.

| Dataset | Site | $AIC_c$ -exponential | $AIC_c$-Pareto | $AIC_c$ -Weibull | $AIC_c$ -METE |
|---|---|---|---|---|---|
| FERP | FERP | 85971.15 | 82823.11 | **81893.76** | 88390.74 |
| ACA | eno-2 | 3047.892 | 3123.951 | **3037.737** | 3048.544 |
| WesternGhats | BSP104 | 8447.378 | 8232.82 | **8147.375** | 8597.933 |
| WesternGhats | BSP11 | 9670.786 | 9737.739 | **9565.319** | 9756.008 |
| WesternGhats | BSP12 | 8072.348 | 7580.985 | **7580.105** | 8005.097 |
| WesternGhats | BSP16 | 6505.854 | 6465.984 | **6371.536** | 6473.227 |
| WesternGhats | BSP27 | 4158.854 | 4352.934 | **4154.657** | 4168.587 |
| WesternGhats | BSP29 | 5200.085 | 5601.832 | **5186.167** | 5246.872 |
| WesternGhats | BSP30 | **5228.032** | 5550.478 | 5229.22 | 5272.148 |
| WesternGhats | BSP36 | 5363.257 | 4997.568 | **4994.507** | 5613.485 |
| WesternGhats | BSP37 | 6648.723 | **5882.951** | 5940.894 | 6702.201 |
| WesternGhats | BSP42 | 4862.353 | 4579.541 | **4572.774** | 4912.597 |
| WesternGhats | BSP5 | 6316.684 | **5868.932** | 5879.056 | 6344.512 |
| WesternGhats | BSP6 | 8362.132 | 8224.467 | **8144.515** | 8368.706 |
| WesternGhats | BSP65 | 10730.14 | 10597.32 | 10418.12 | **10323.55** |
| WesternGhats | BSP66 | 6127.039 | 6078.716 | **5969.159** | 6118.758 |
| WesternGhats | BSP67 | 5733.979 | 6116.641 | **5713.447** | 5970.901 |
| WesternGhats | BSP69 | 9639.039 | 9839.743 | **9566.506** | 9677.272 |
| WesternGhats | BSP70 | 7568.366 | 7643.62 | 7475.877 | **7471.337** |
| WesternGhats | BSP73 | **13866.8** | 14638.34 | 13867.97 | 14056.6 |
| WesternGhats | BSP74 | 10384.88 | 10164.99 | **10043.66** | 10178.07 |
| WesternGhats | BSP75 | **3828.718** | 4032.776 | 3830.225 | 3844.366 |
| WesternGhats | BSP79 | 10012.15 | 10192.38 | **9943.069** | 10014.63 |
| WesternGhats | BSP80 | 10351.04 | 10721.97 | **10333.53** | 10392.1 |
| WesternGhats | BSP82 | 7775.241 | 8109.038 | **7766.727** | 7779.842 |
| WesternGhats | BSP83 | **10080.84** | 10603.67 | 10082.84 | 10184.62 |
| WesternGhats | BSP84 | 9941.77 | 10676.22 | **9906.56** | 10087.81 |
| WesternGhats | BSP85 | 4090.759 | 4051.023 | **3986.417** | 4092.965 |
| WesternGhats | BSP88 | 9539.878 | 10007.25 | 9532.9 | **9468.538** |
| WesternGhats | BSP89 | 7758.469 | 8040.773 | **7746.257** | 7749.632 |
| WesternGhats | BSP90 | 7802.77 | 8287.765 | **7800.707** | 7891.673 |
| WesternGhats | BSP91 | 8443.673 | 9081.623 | **8392.871** | 8709.277 |
| WesternGhats | BSP92 | 5010.321 | 5156.128 | **4980.47** | 5037.136 |
| WesternGhats | BSP94 | 4995.435 | 5113.566 | **4949.09** | 4997.738 |
| WesternGhats | BSP98 | 6338.305 | 6535.699 | **6312.535** | 6336.033 |
| WesternGhats | BSP99 | 8329.191 | 8461.831 | **8238.427** | 8268.363 |
| BCI | bci | 1663761 | 1595835 | **1580094** | 1616953 |
| BVSF | BVPlot | 2801.075 | 2851.043 | **2790.895** | 2792.688 |
| BVSF | SFPlot | 2452.828 | 2427.723 | **2409.388** | 2413.466 |
| Cocoli | cocoli | 73752.32 | 68152.93 | **67835.59** | 75938.32 |
| Lahei | heath1 | 9947.228 | 9966.227 | **9841.178** | 9888.052 |
| Lahei | heath2 | 9795.598 | 9650.197 | **9595.179** | 9618.001 |
| Lahei | peat | 9183.332 | 9040.189 | **8961.699** | 9030.188 |
| LaSelva | 1 | 5518.14 | 5434.672 | **5376.494** | 5555.8 |
| LaSelva | 2 | 5504.011 | 5548.332 | **5444.005** | 5489.366 |
| LaSelva | 3 | 6337.174 | 6328.63 | **6237.519** | 6294.73 |
| LaSelva | 4 | 5445.745 | 5527.303 | **5402.815** | 5409.85 |
| LaSelva | 5 | 4410.166 | 4318.777 | **4281.463** | 4440.427 |
| Luquillo | lfdp | 534427.2 | 515126.9 | **509926.5** | 525725.7 |
| NC | 12 | 45716.48 | 44860.83 | **44212.08** | 45592.31 |
| NC | 13 | 36251.18 | 34948.55 | **34539.55** | 36220.19 |
| NC | 14 | 56695.06 | 52506.98 | **52273.61** | 55964.15 |
| NC | 4 | 36203.17 | 36553.64 | **35587.05** | 36447.78 |
| NC | 93 | 34667.37 | 33277.48 | **32934.38** | 34730.18 |
| Oosting | Oosting | 74293.18 | 69837.5 | **69718.9** | 74739.21 |
| Serimbu | S-1 | 7887.232 | 7471.463 | **7463.06** | 7981.97 |
| Serimbu | S-2 | 8507.118 | 8123.406 | **8102.843** | 8614.922 |
| Shirakami | Akaishizawa | 3105.173 | 3104.759 | **3057.59** | 3188.967 |
| Shirakami | Kumagera | **3473.692** | 3680.852 | 3473.805 | 3597.692 |
| Sherman | sherman | 191735.8 | 188206 | **185424** | 190339.9 |

**Figure B1.**

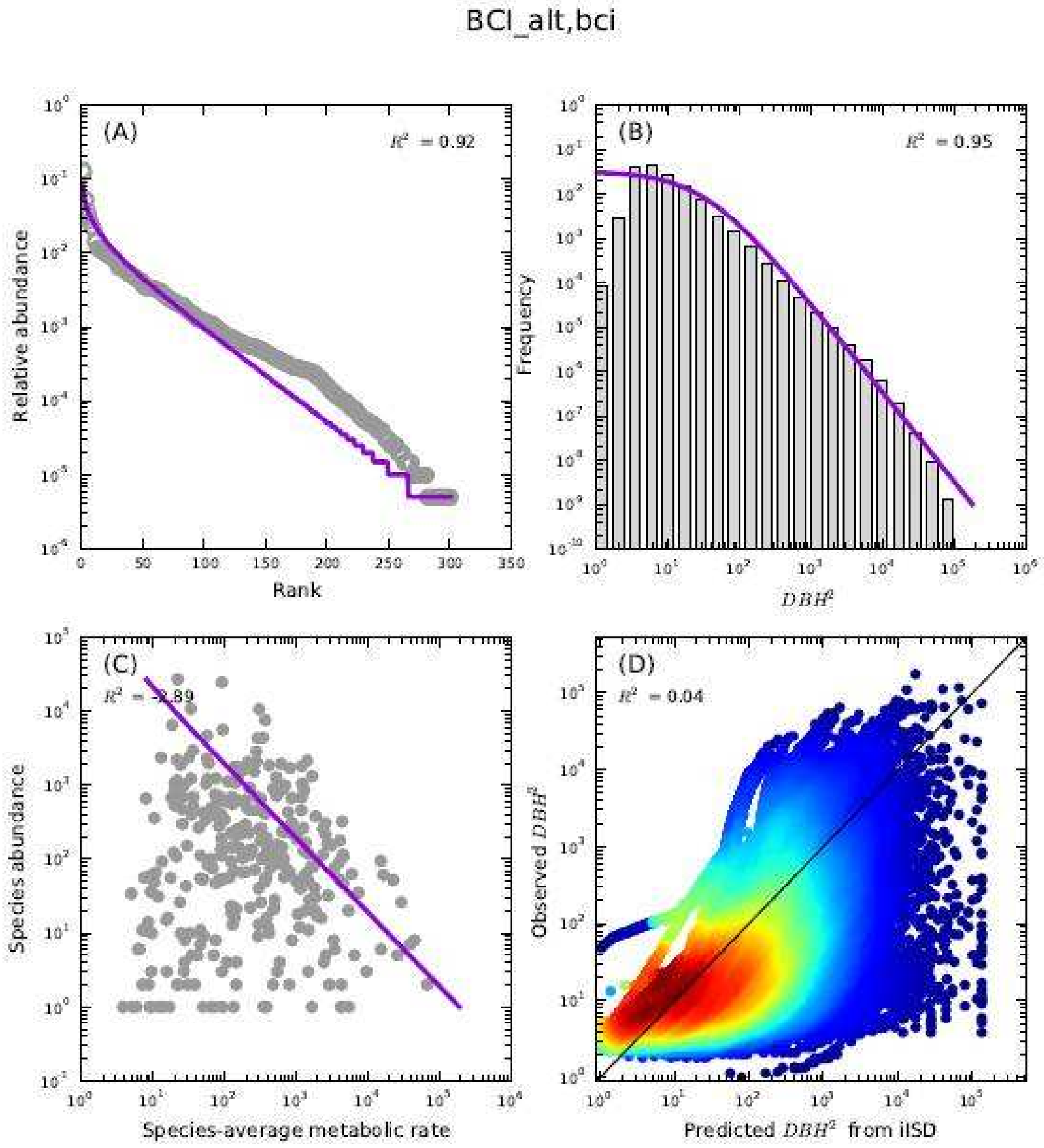

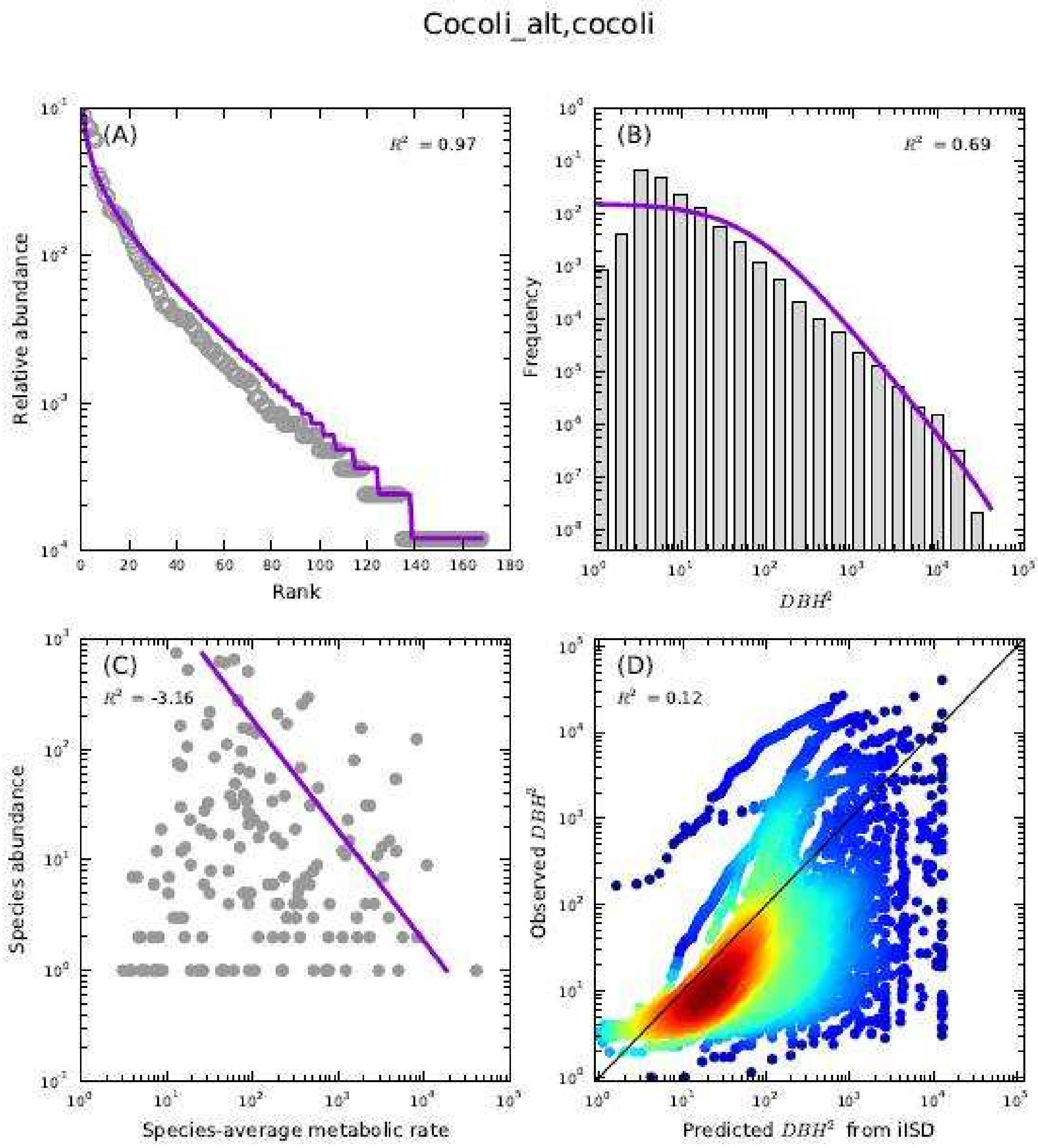

LaSelva_alt,4

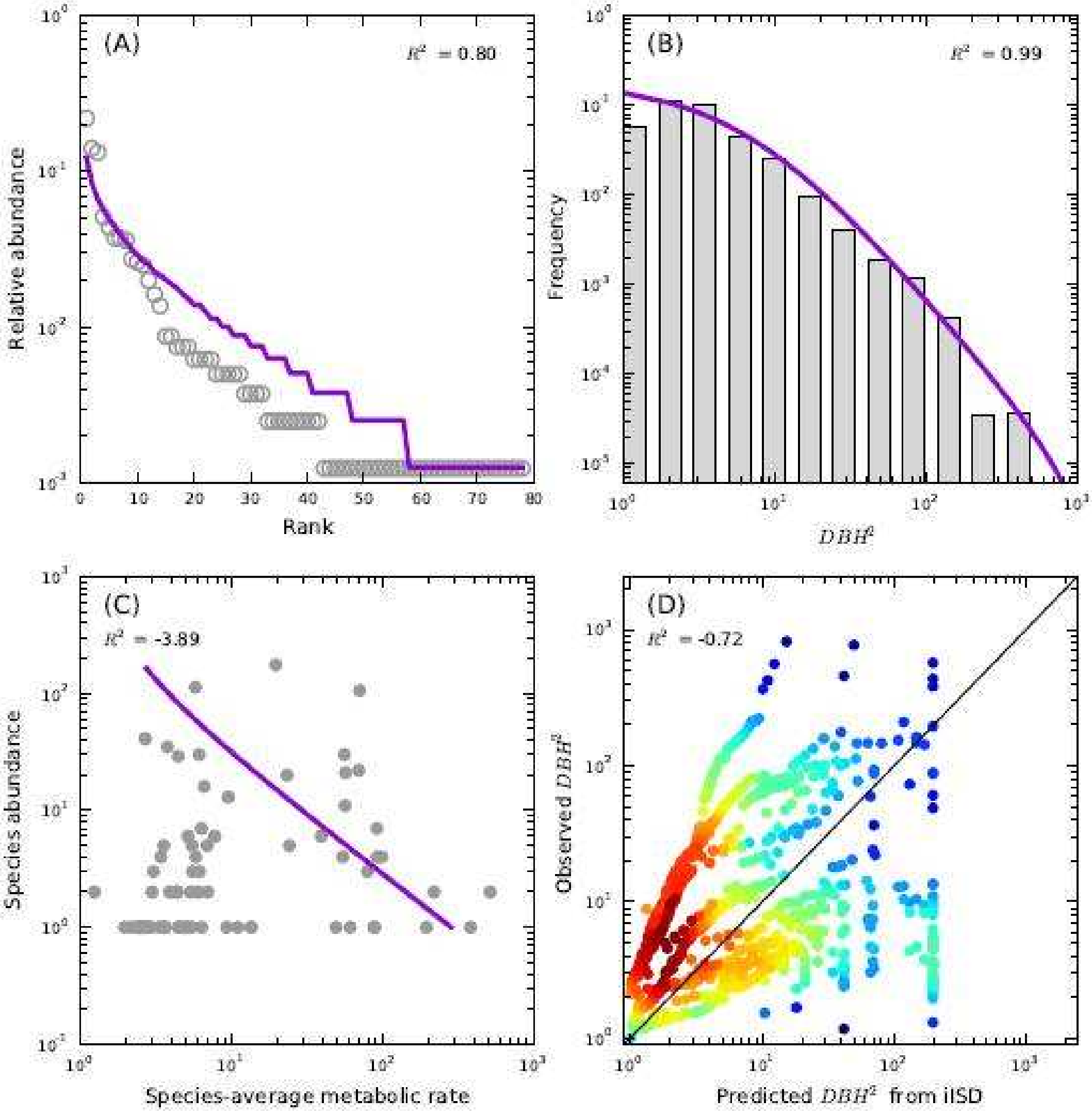

LaSelva_alt,5

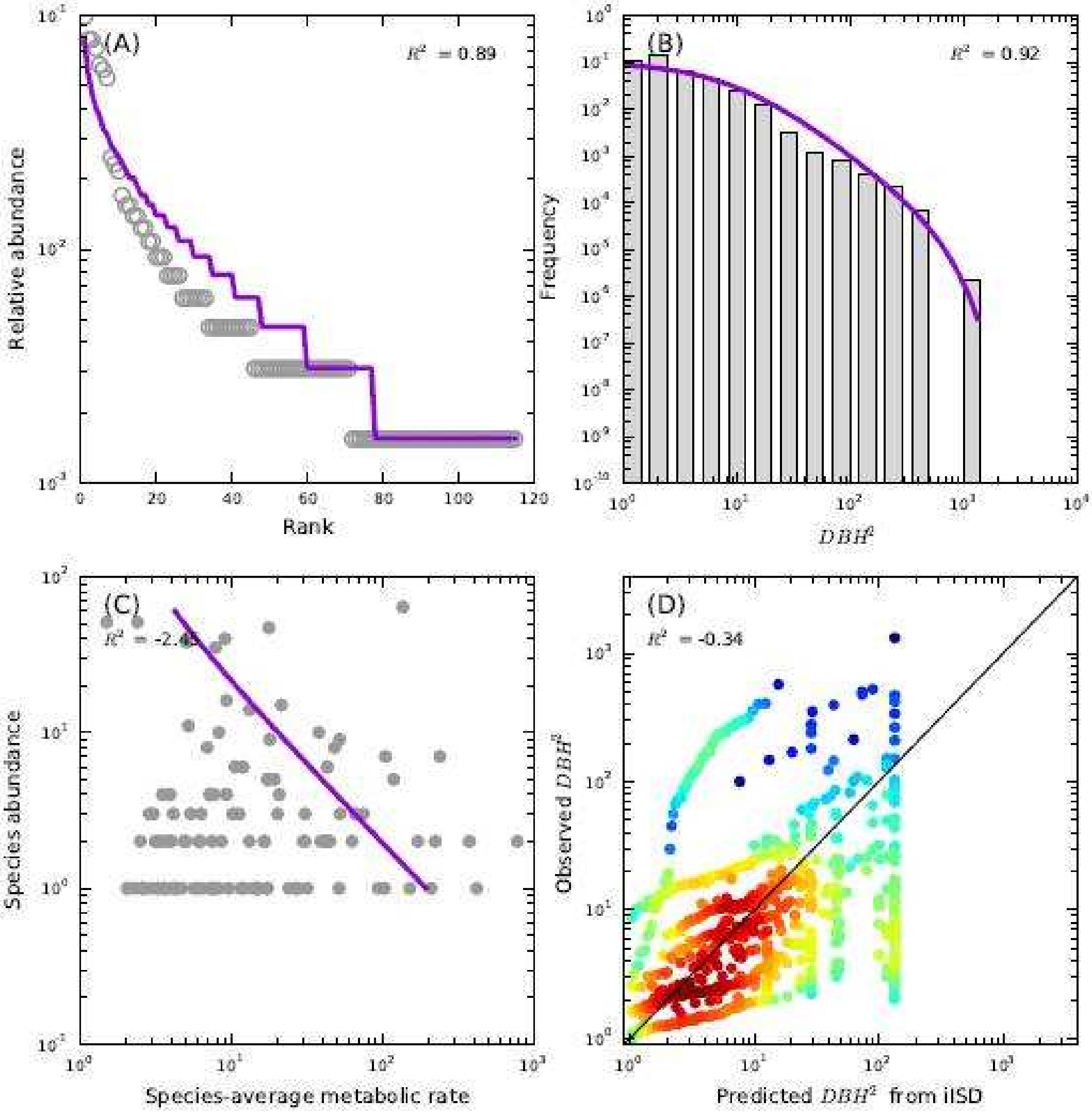

Luquillo_alt,lfdp

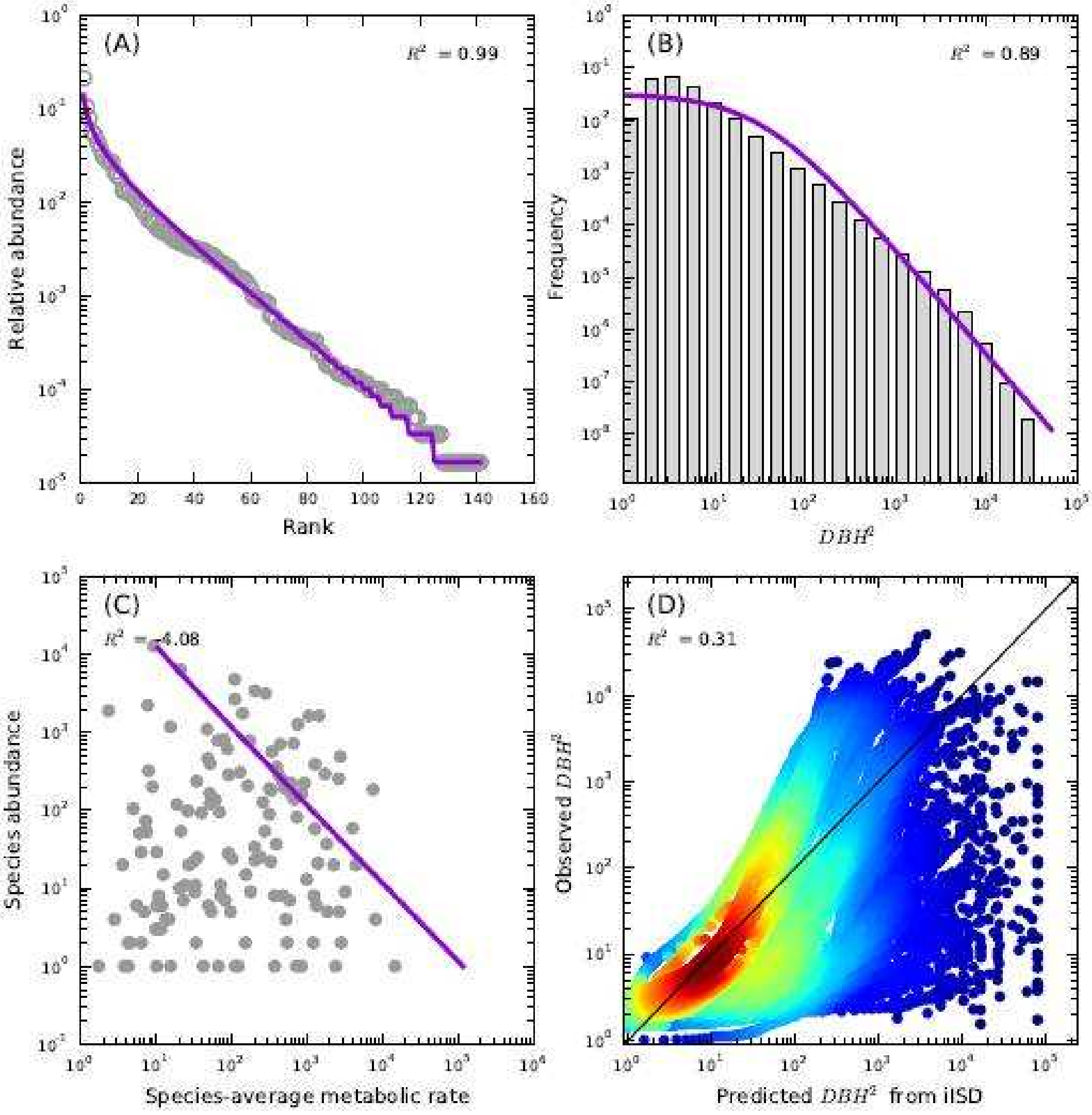

**Figure D1.**

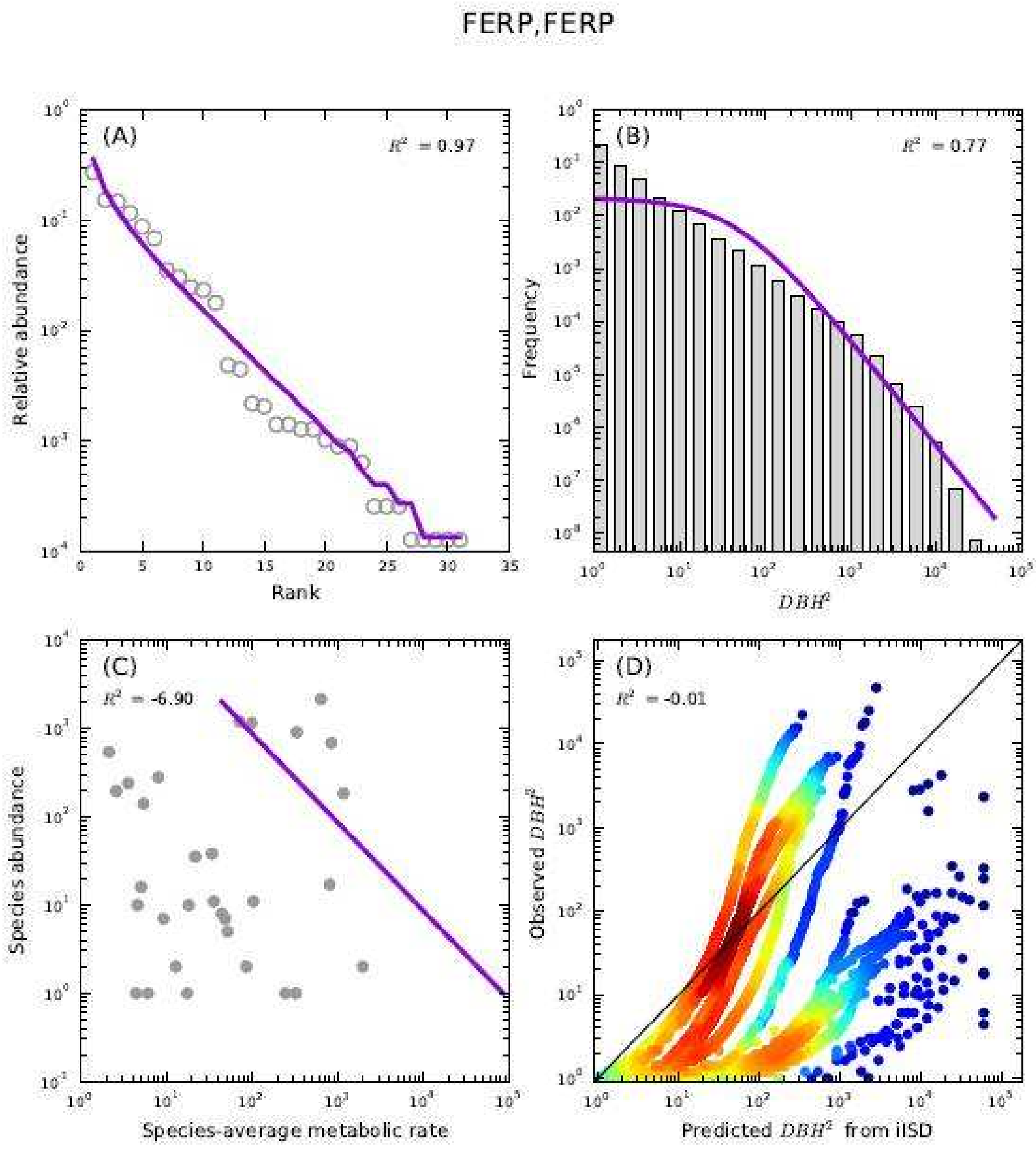

# ACA,eno-2

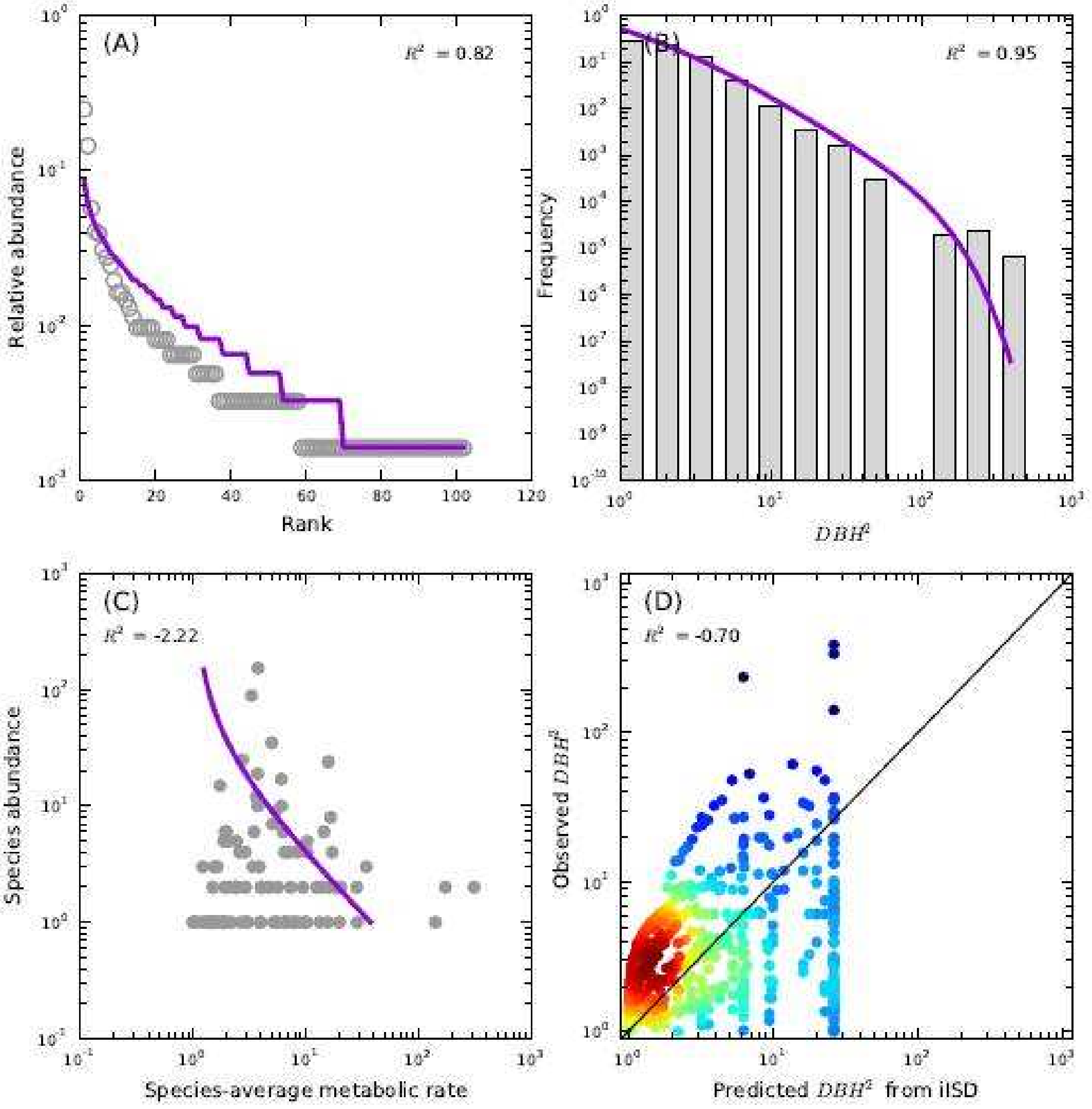

WesternGhats,BSP104

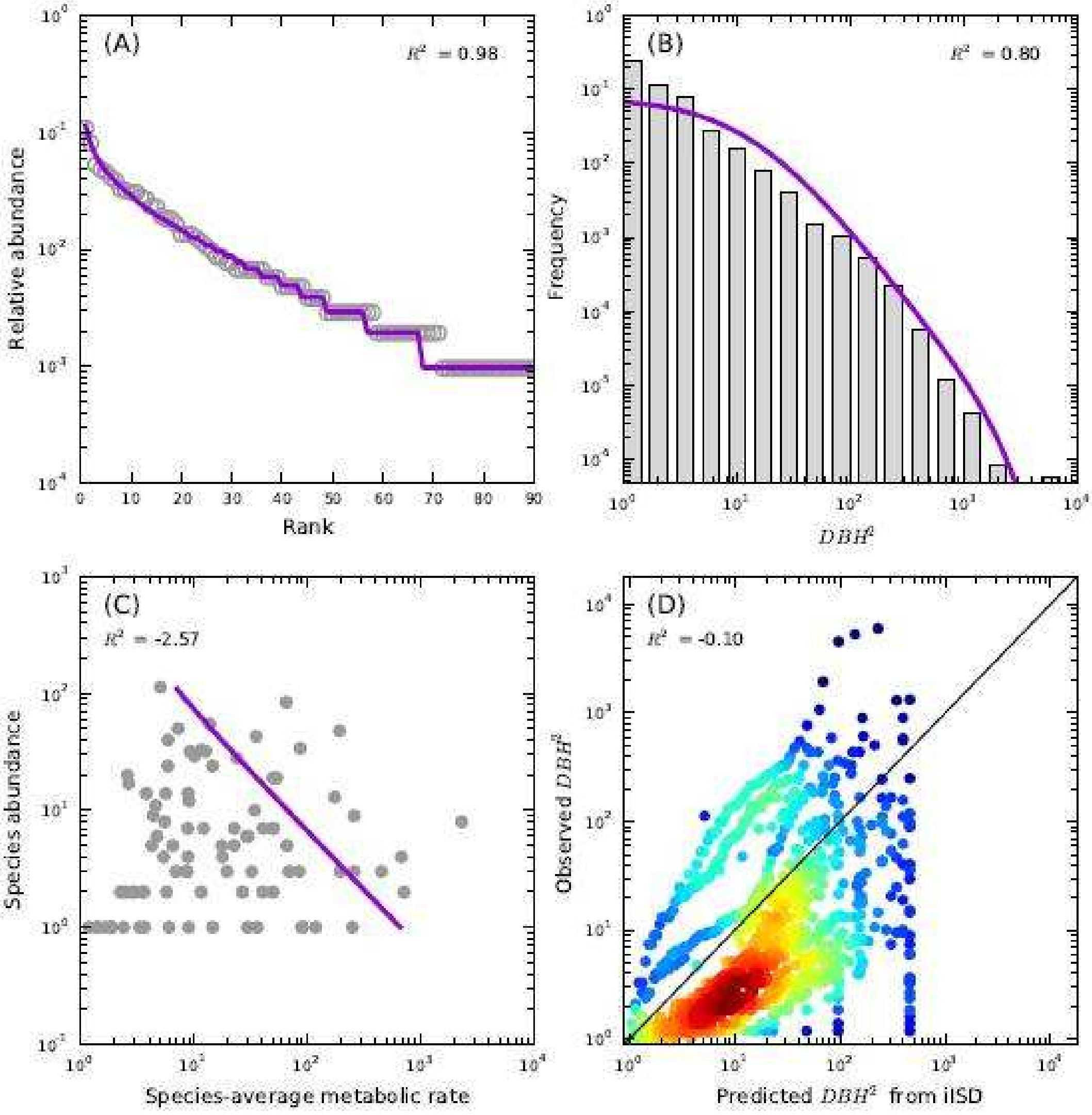

WesternGhats,BSP11

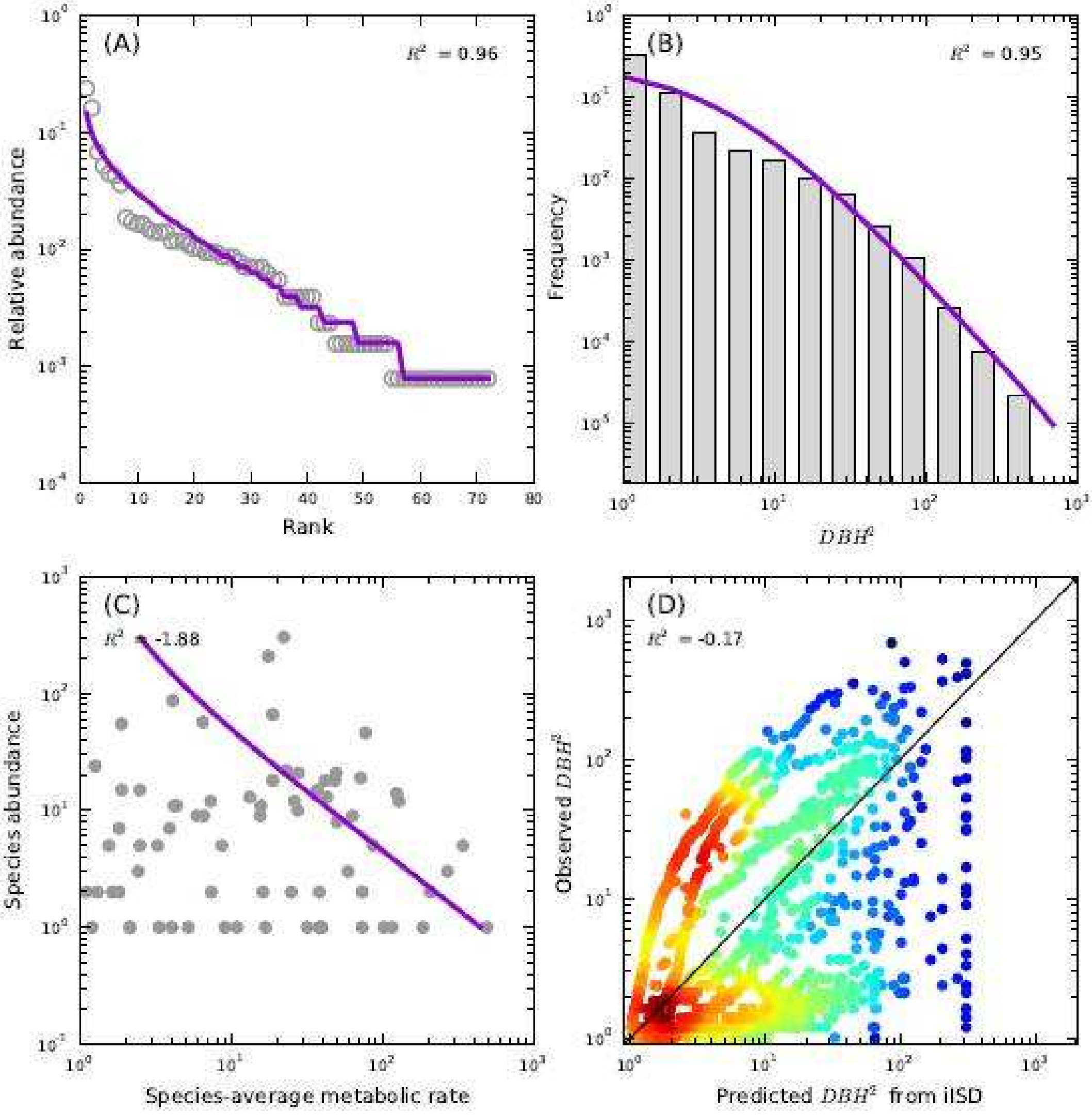

WesternGhats,BSP12

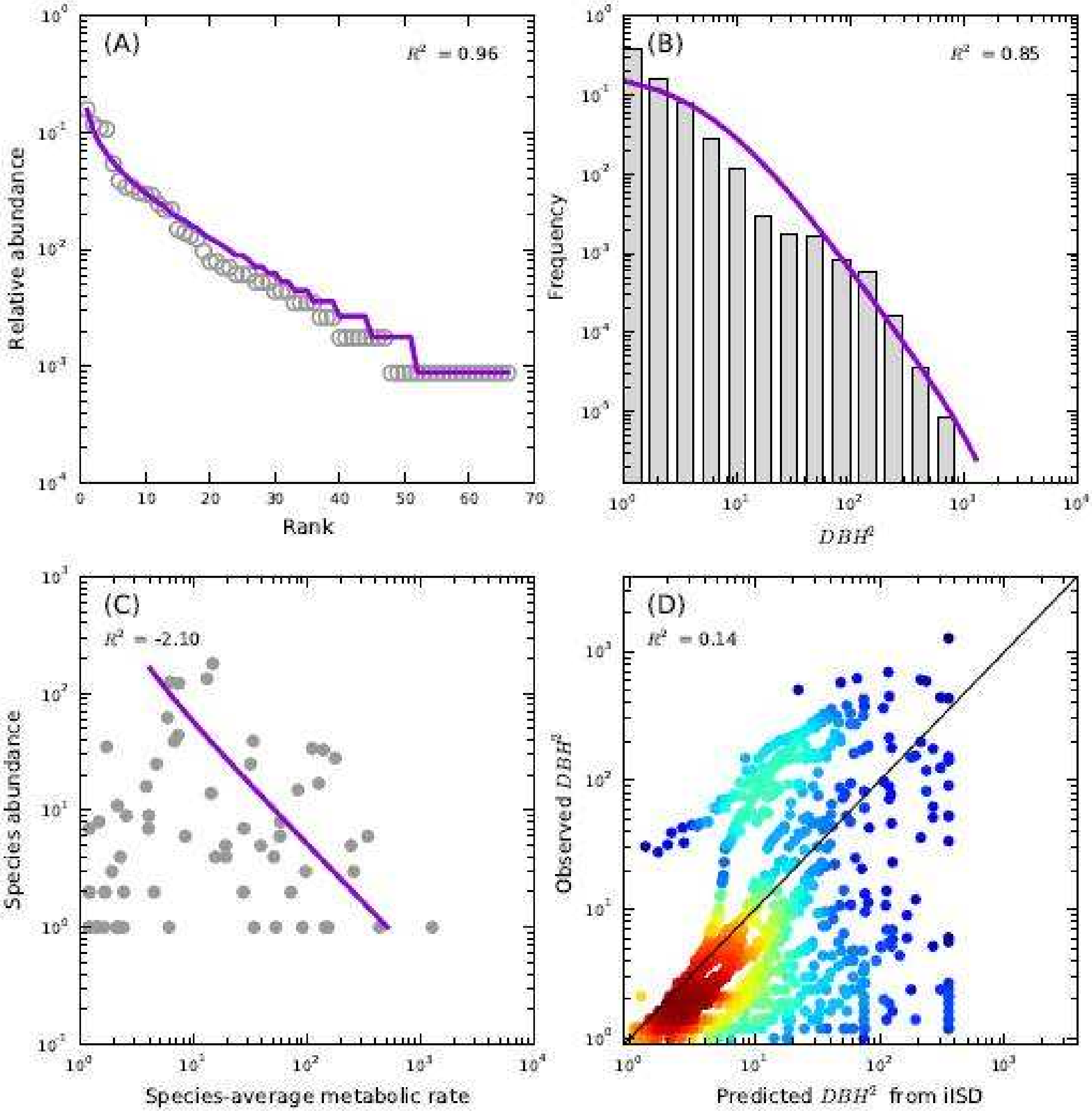

WesternGhats,BSP16

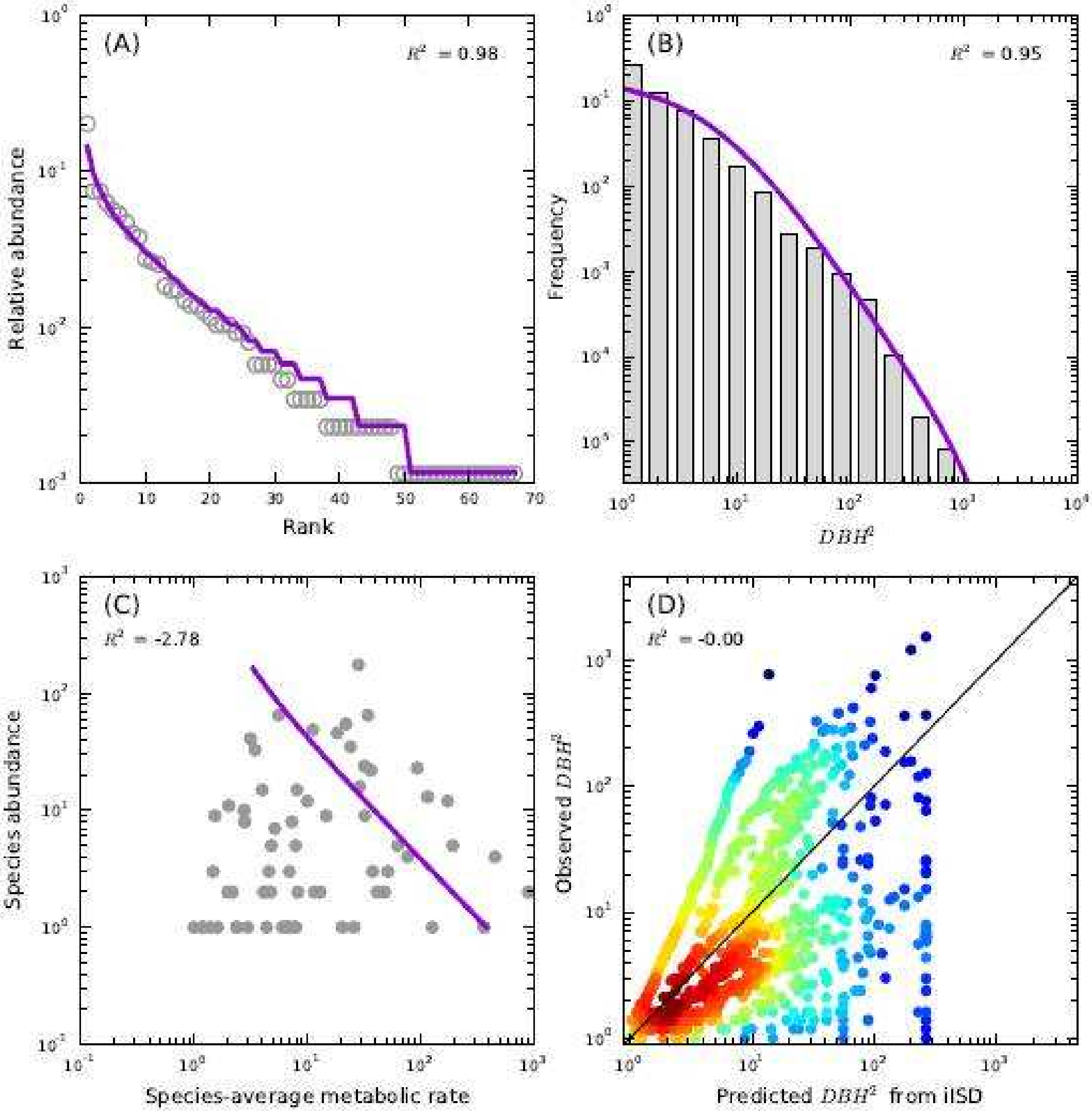

# WesternGhats,BSP27

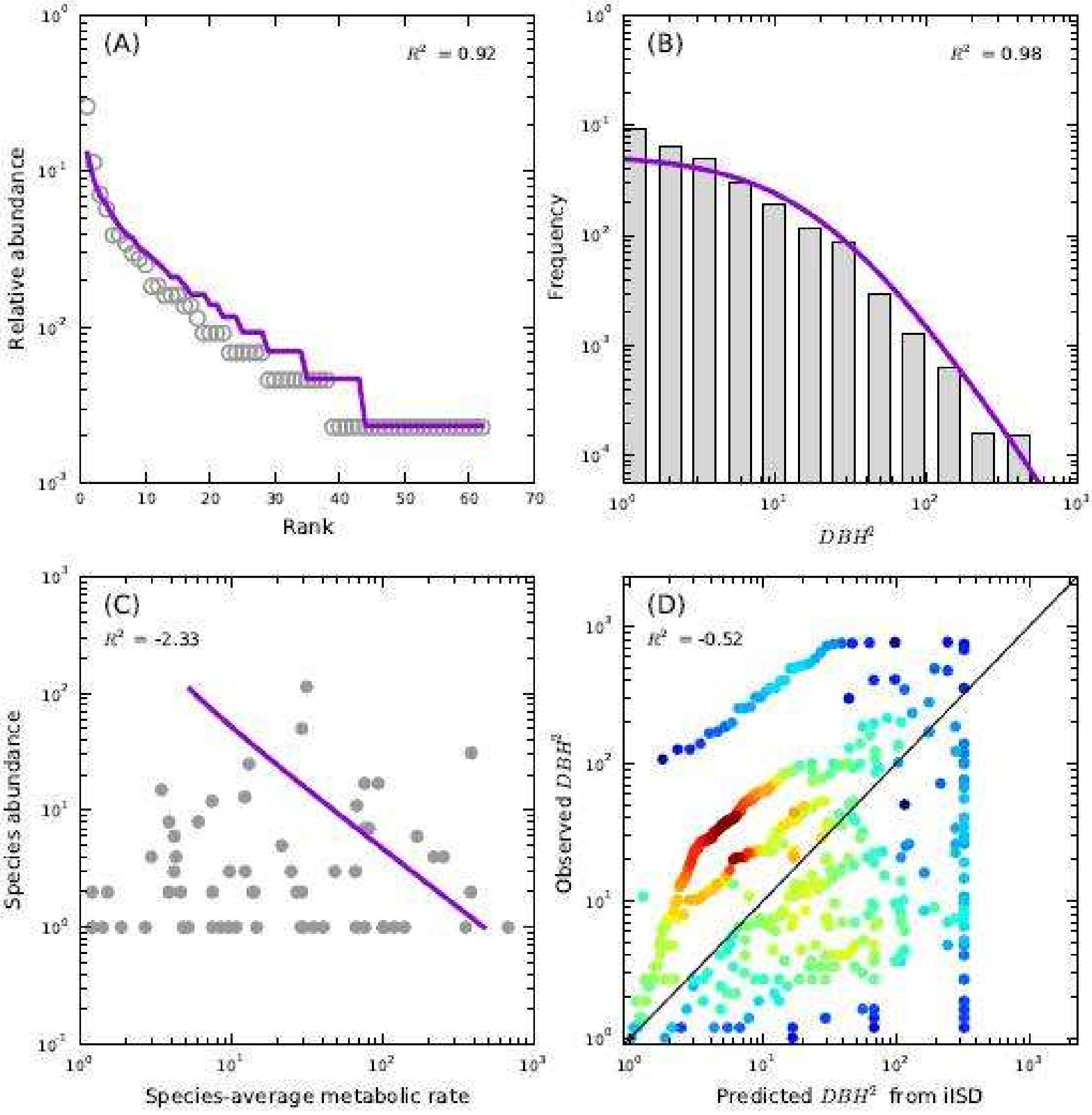

WesternGhats,BSP29

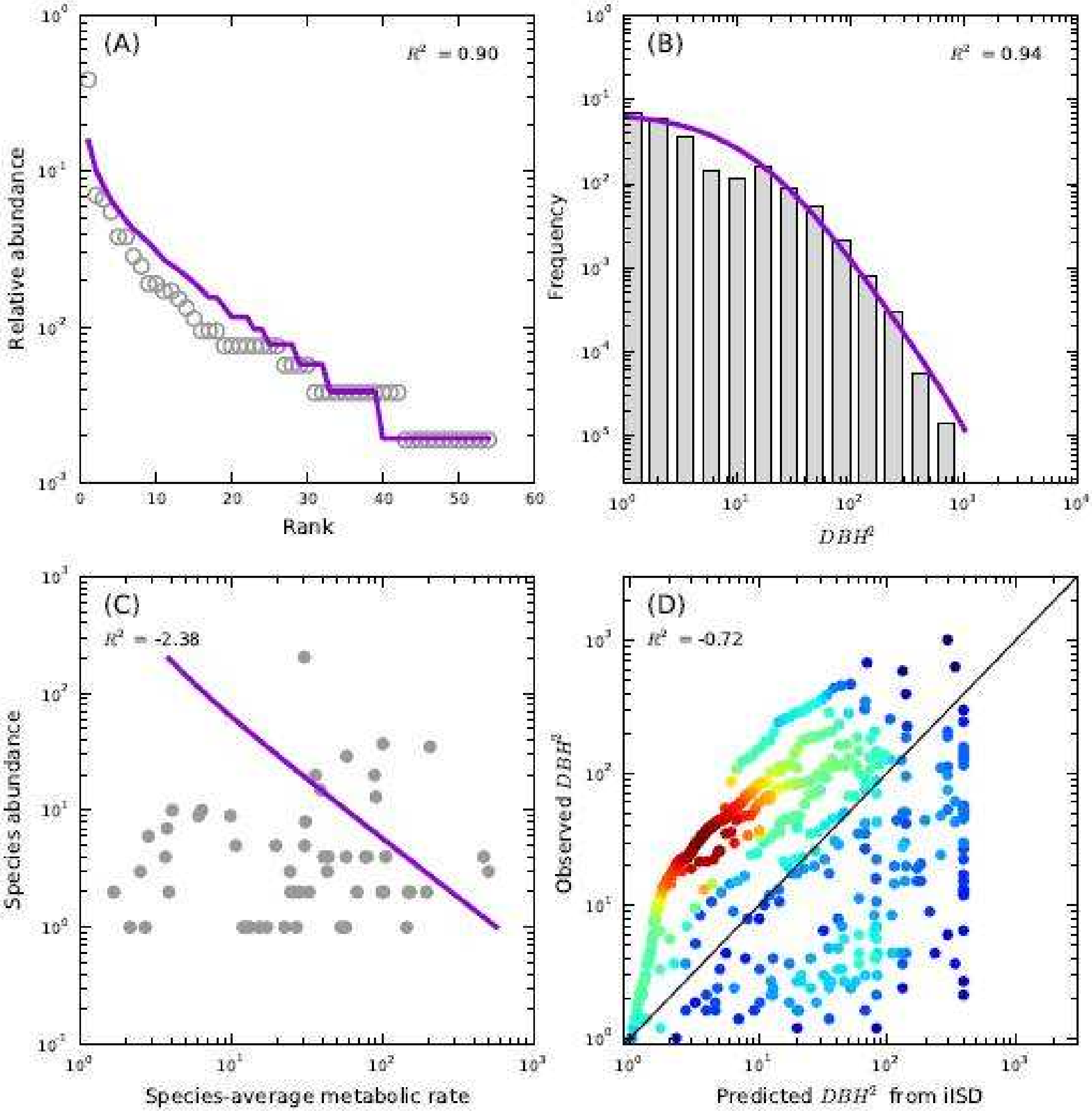

WesternGhats,BSP30

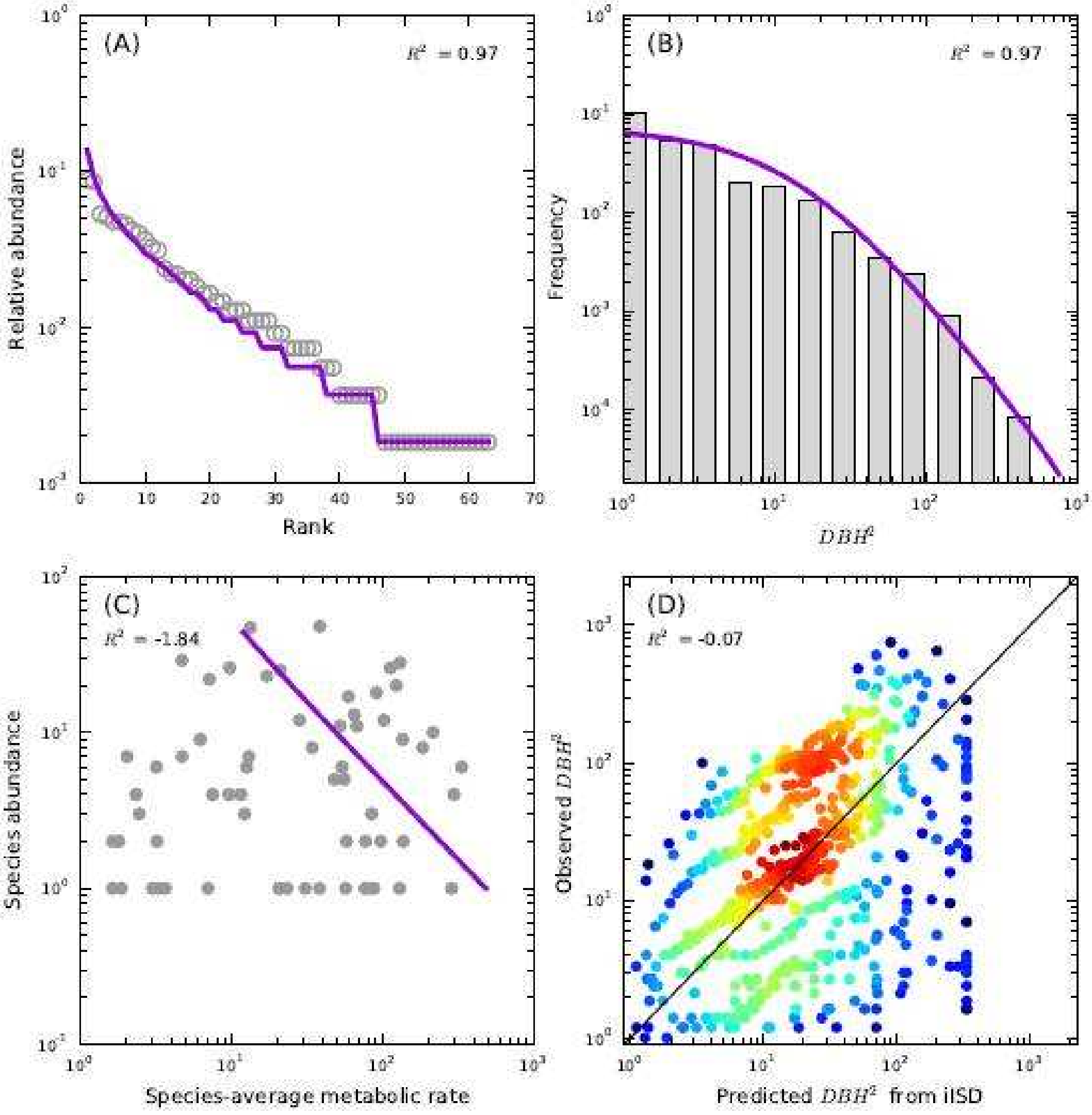

WesternGhats,BSP36

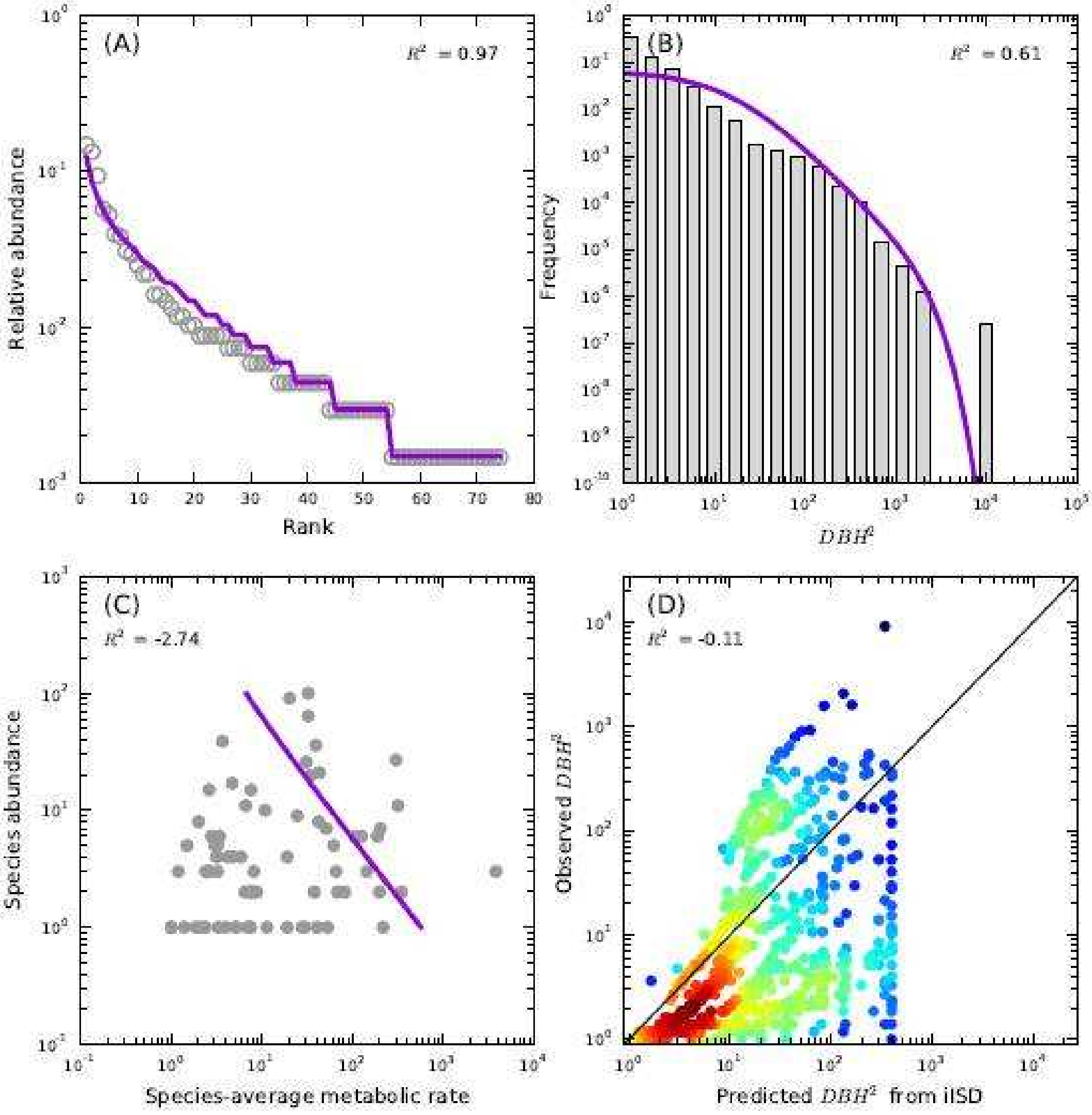

WesternGhats,BSP37

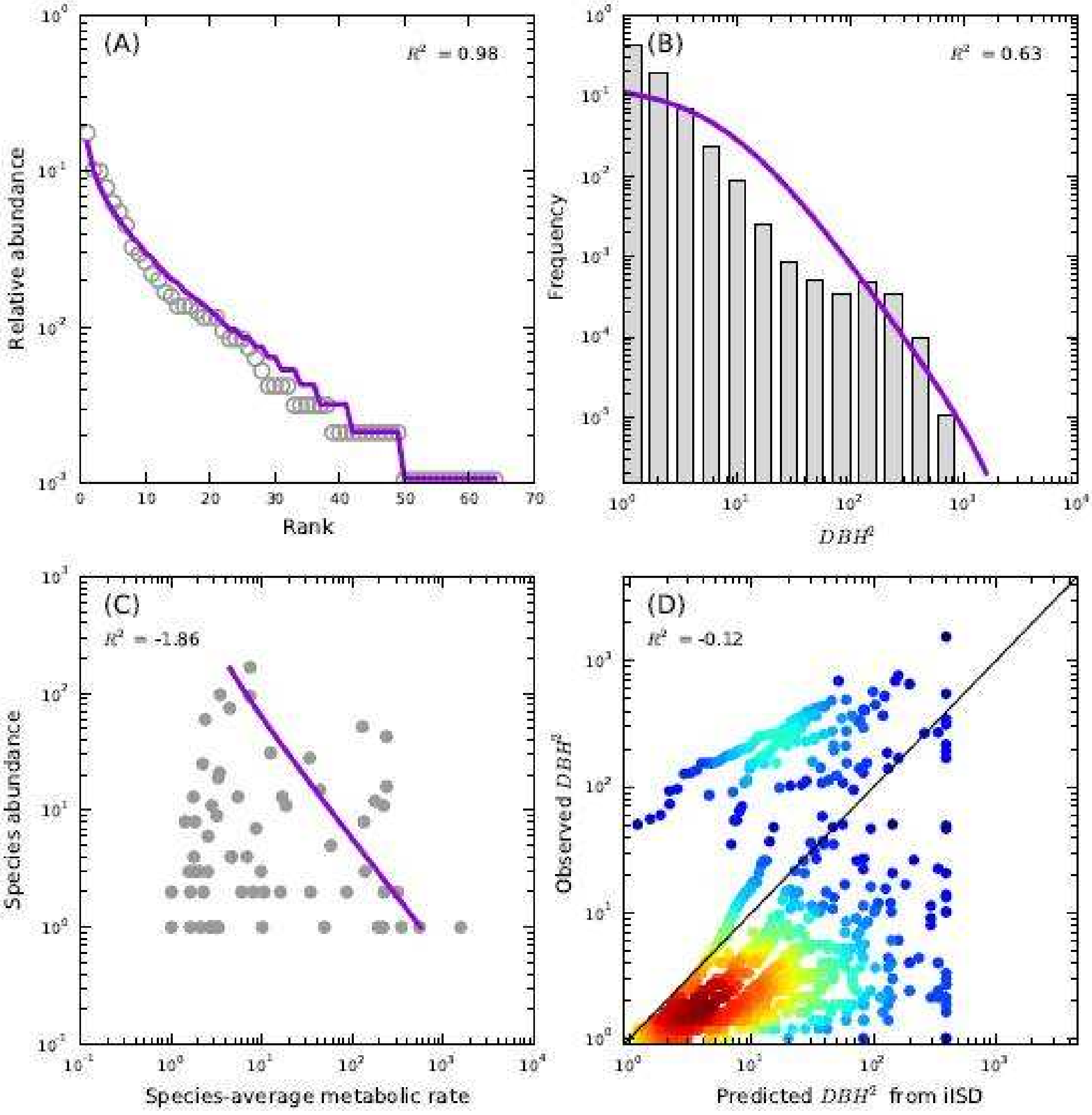

WesternGhats,BSP42

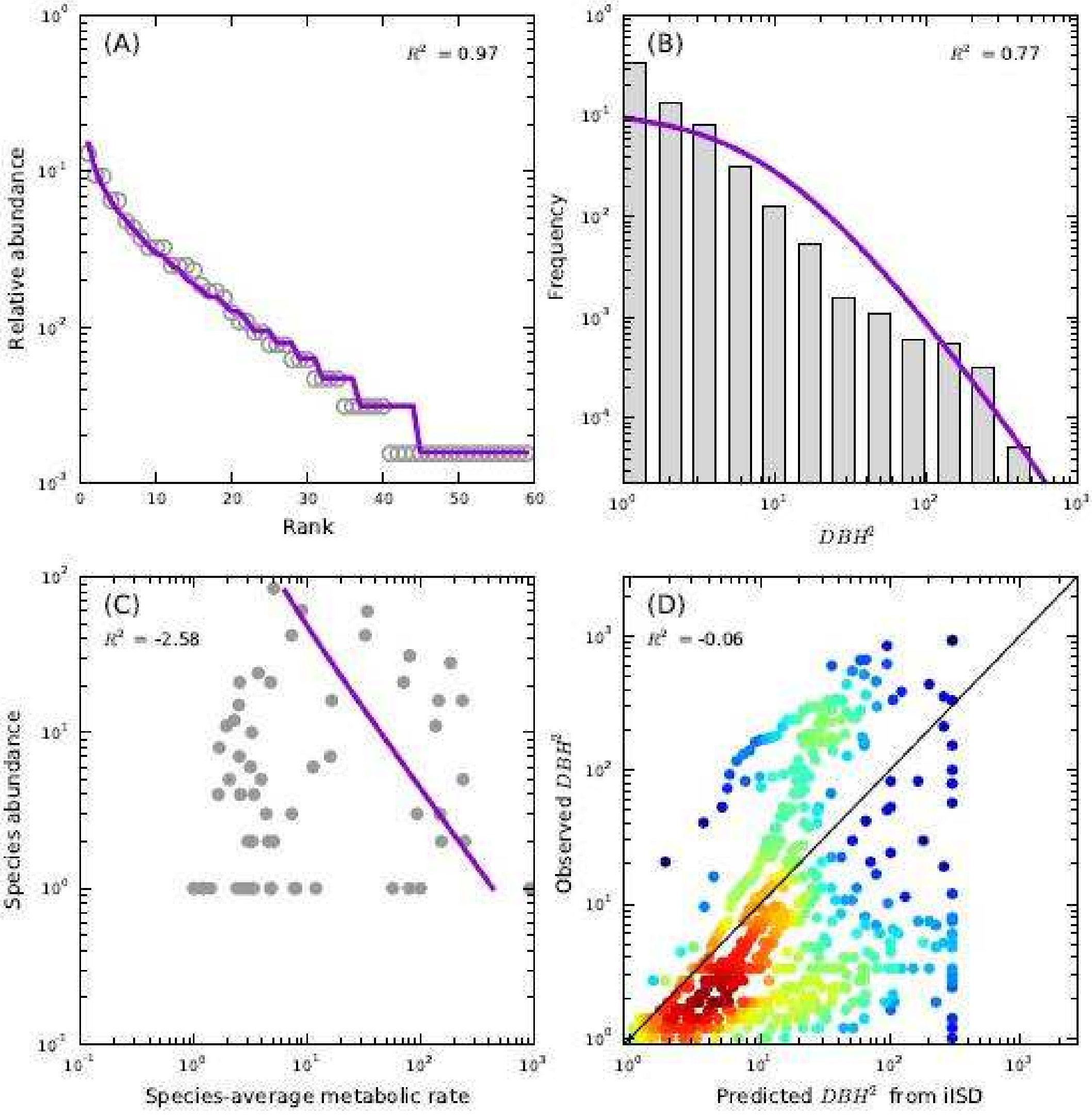

WesternGhats,BSP5

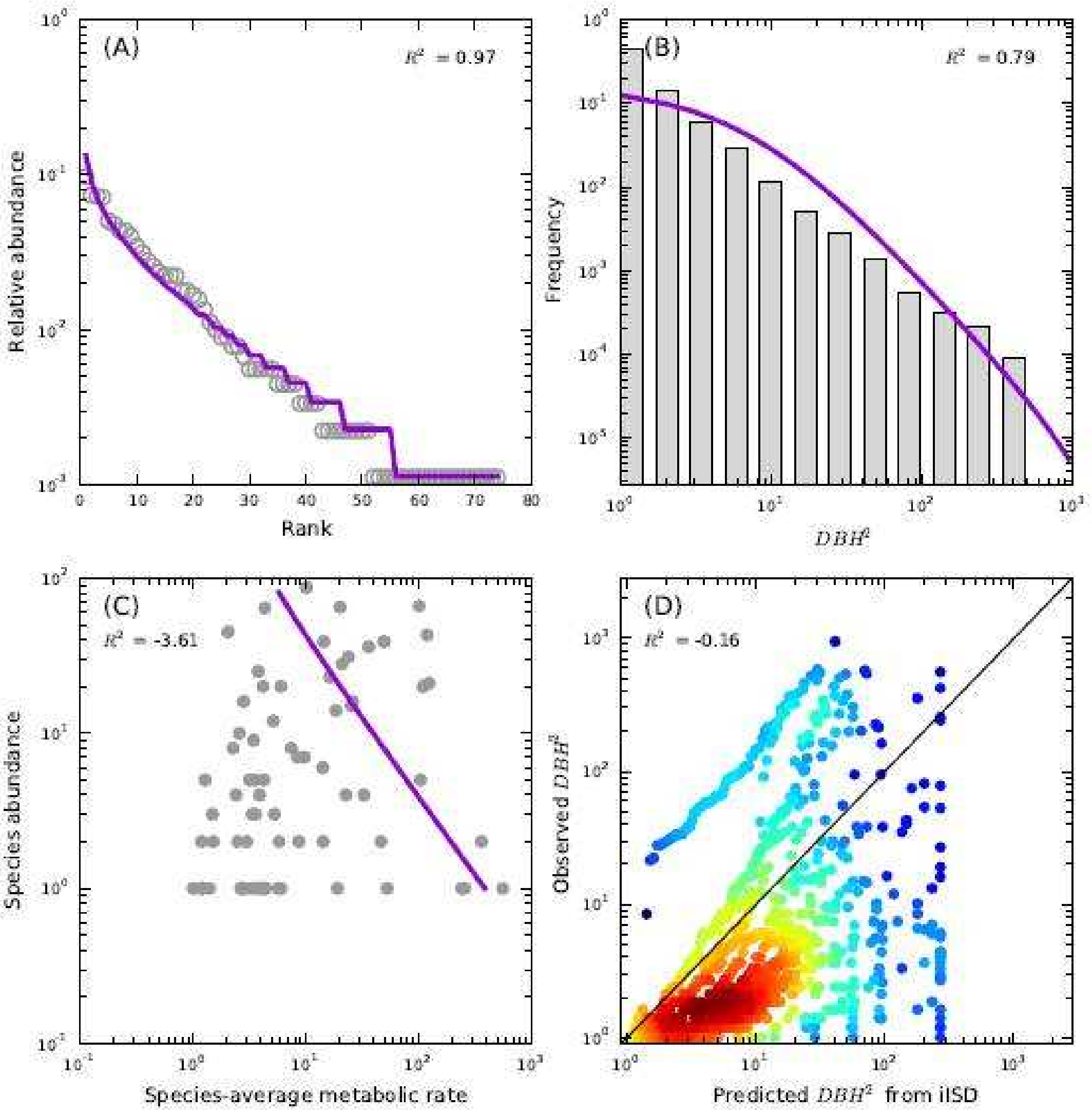

WesternGhats,BSP6

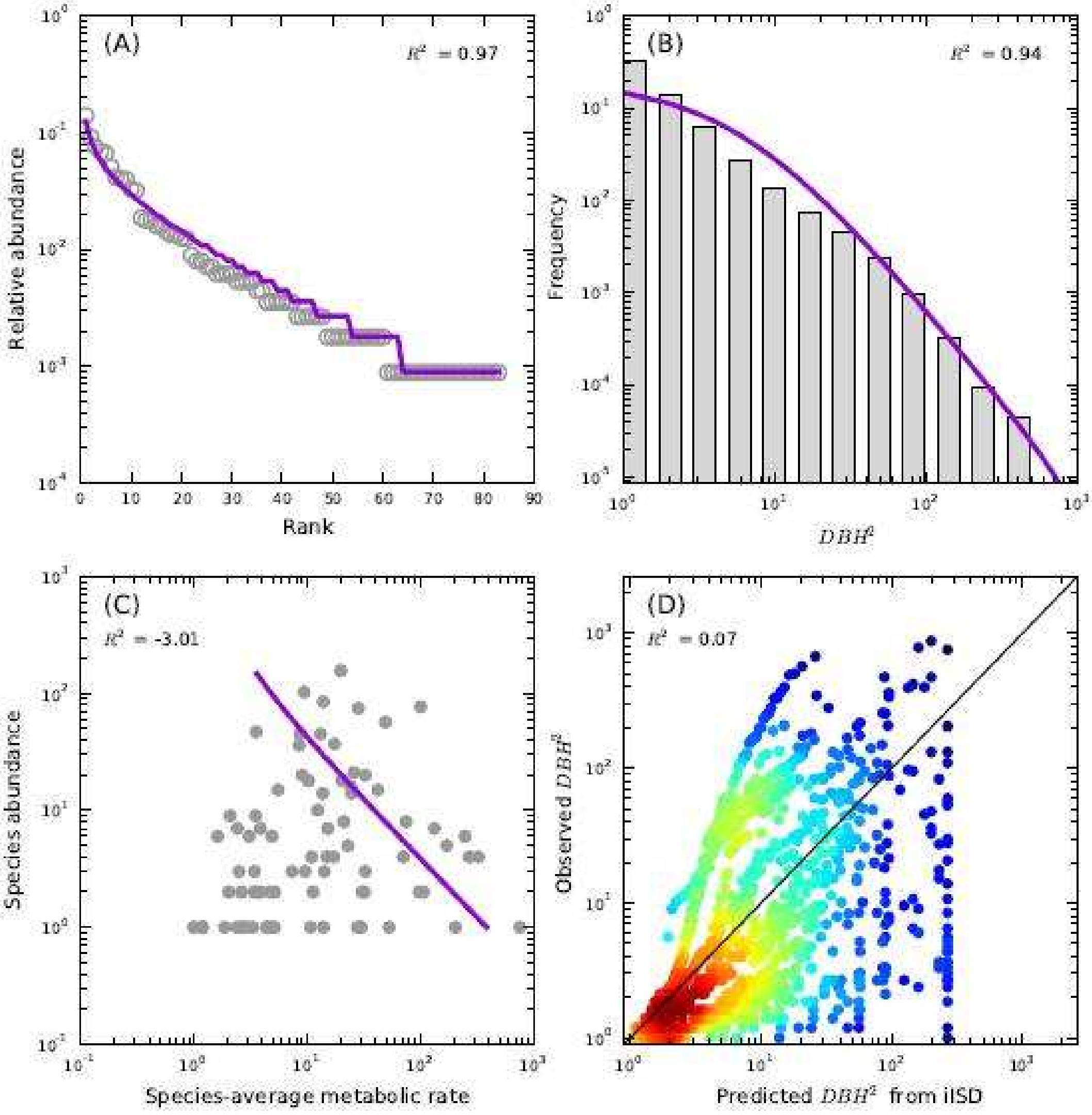

WesternGhats,BSP65

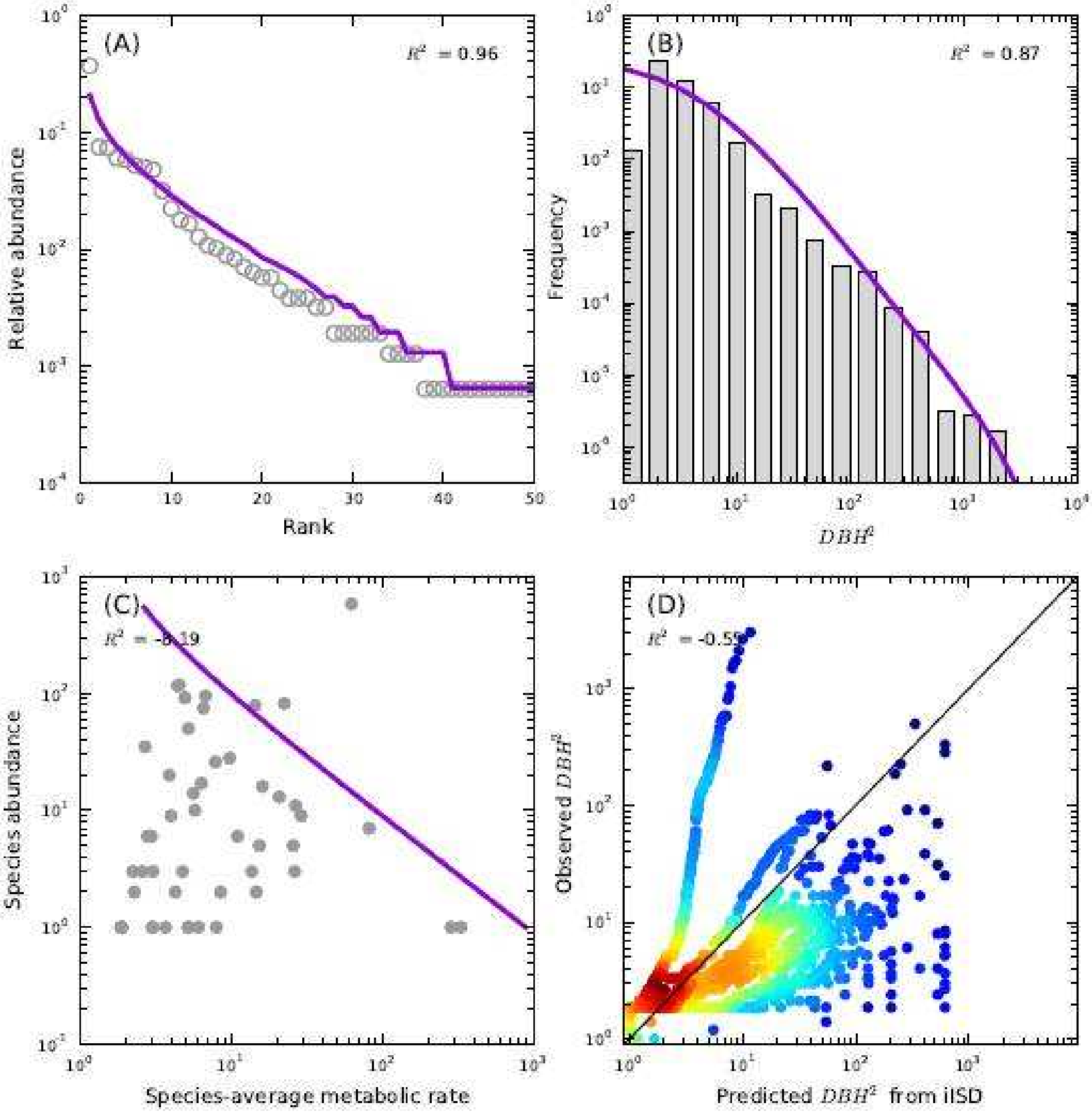

WesternGhats,BSP66

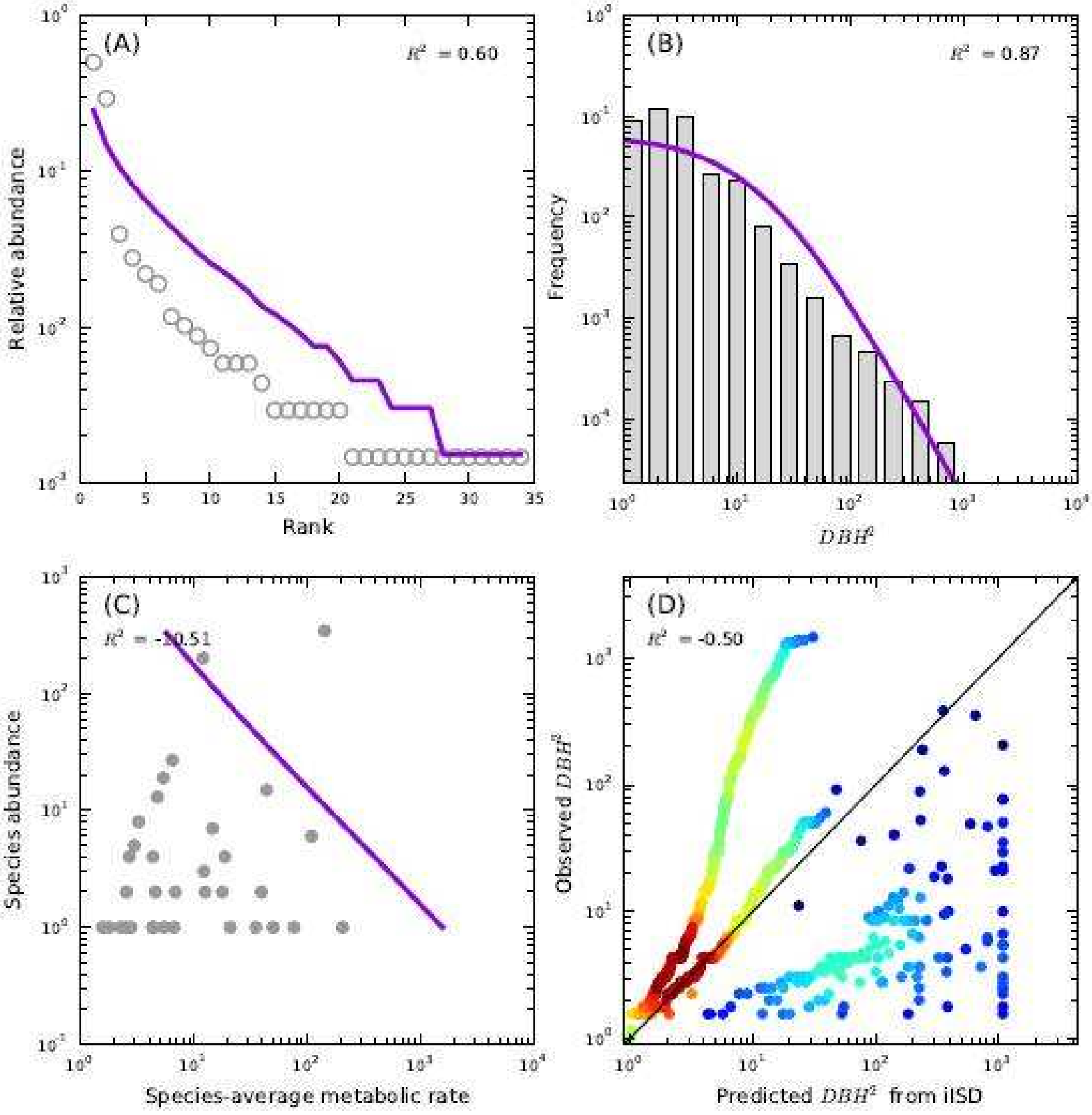

WesternGhats,BSP67

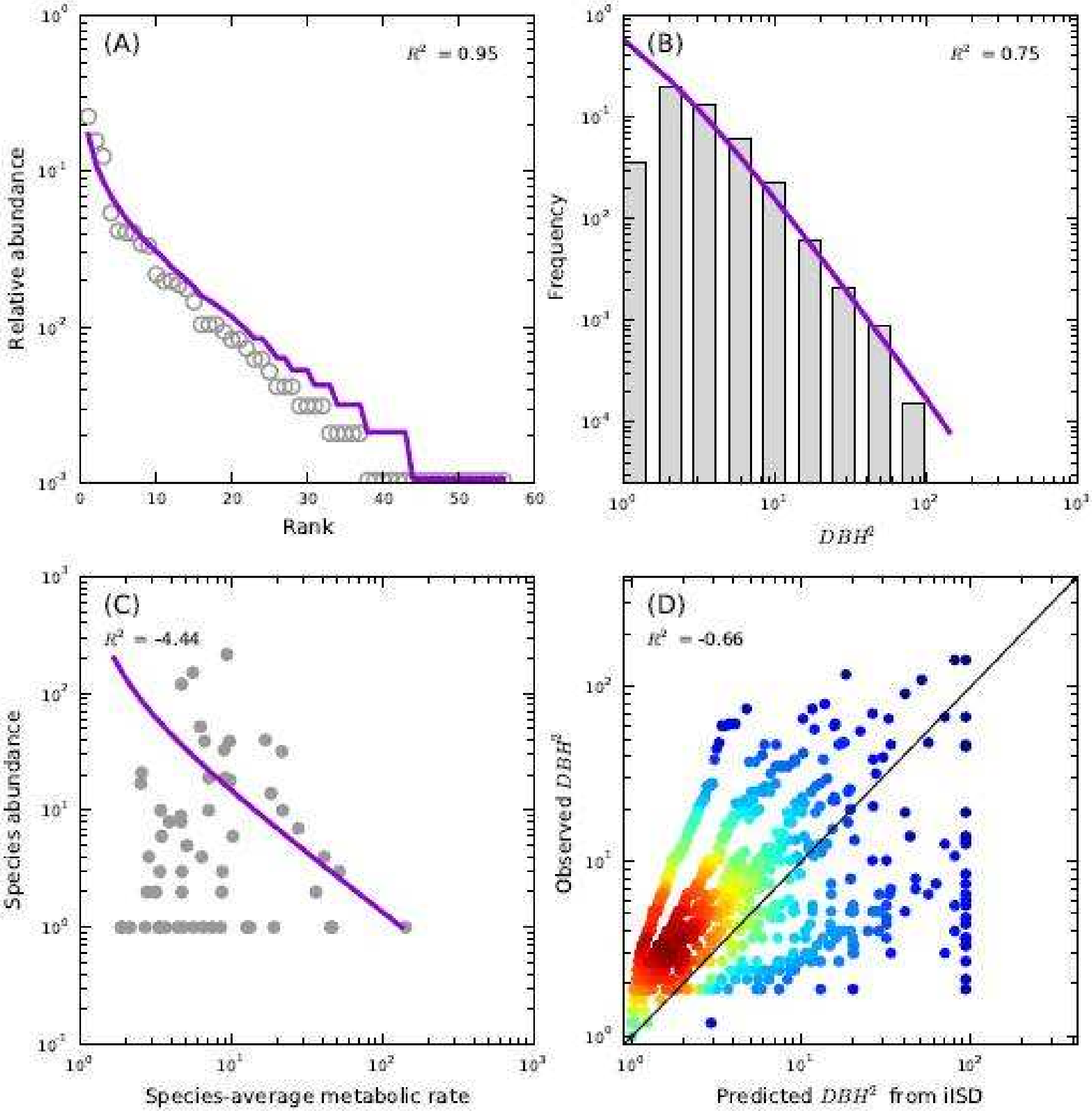

WesternGhats,BSP69

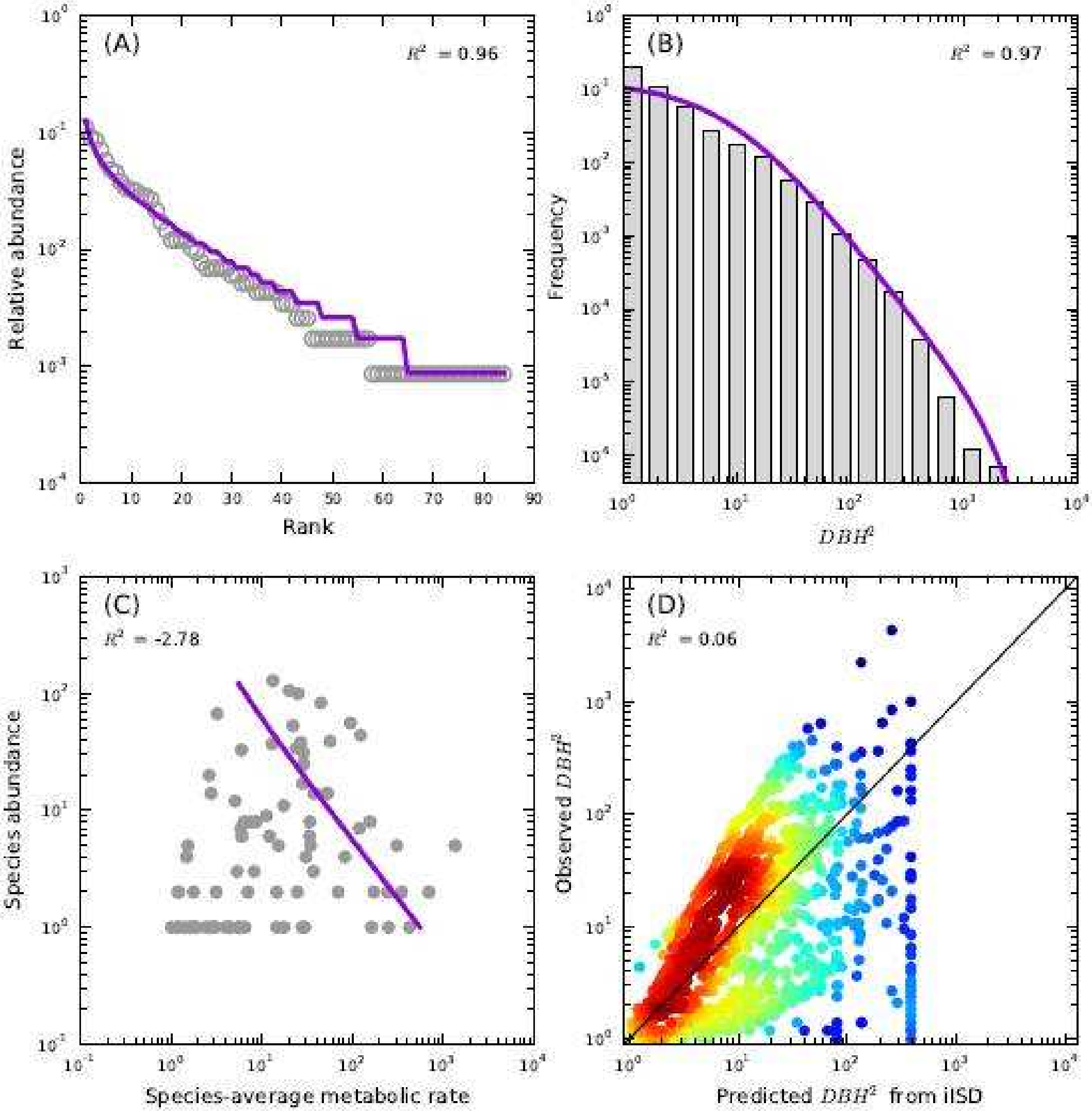

WesternGhats,BSP70

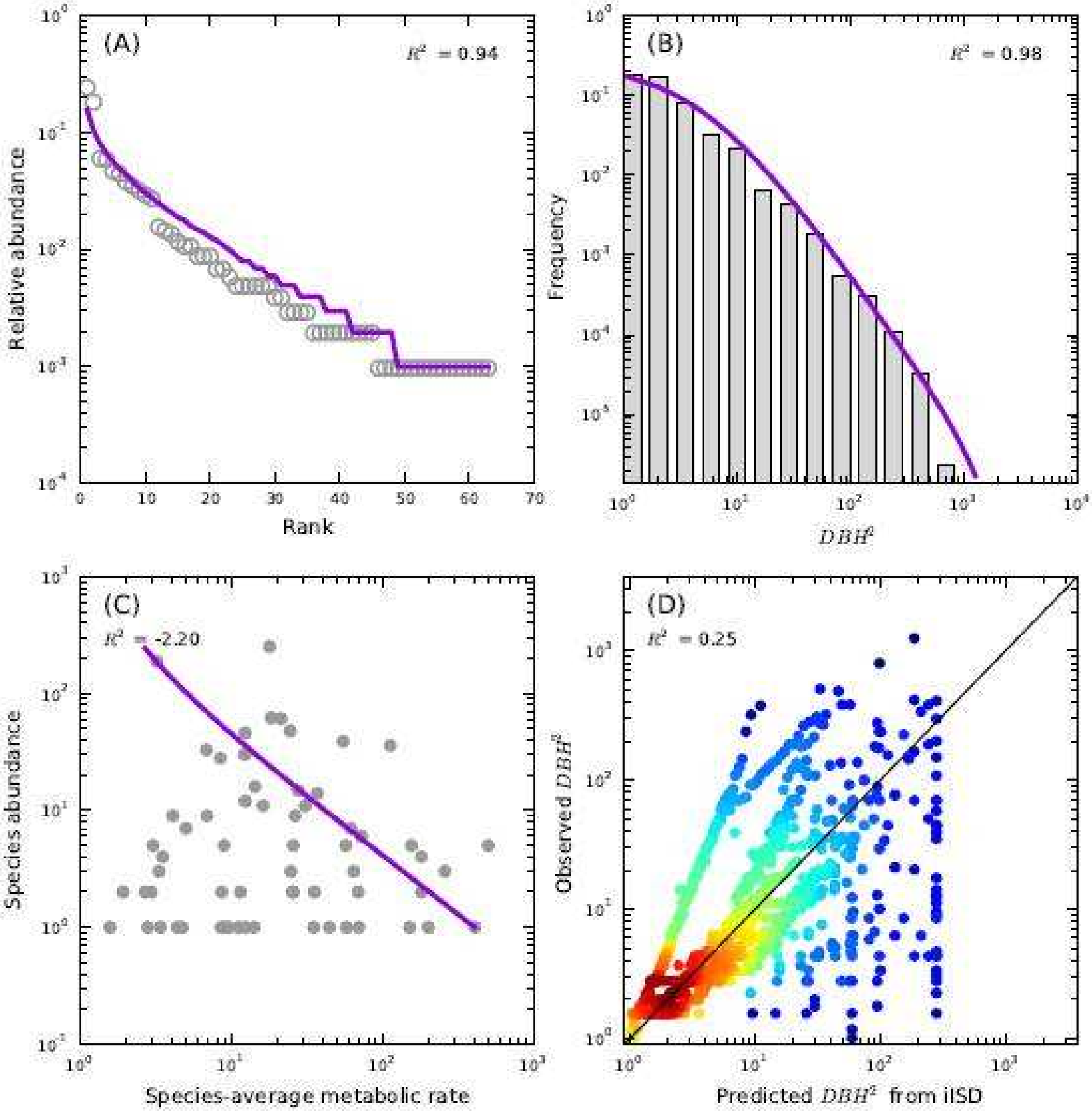

WesternGhats,BSP73

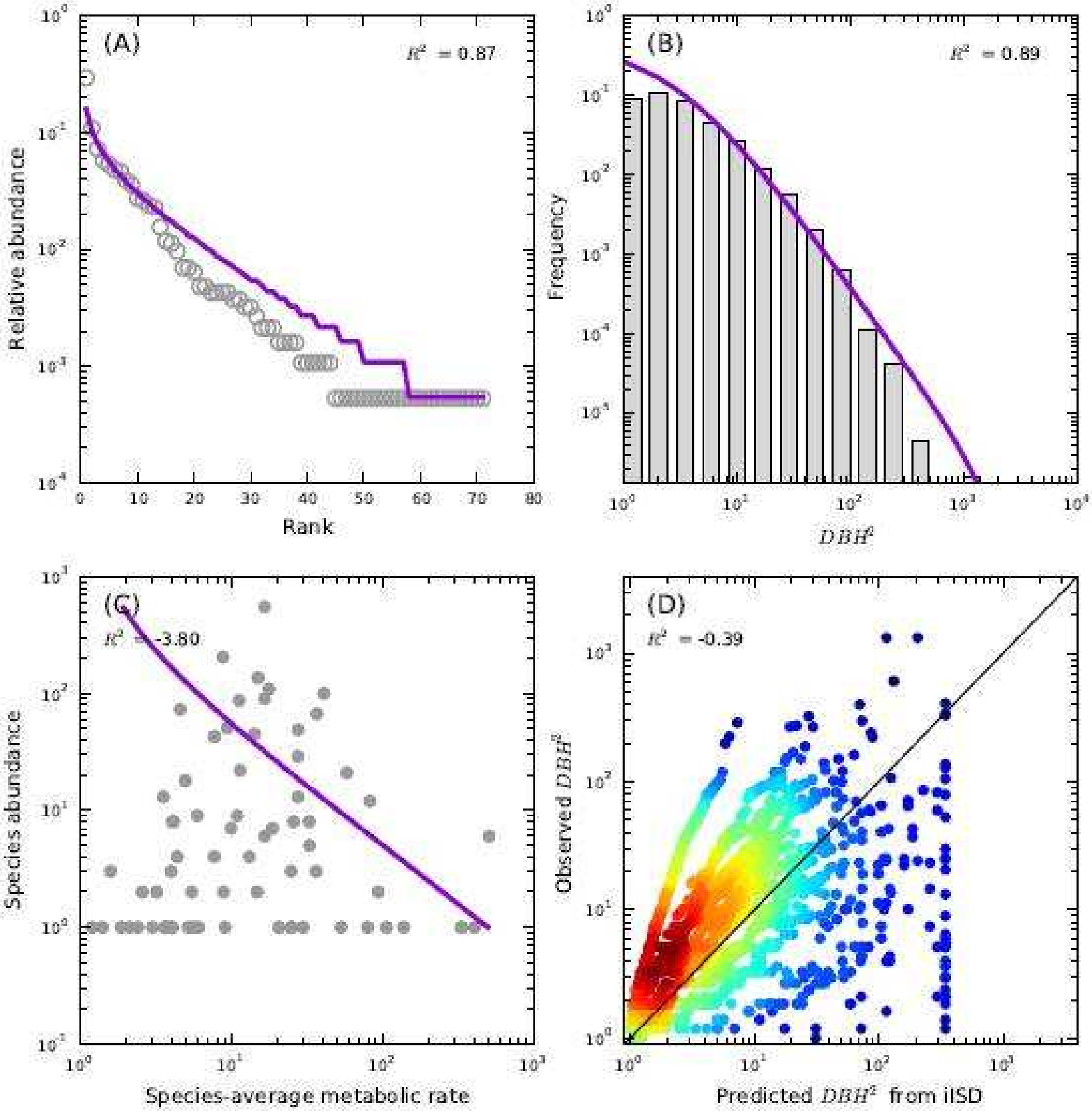

WesternGhats,BSP74

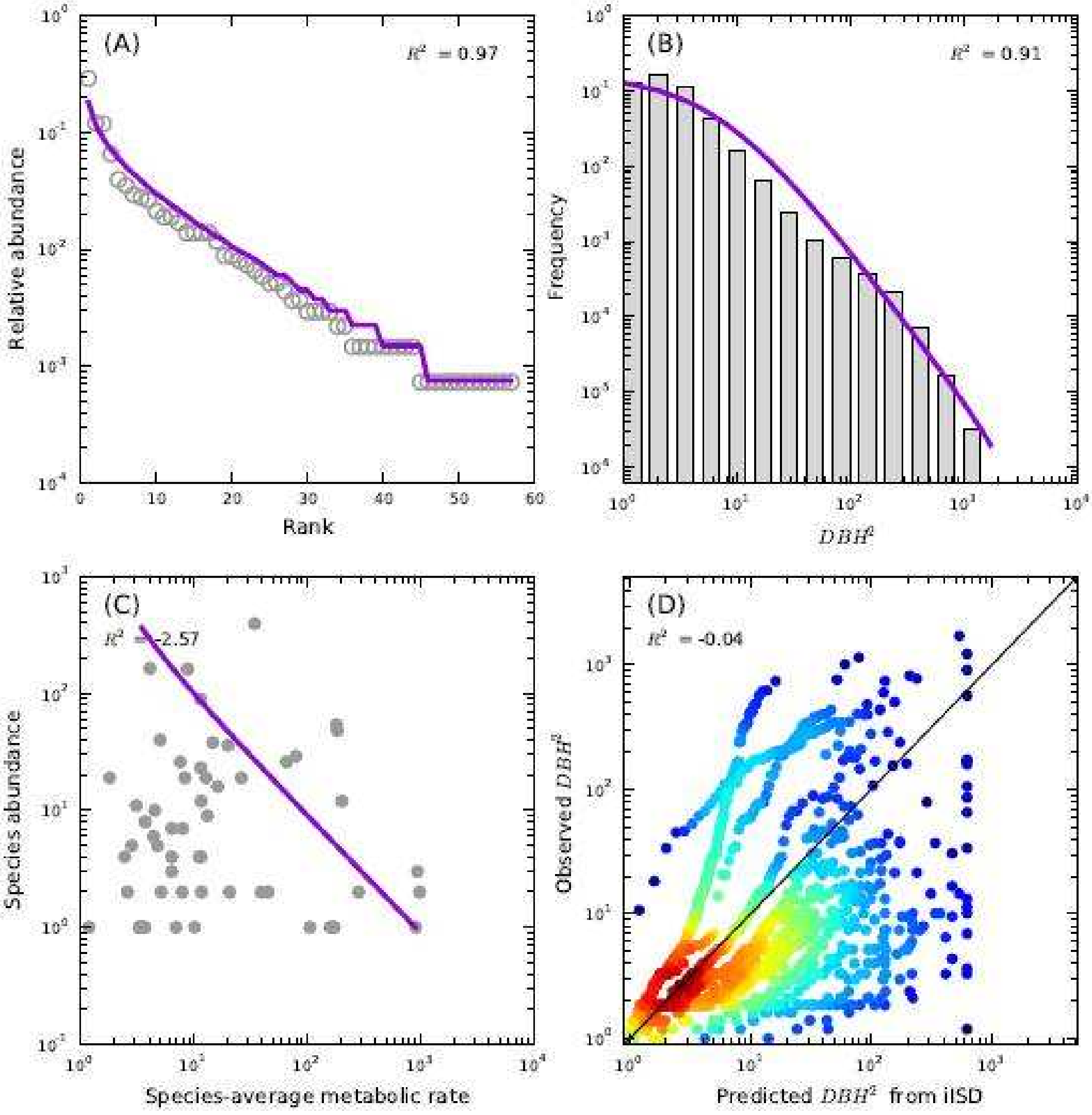

WesternGhats,BSP75

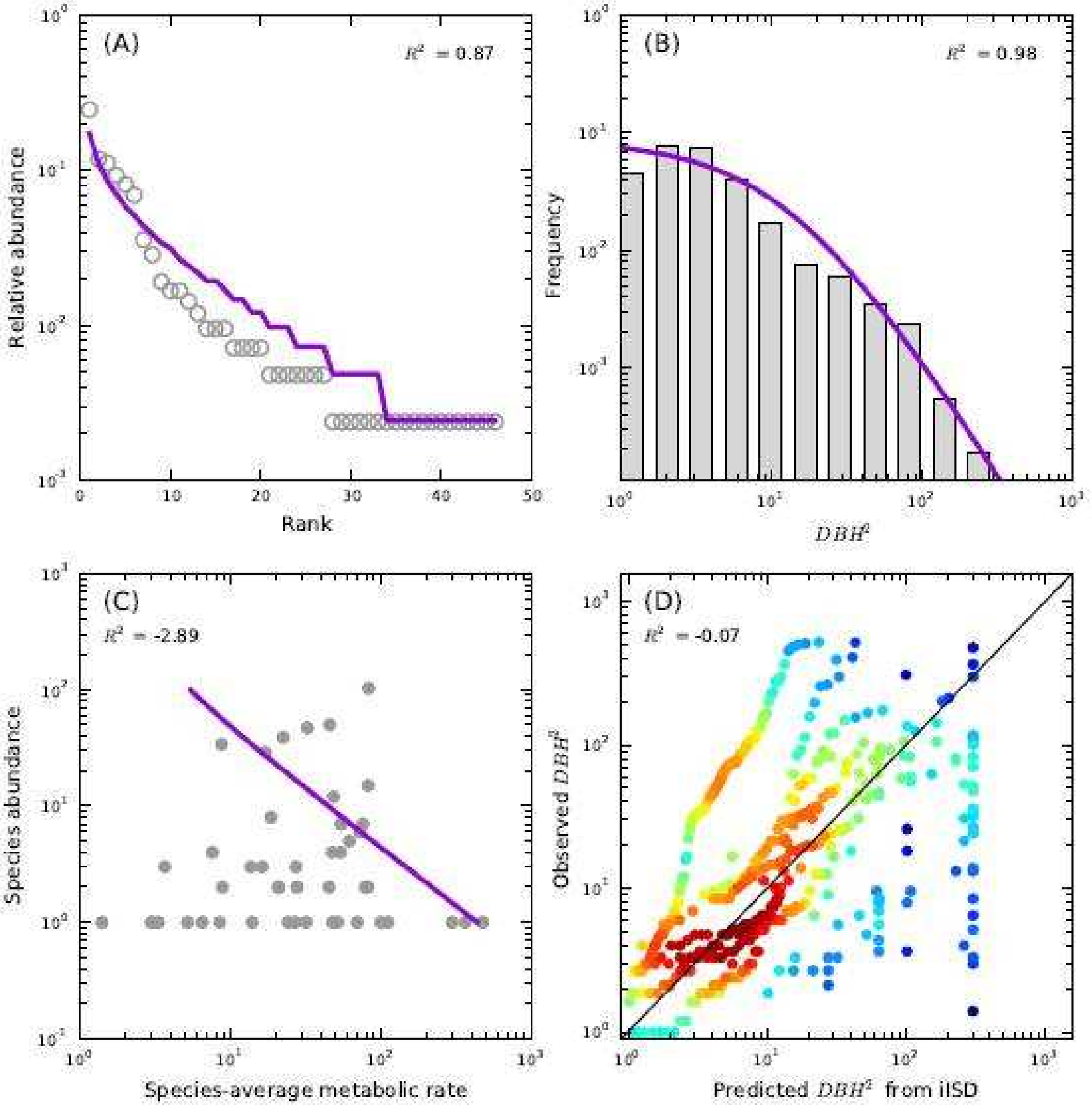

# WesternGhats,BSP79

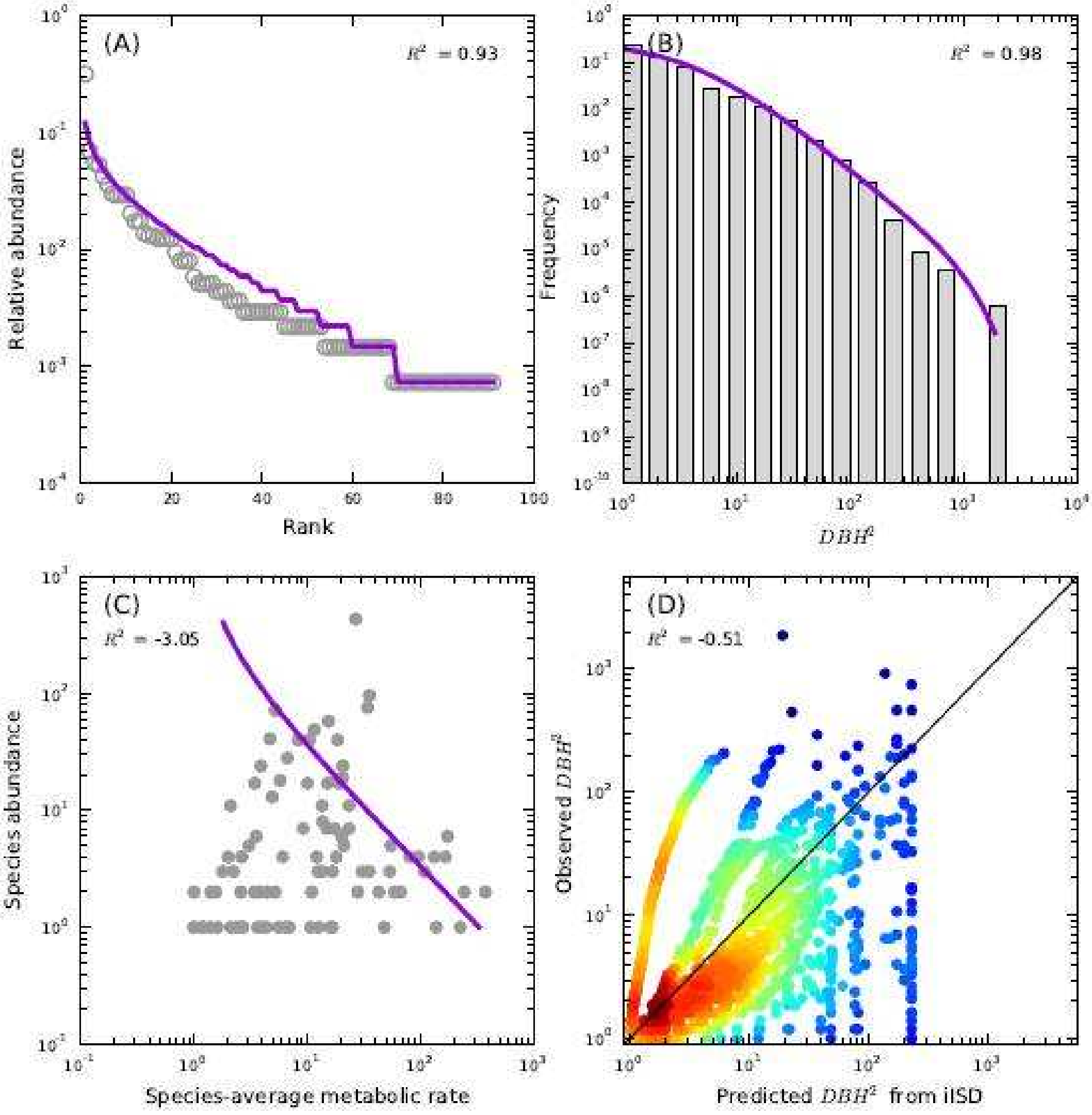

WesternGhats,BSP80

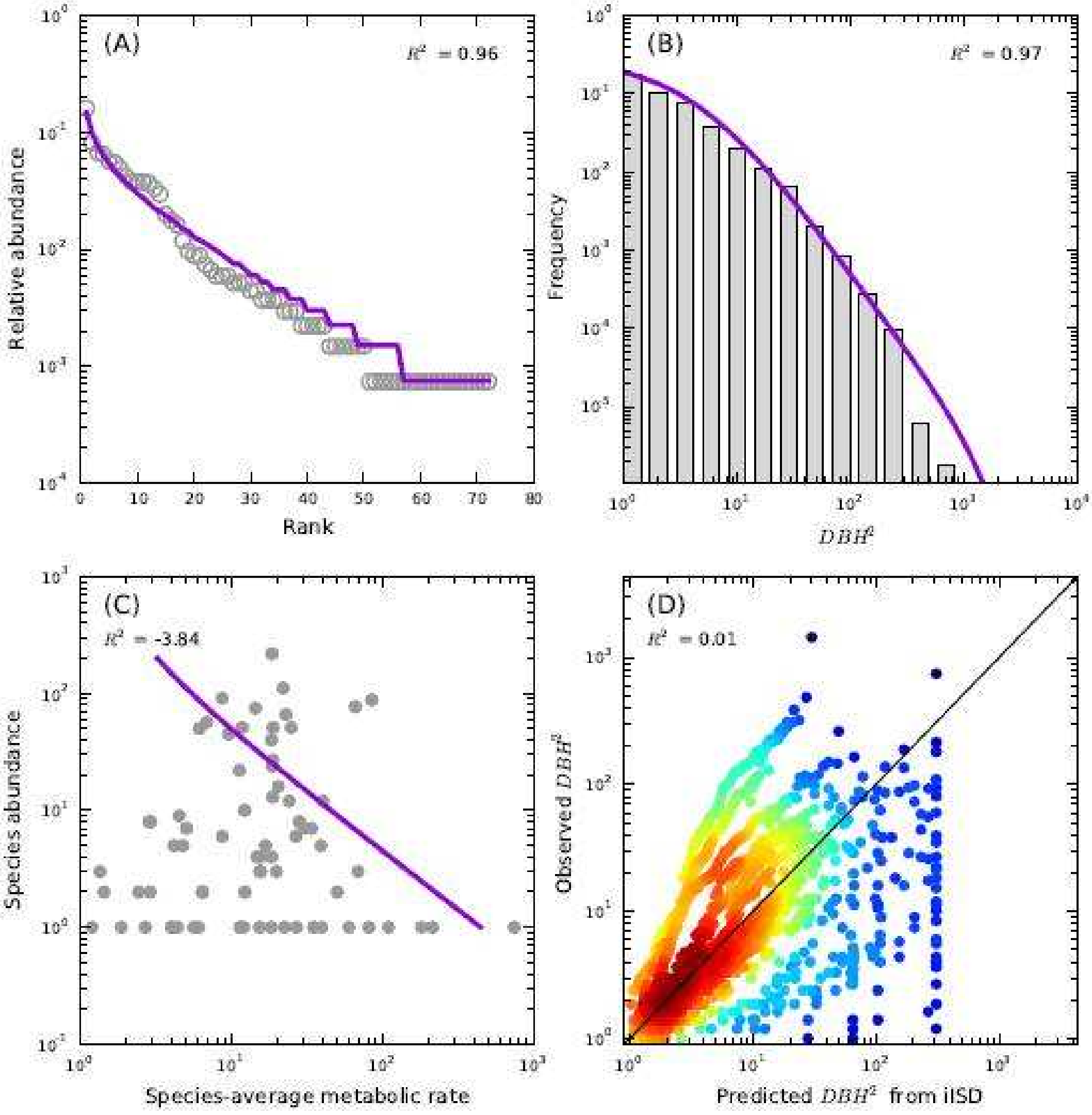

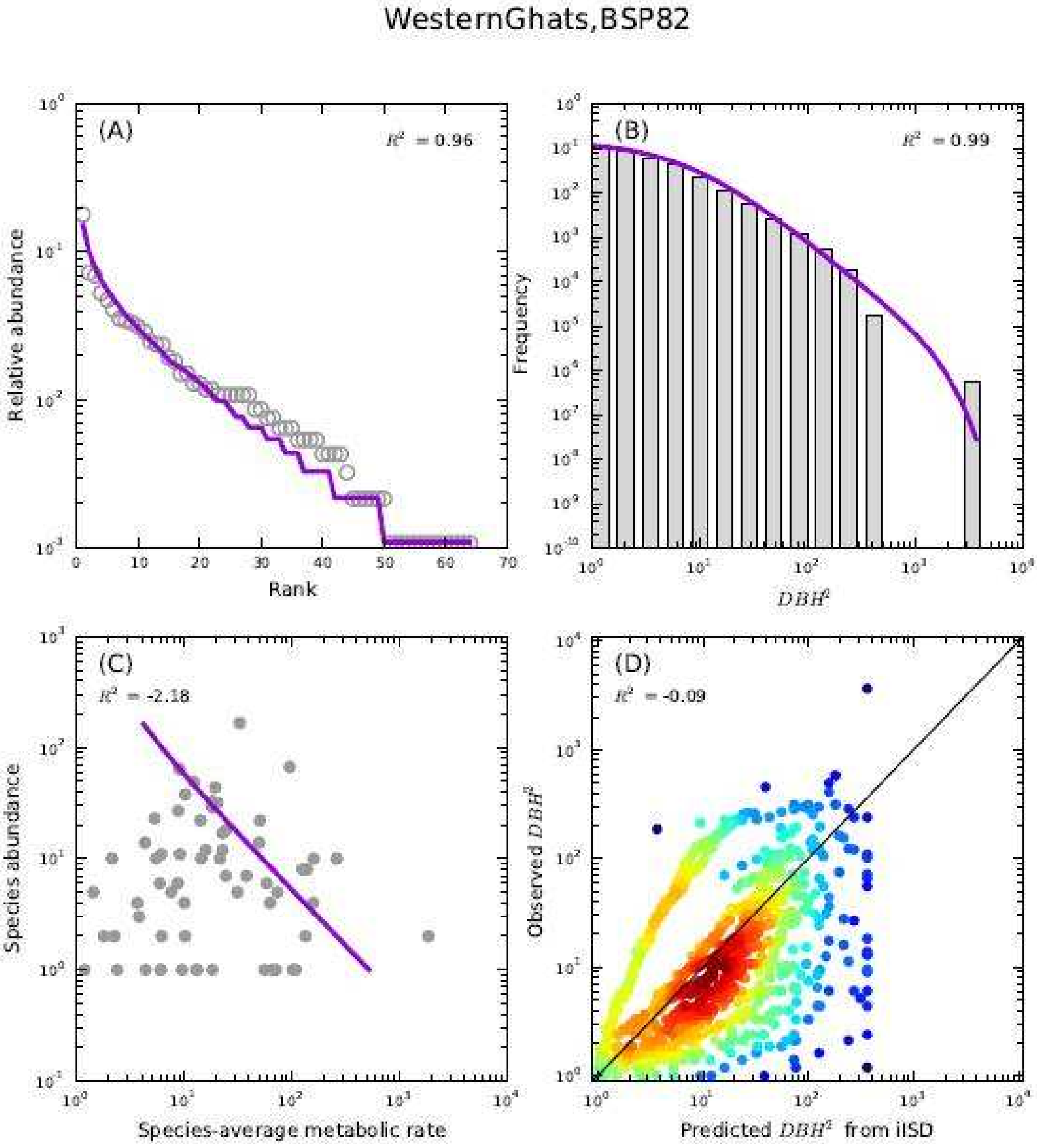

WesternGhats,BSP83

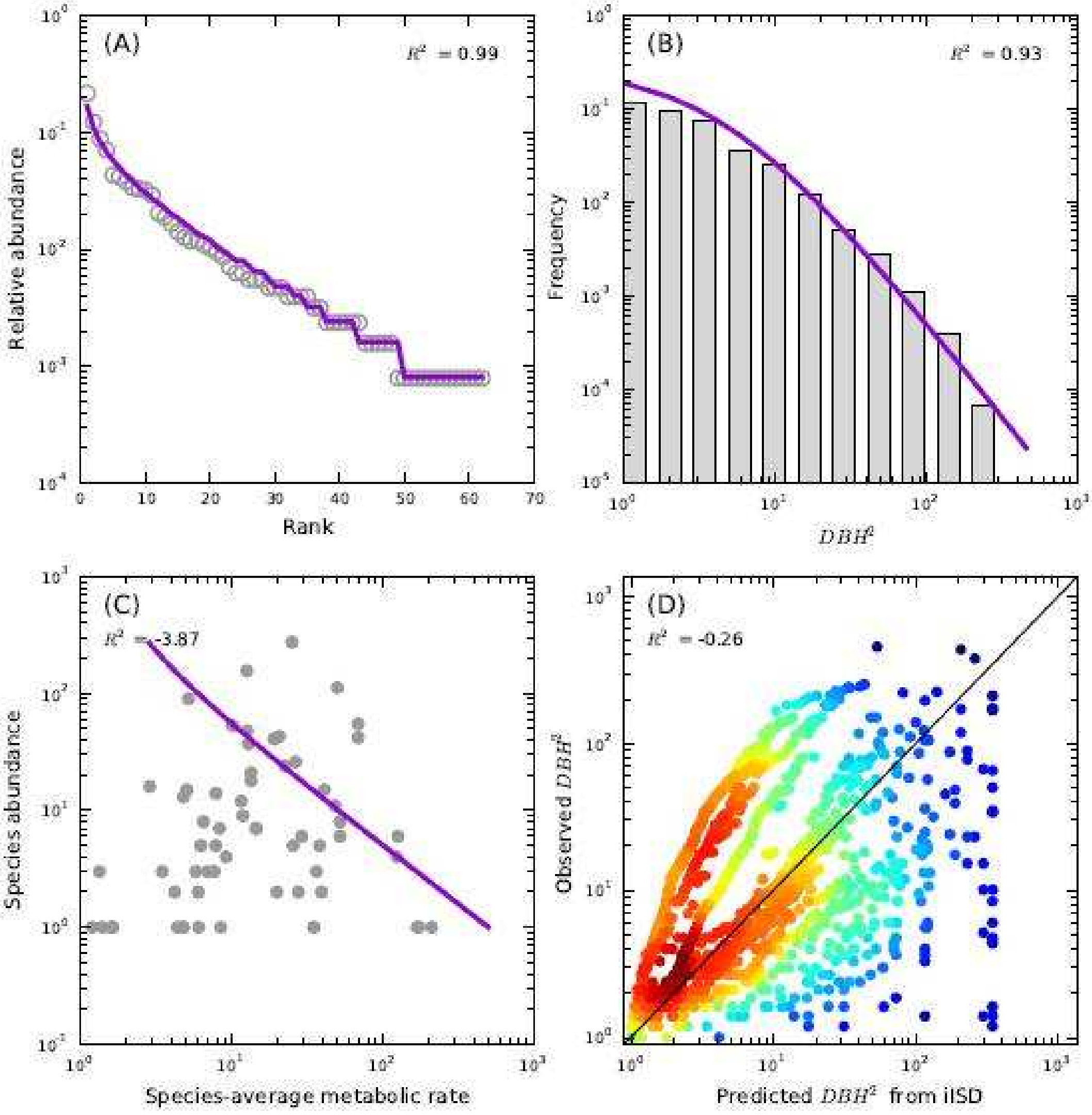

WesternGhats,BSP84

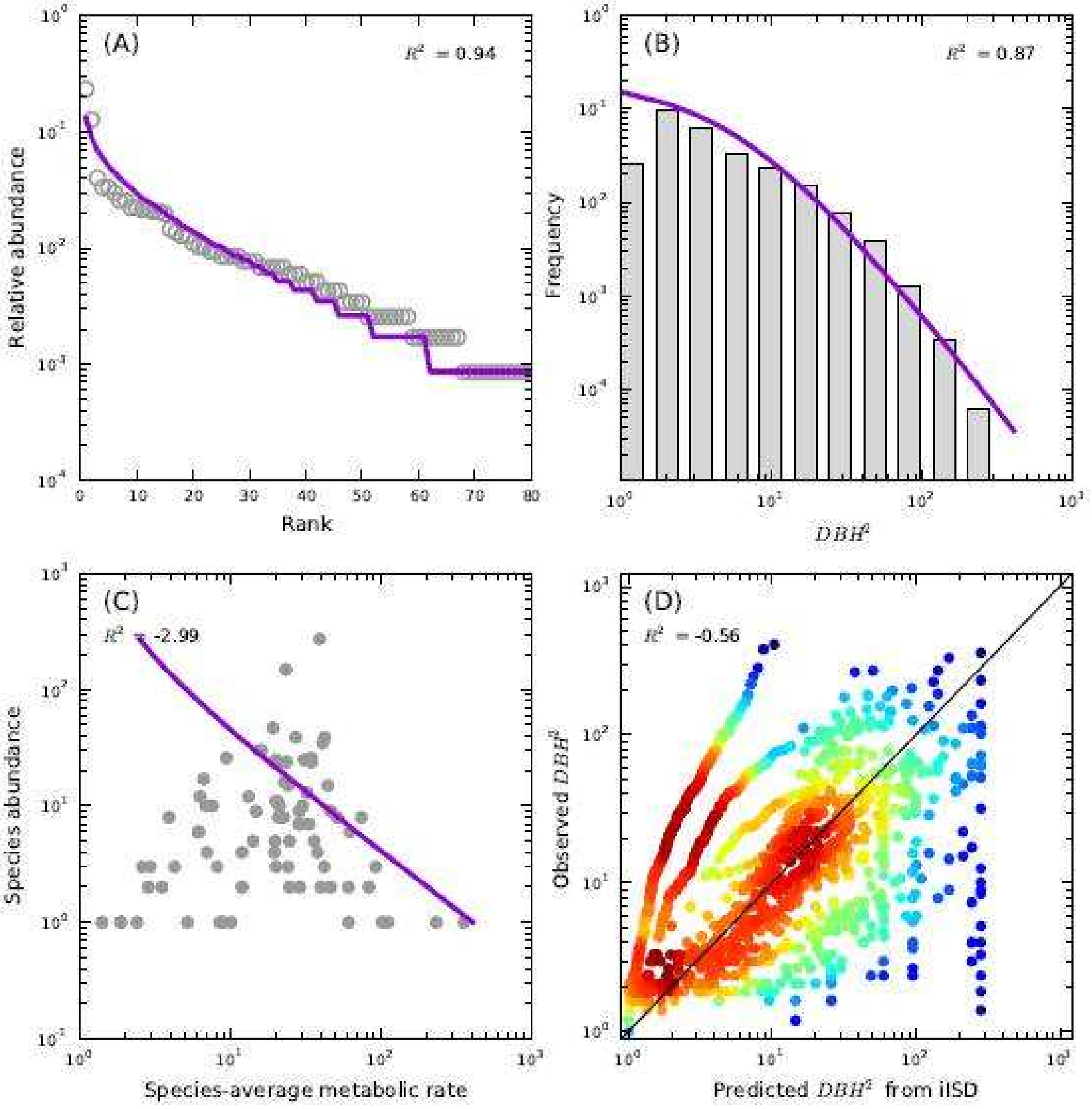

WesternGhats,BSP85

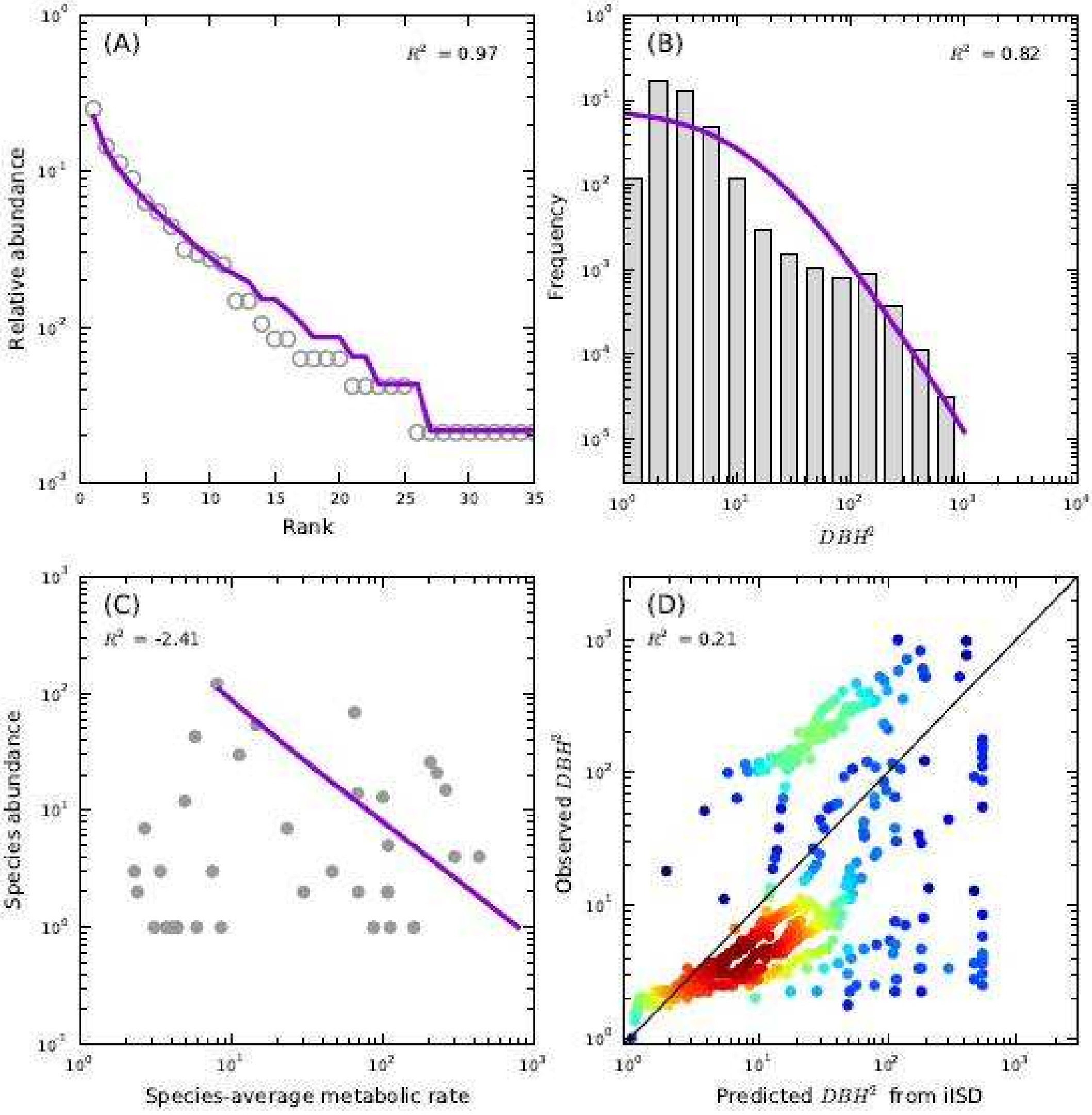

WesternGhats,BSP88

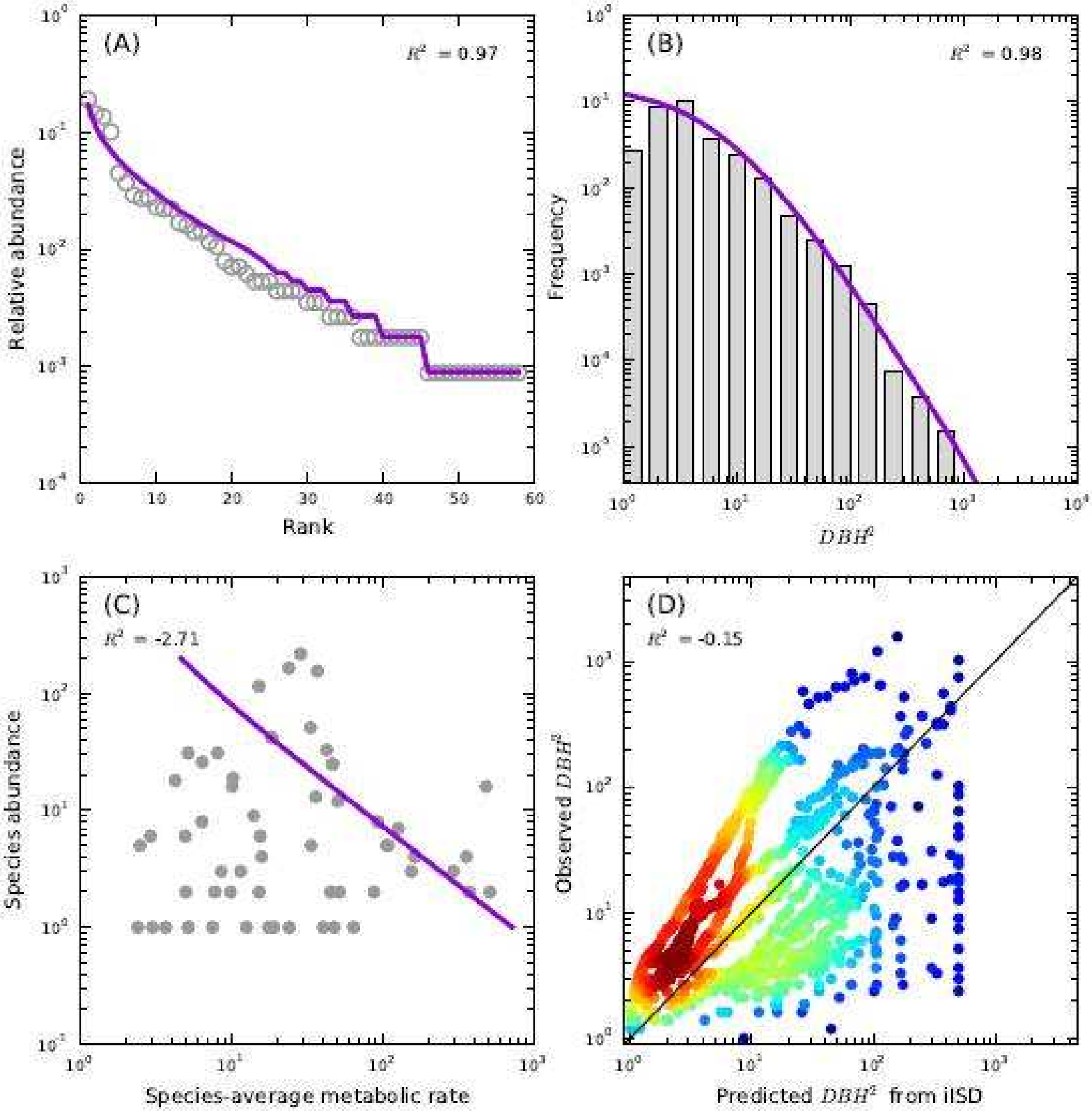

WesternGhats,BSP89

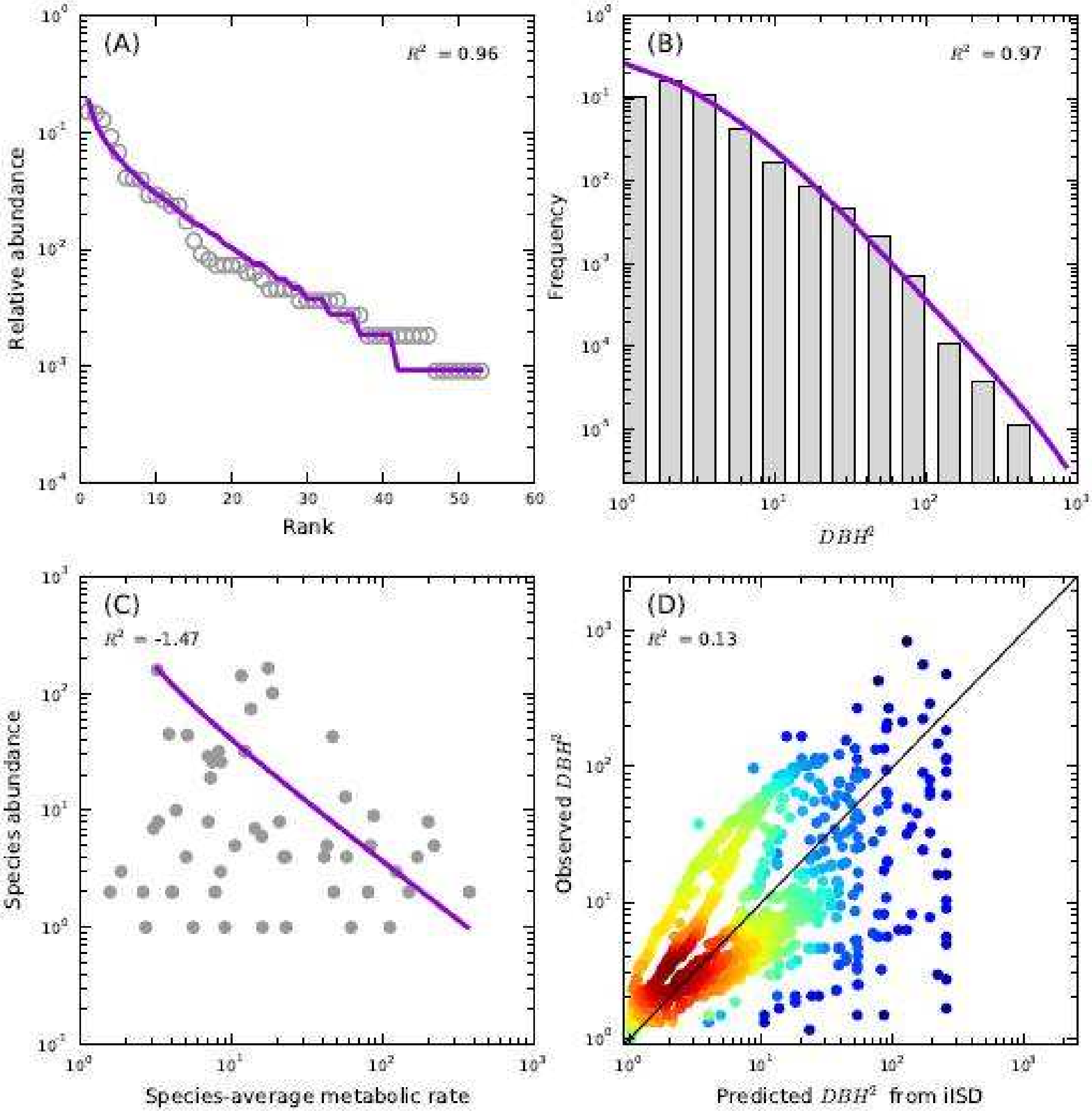

WesternGhats,BSP90

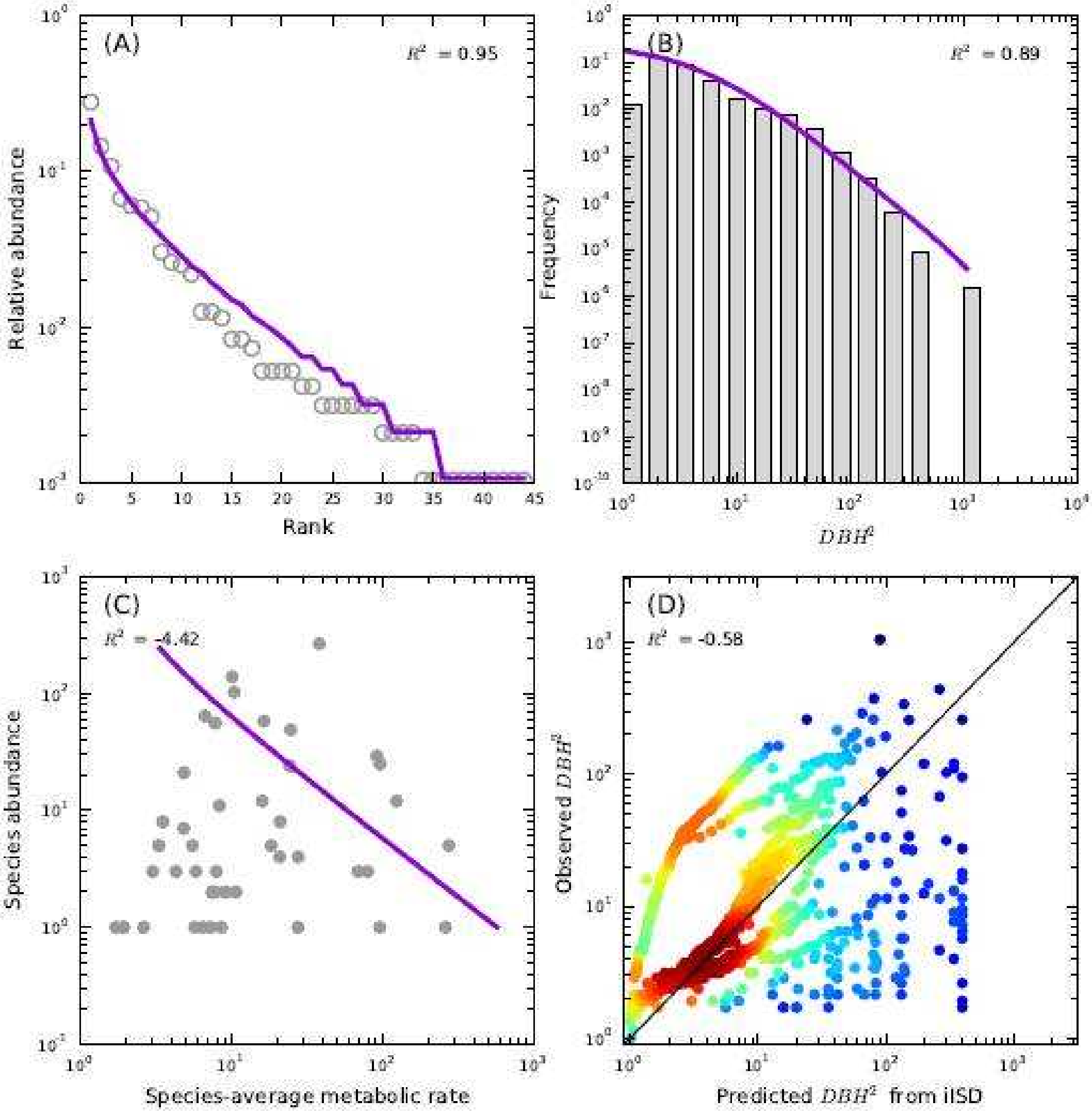

WesternGhats,BSP91

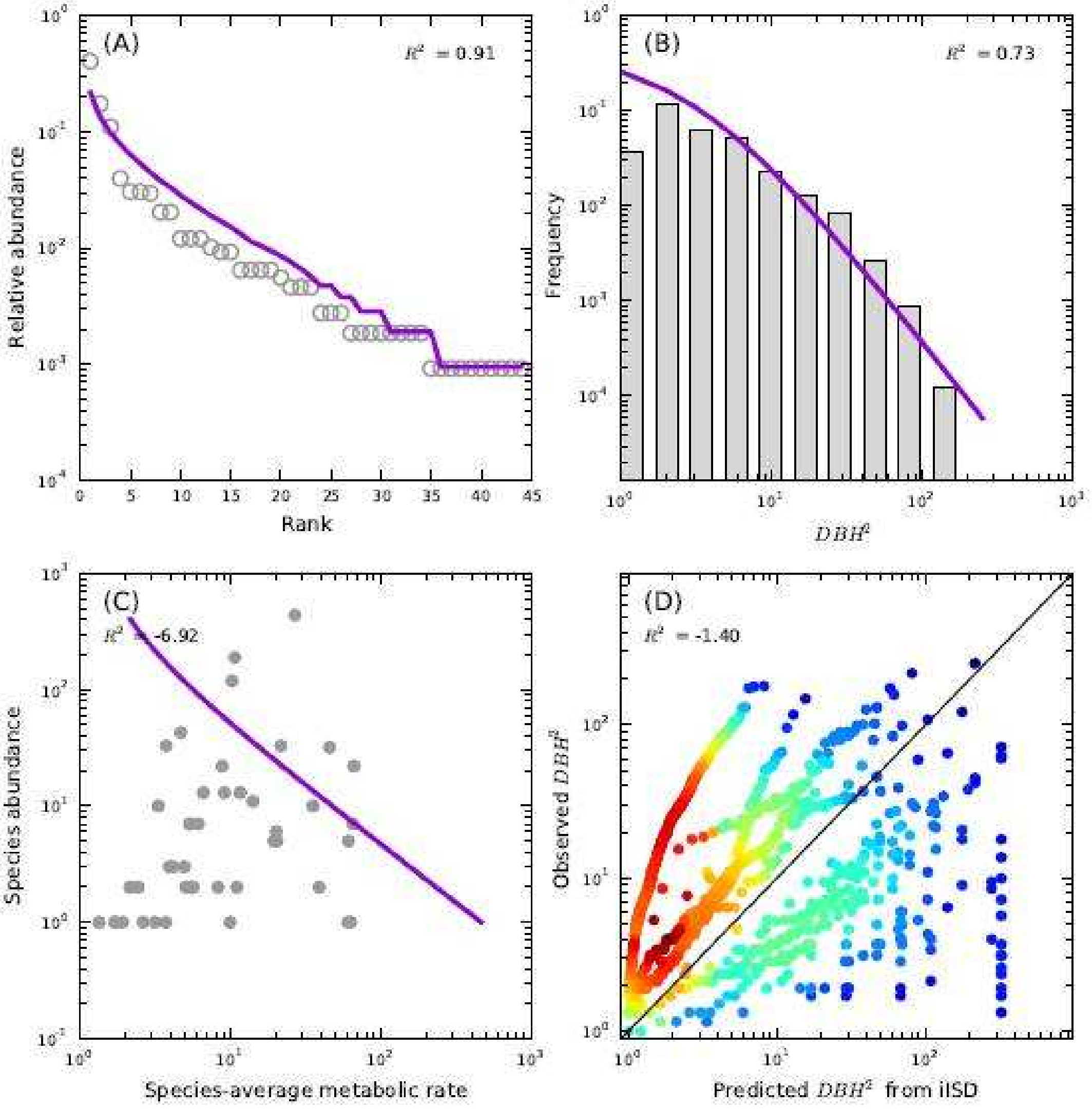

WesternGhats,BSP92

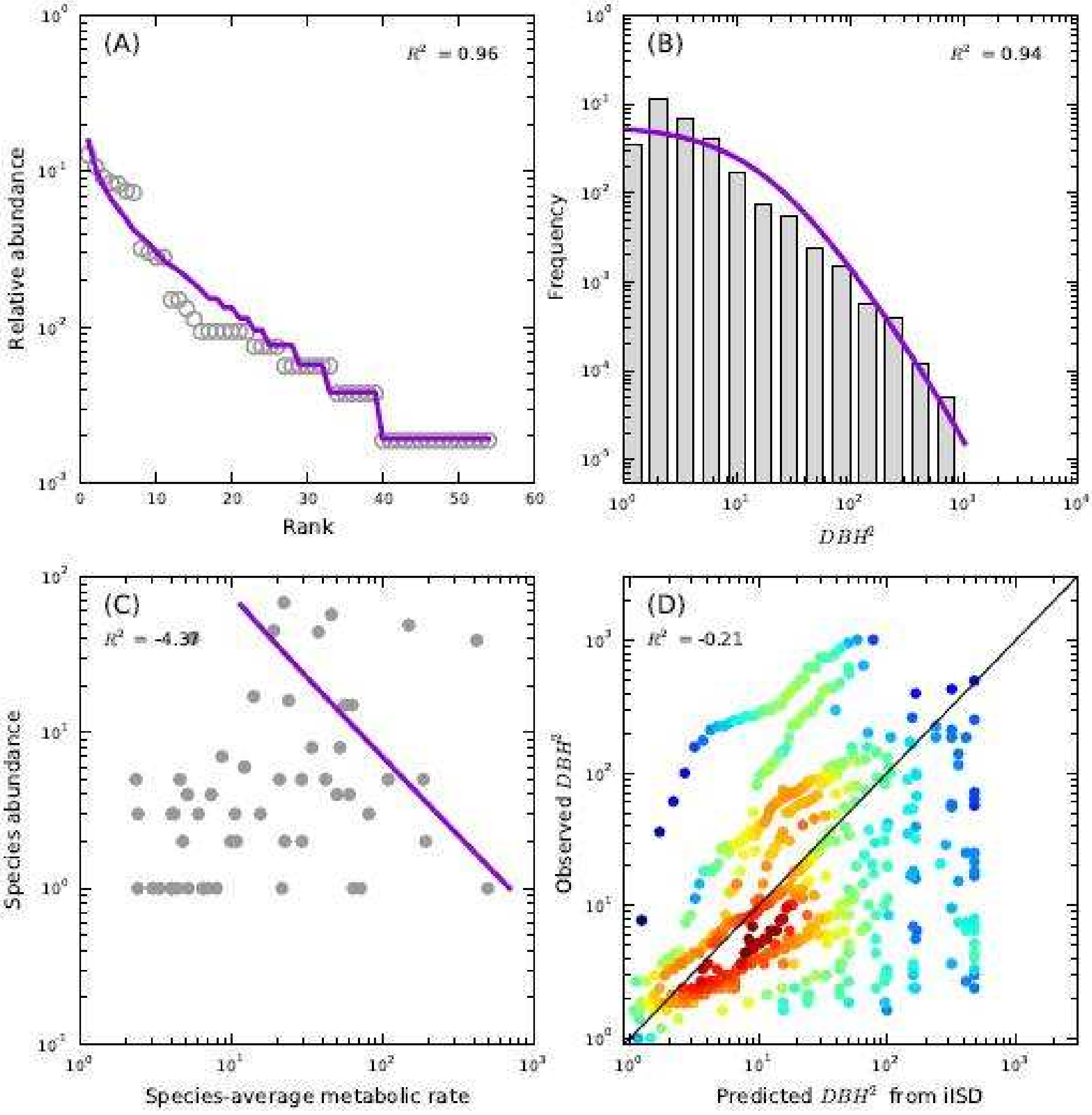

WesternGhats,BSP94

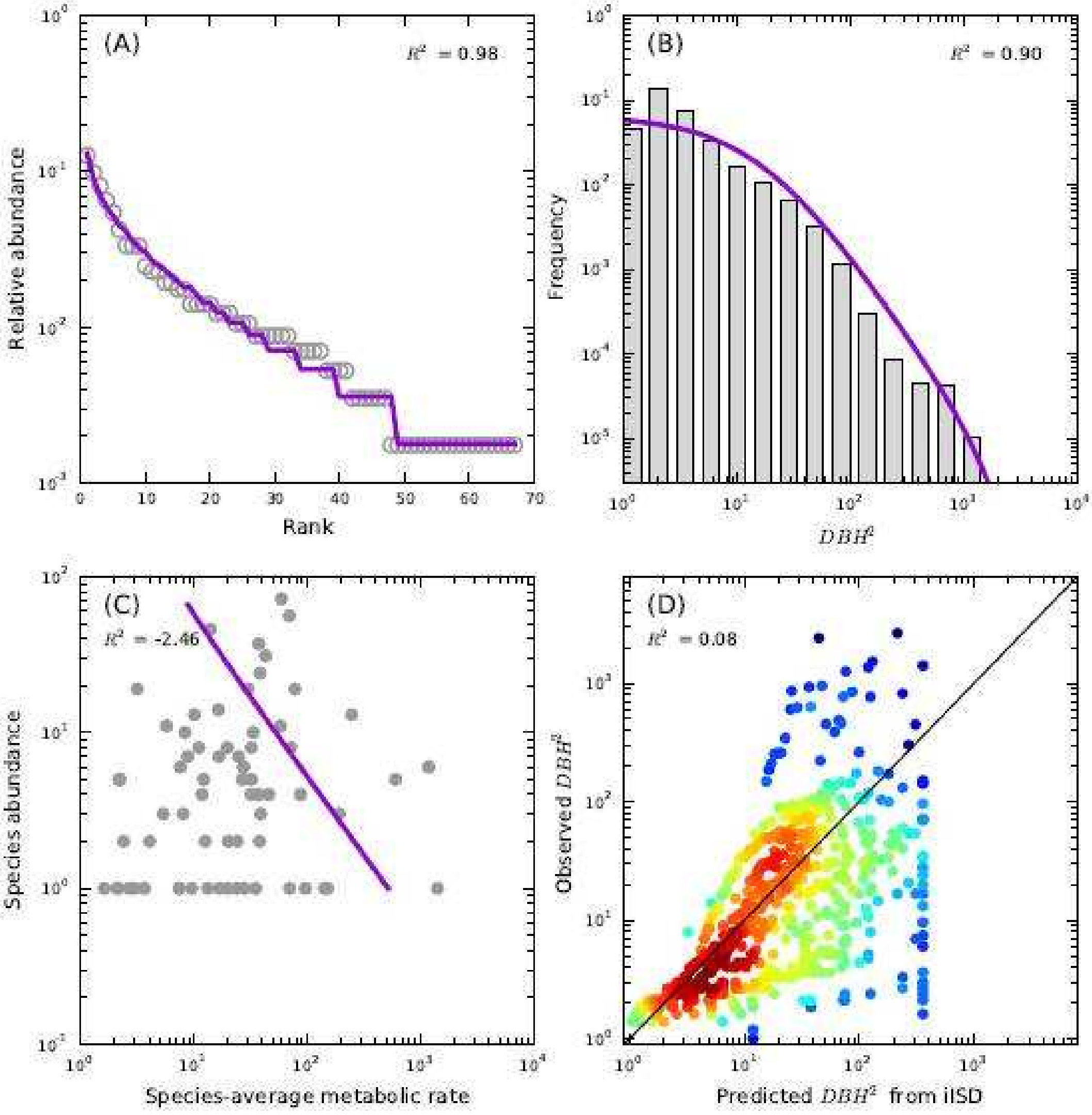

WesternGhats,BSP98

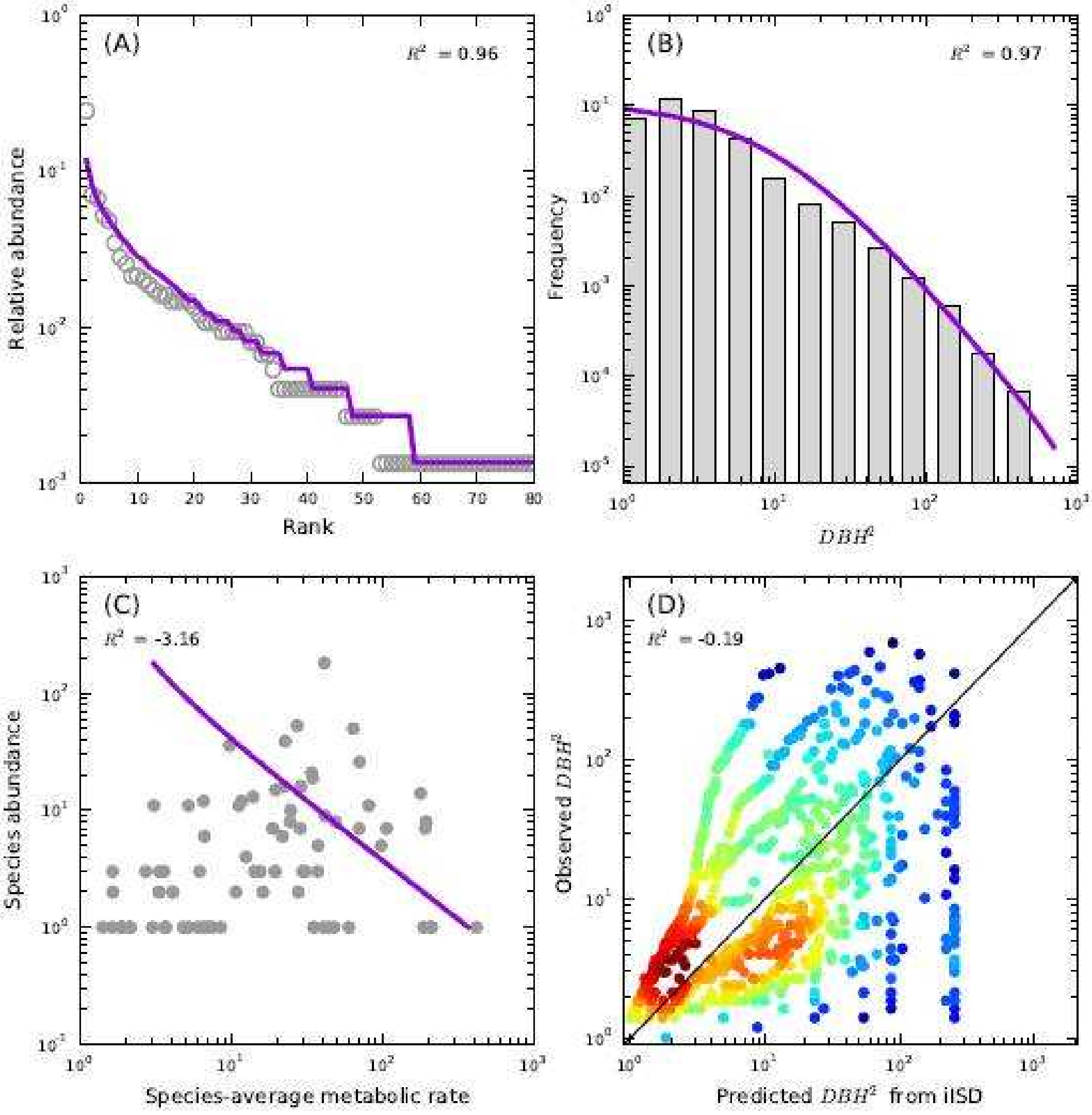

WesternGhats,BSP99

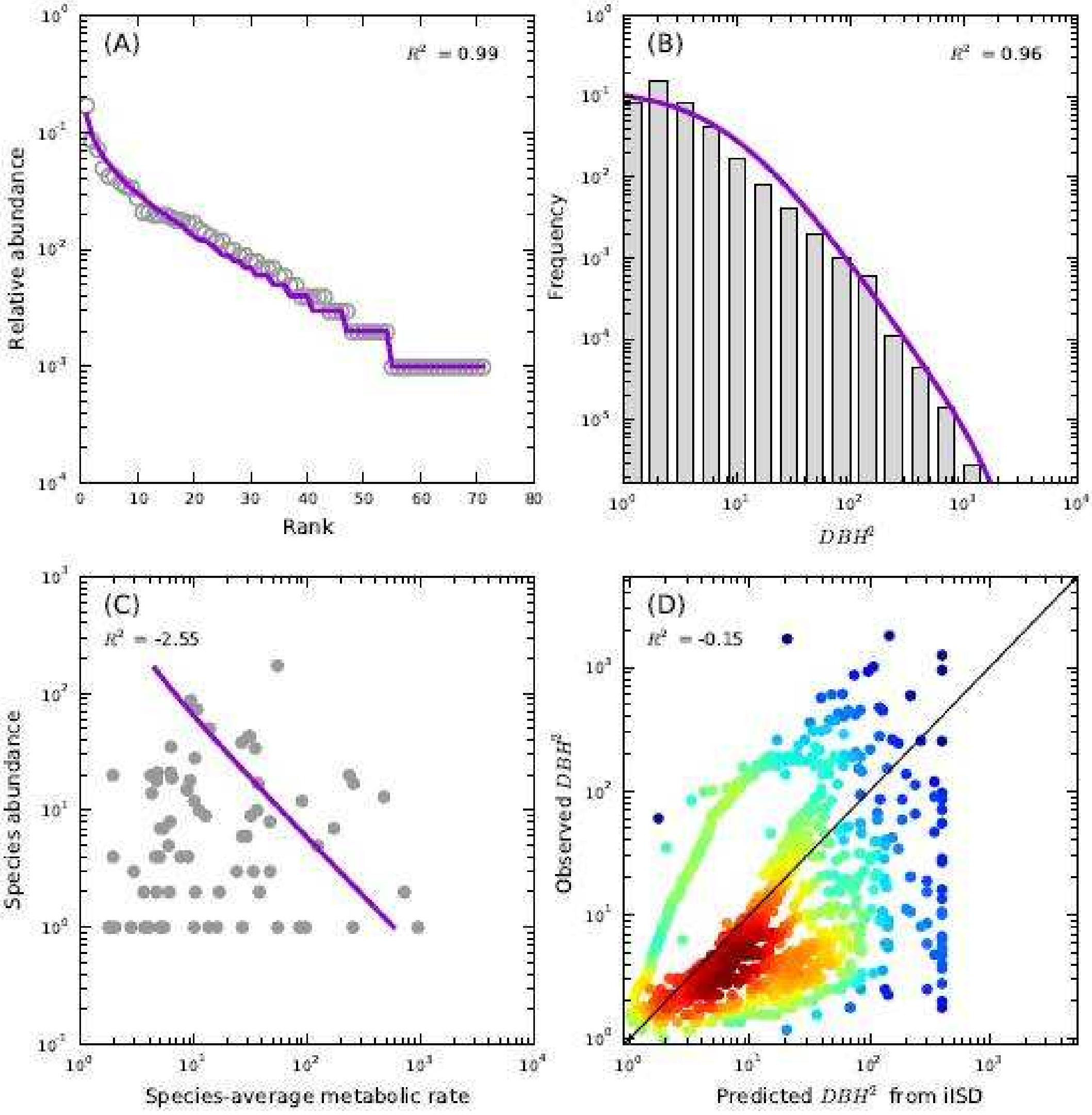

BCI,bci

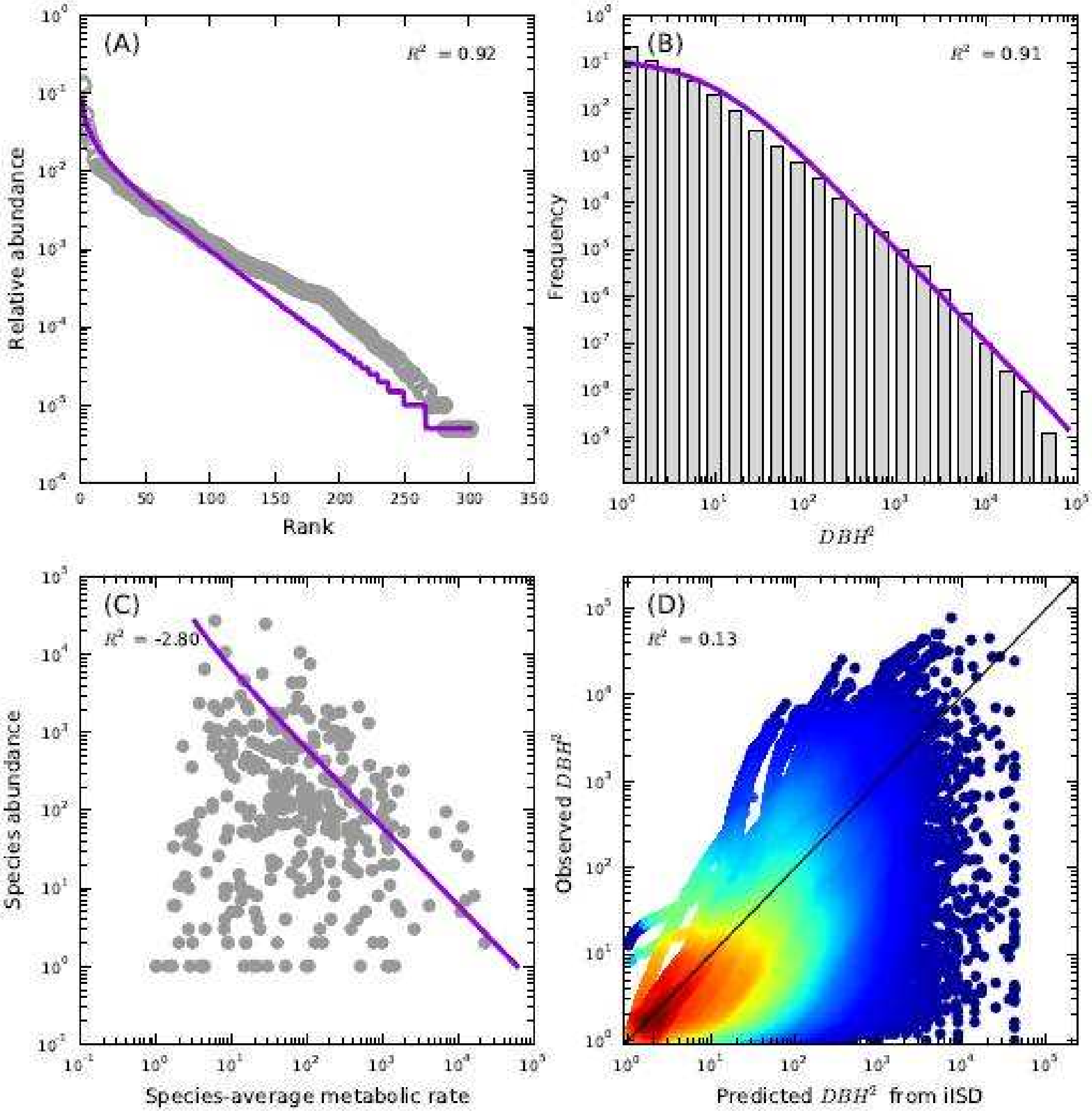

BVSF,BVPlot

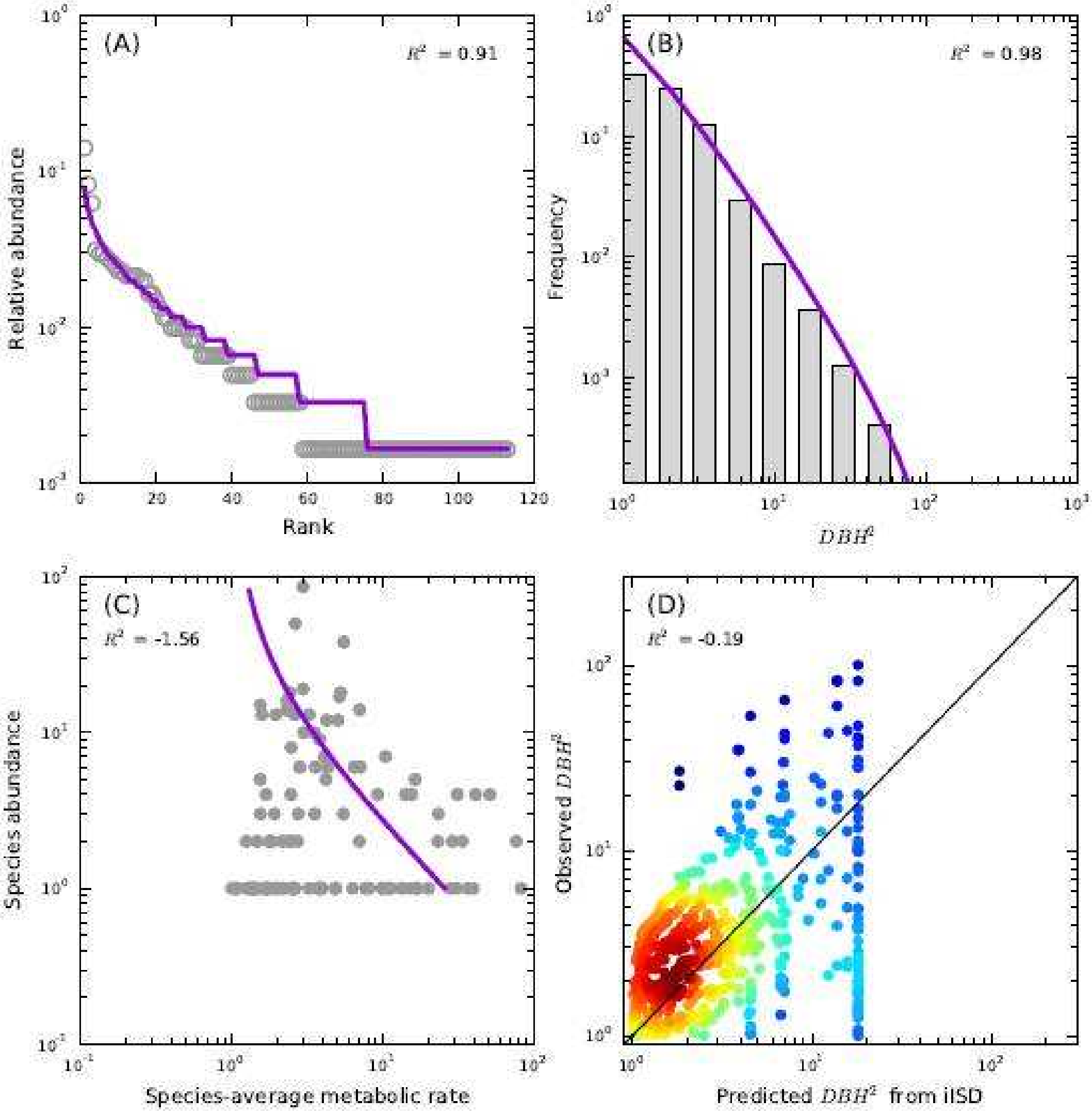

BVSF,SFPlot

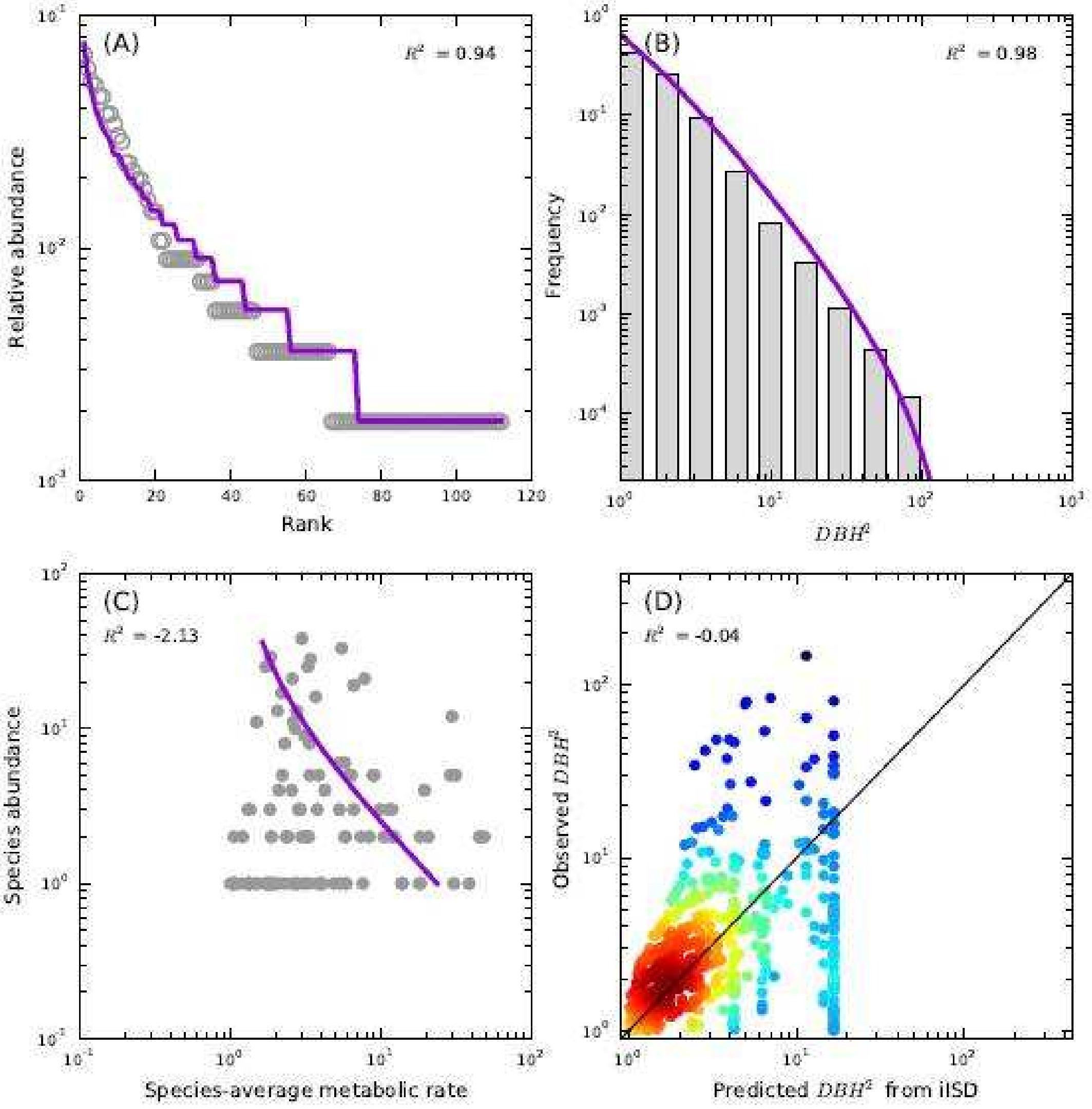

Cocoli,cocoli

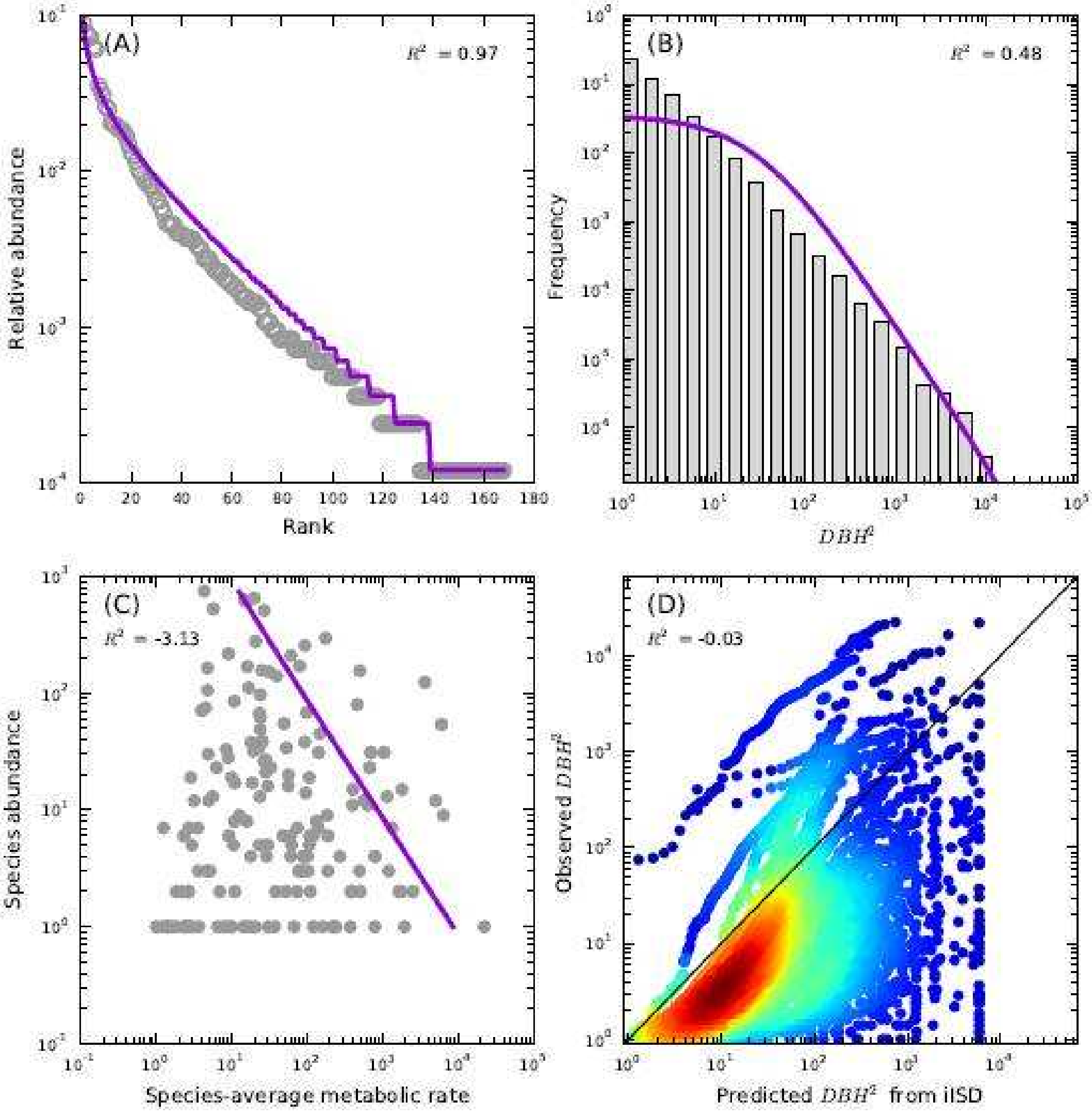

Lahei,heath1

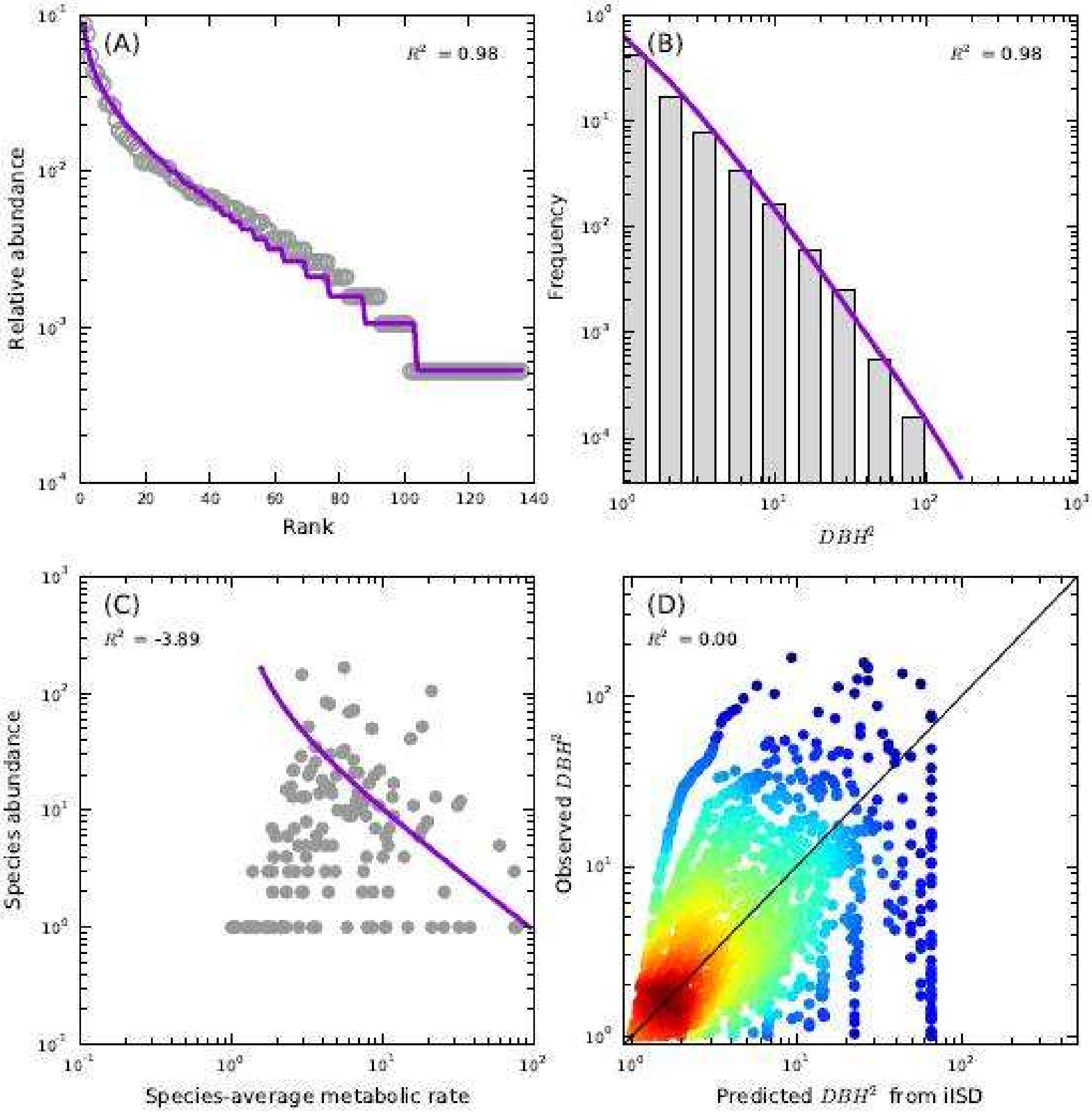

Lahei,heath2

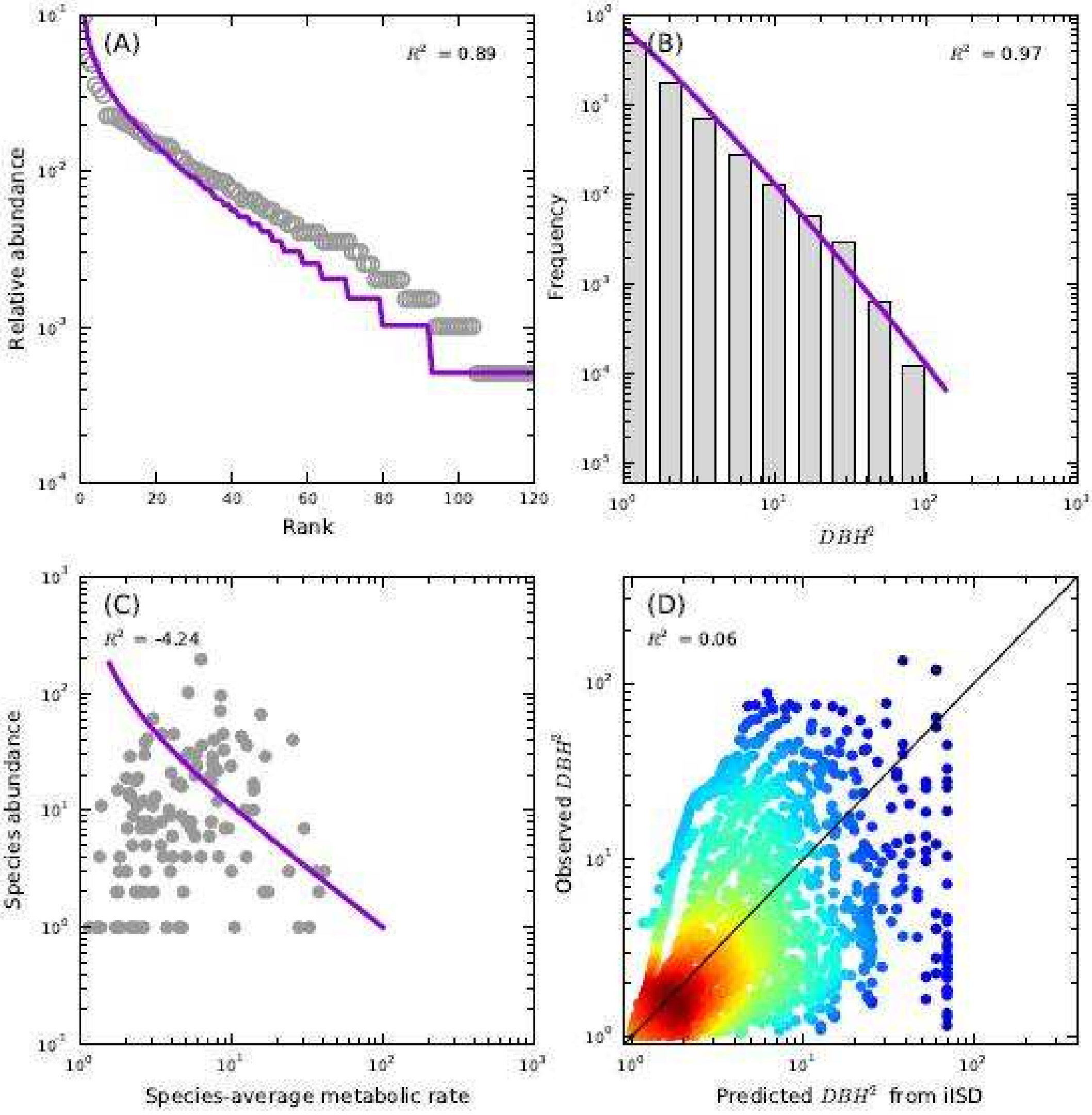

Lahei,peat

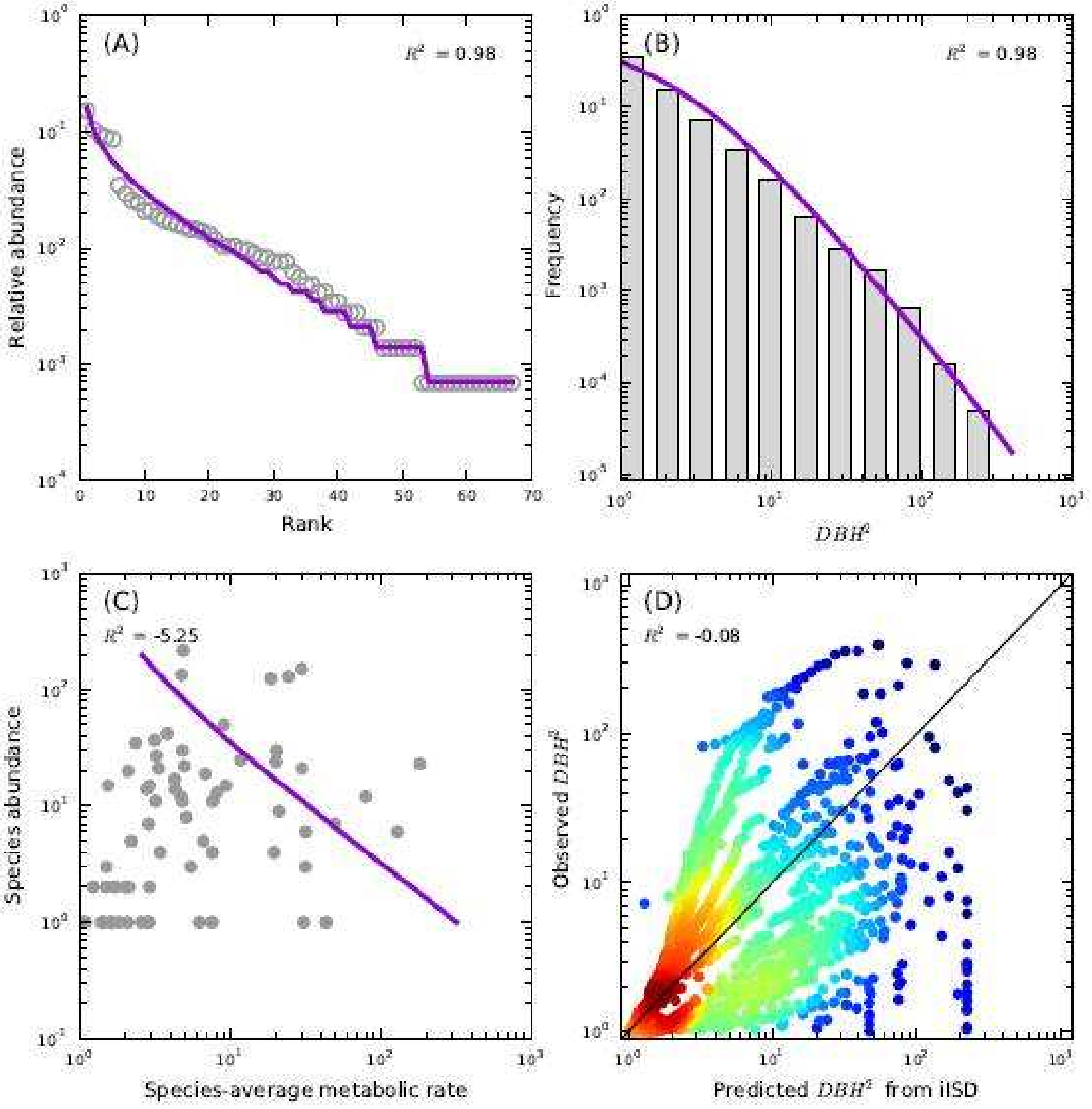

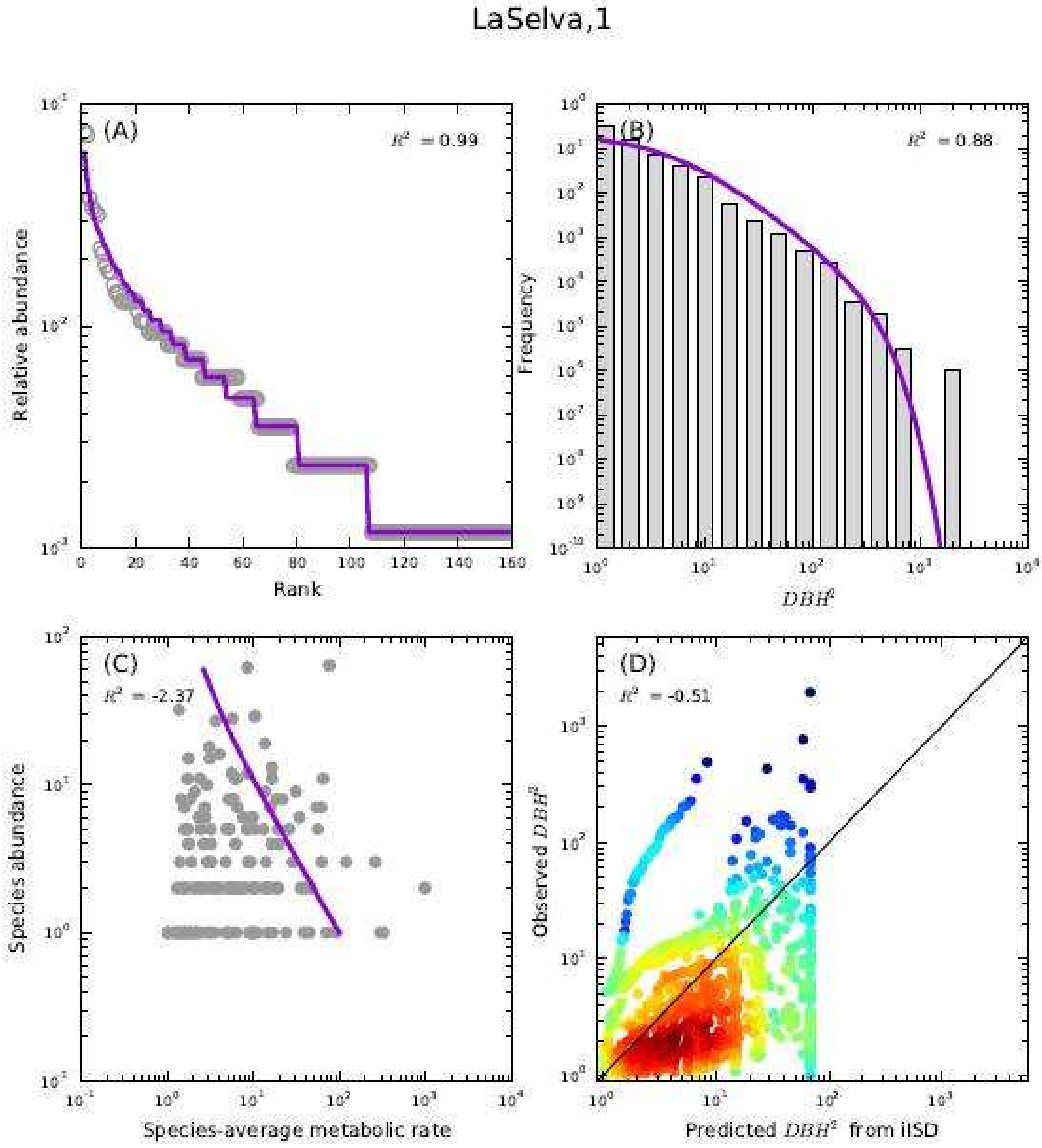

LaSelva,2

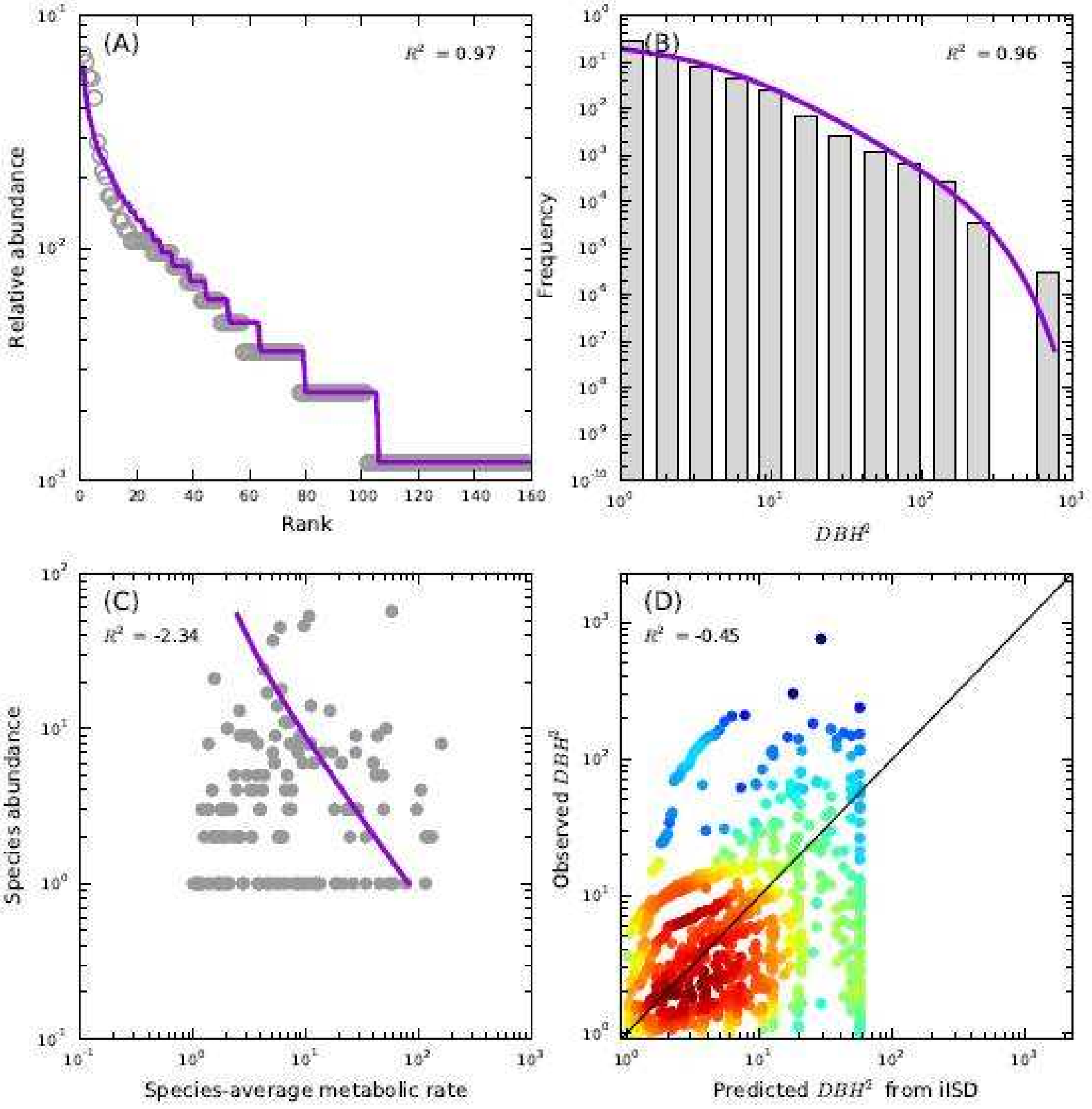

LaSelva,3

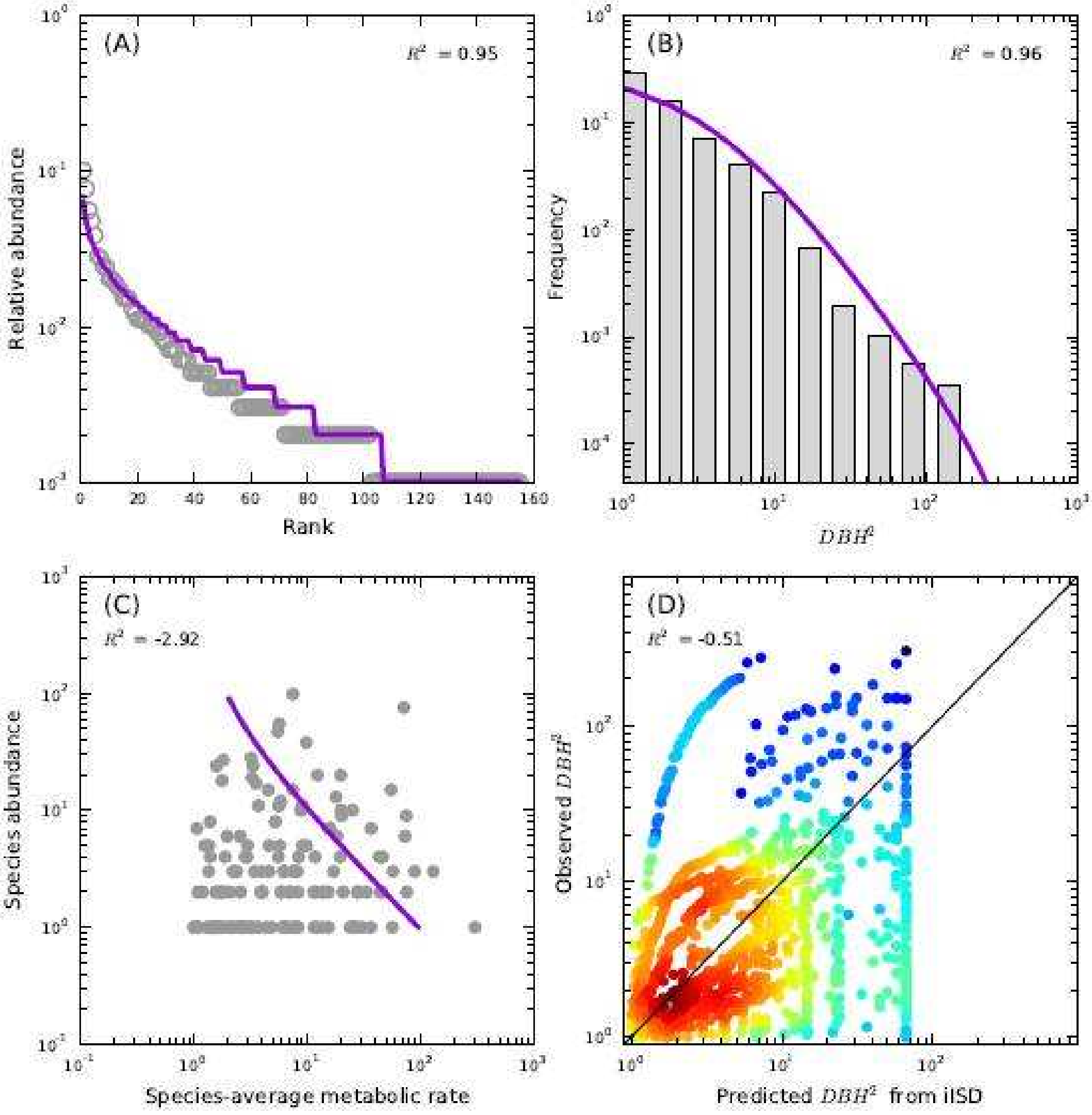

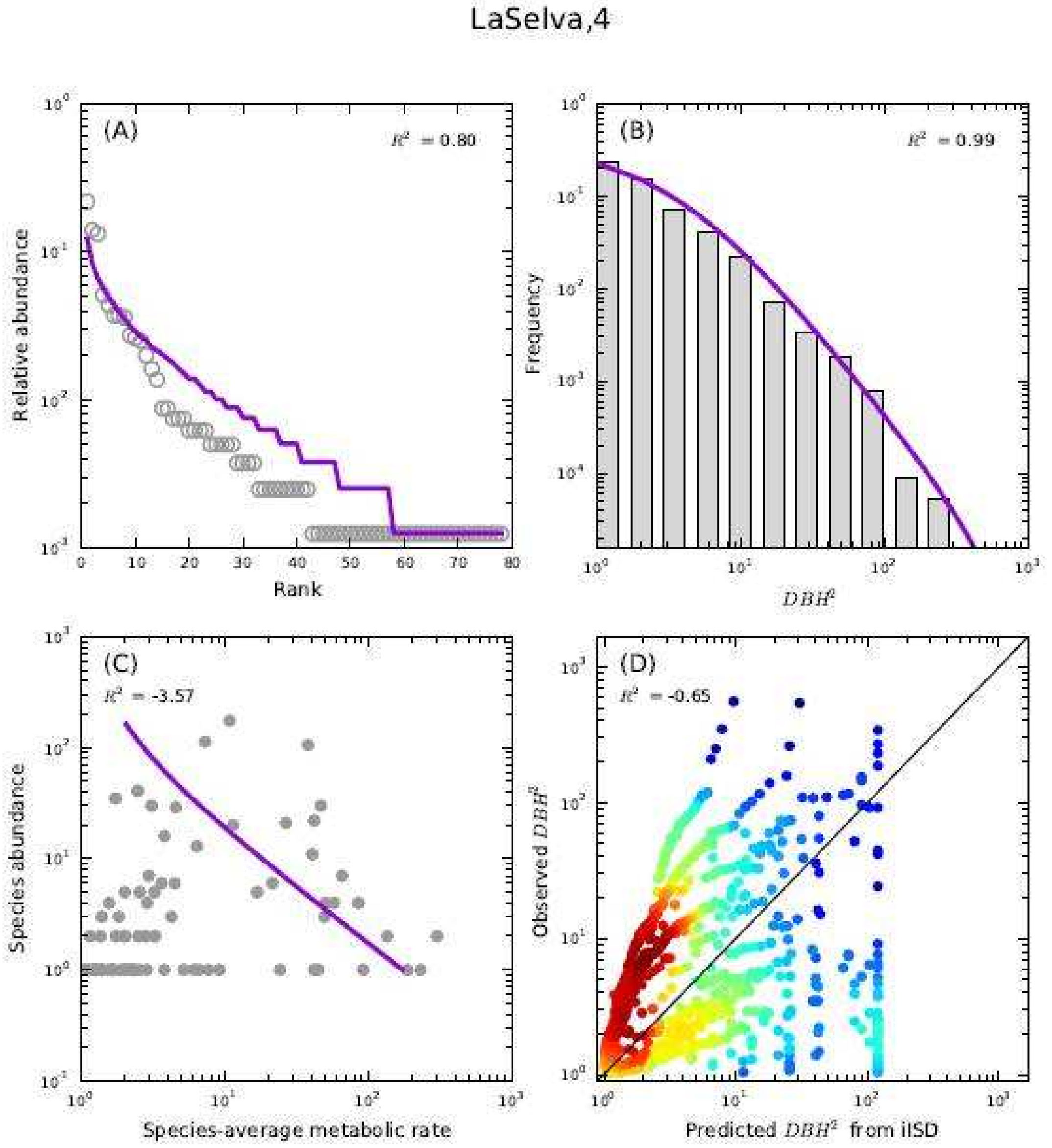

LaSelva,5

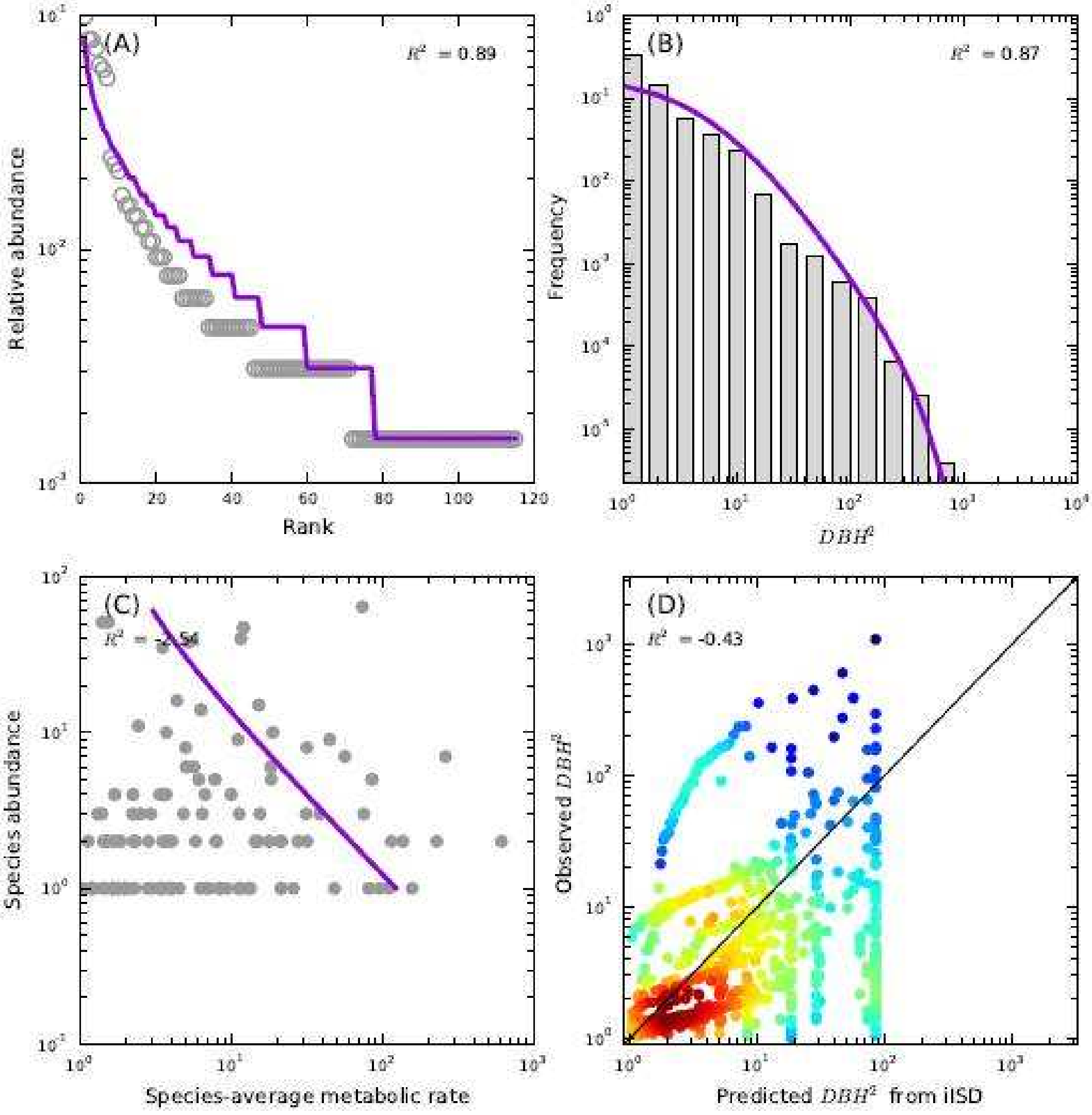

Luquillo,lfdp

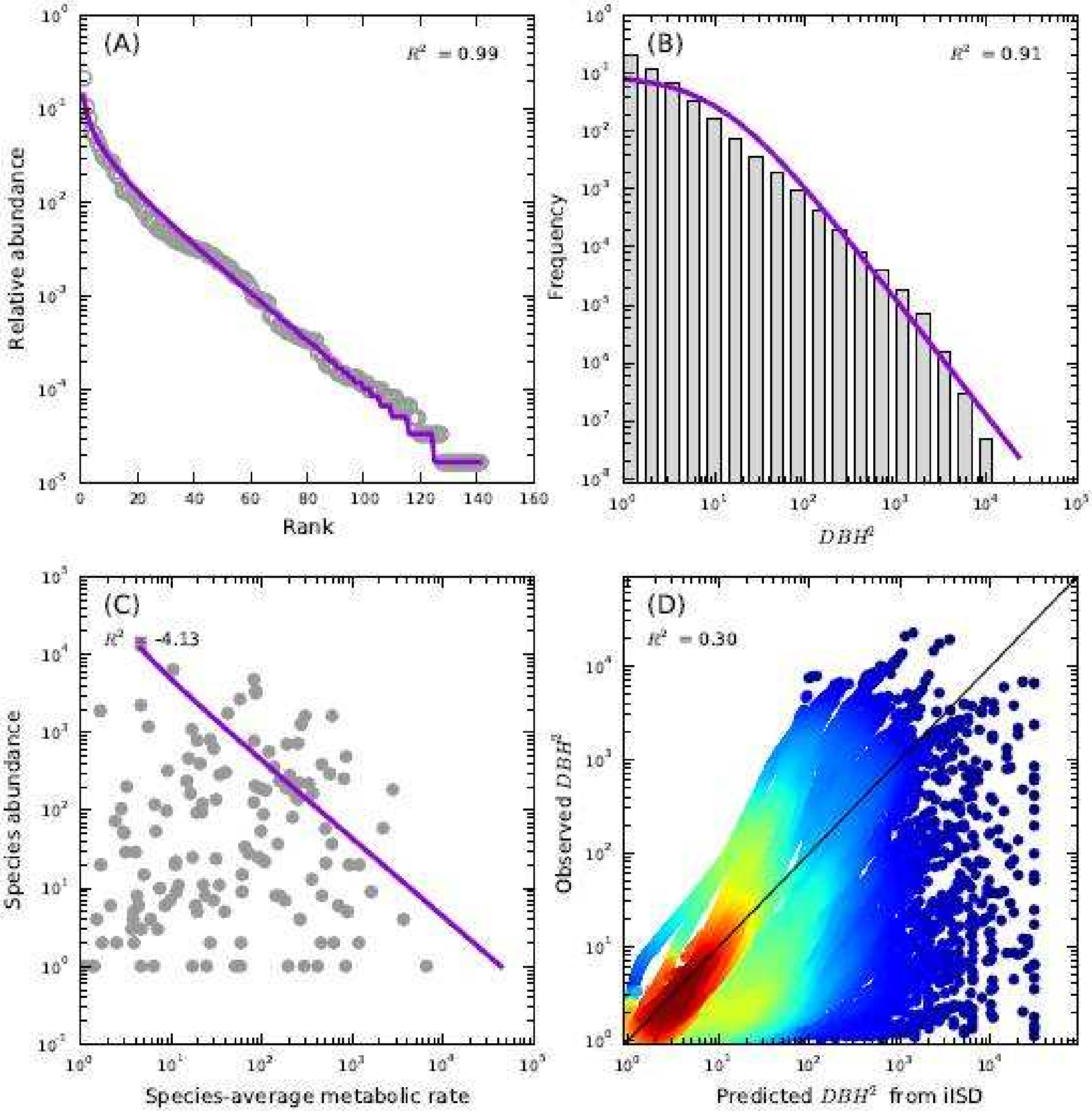

NC,12

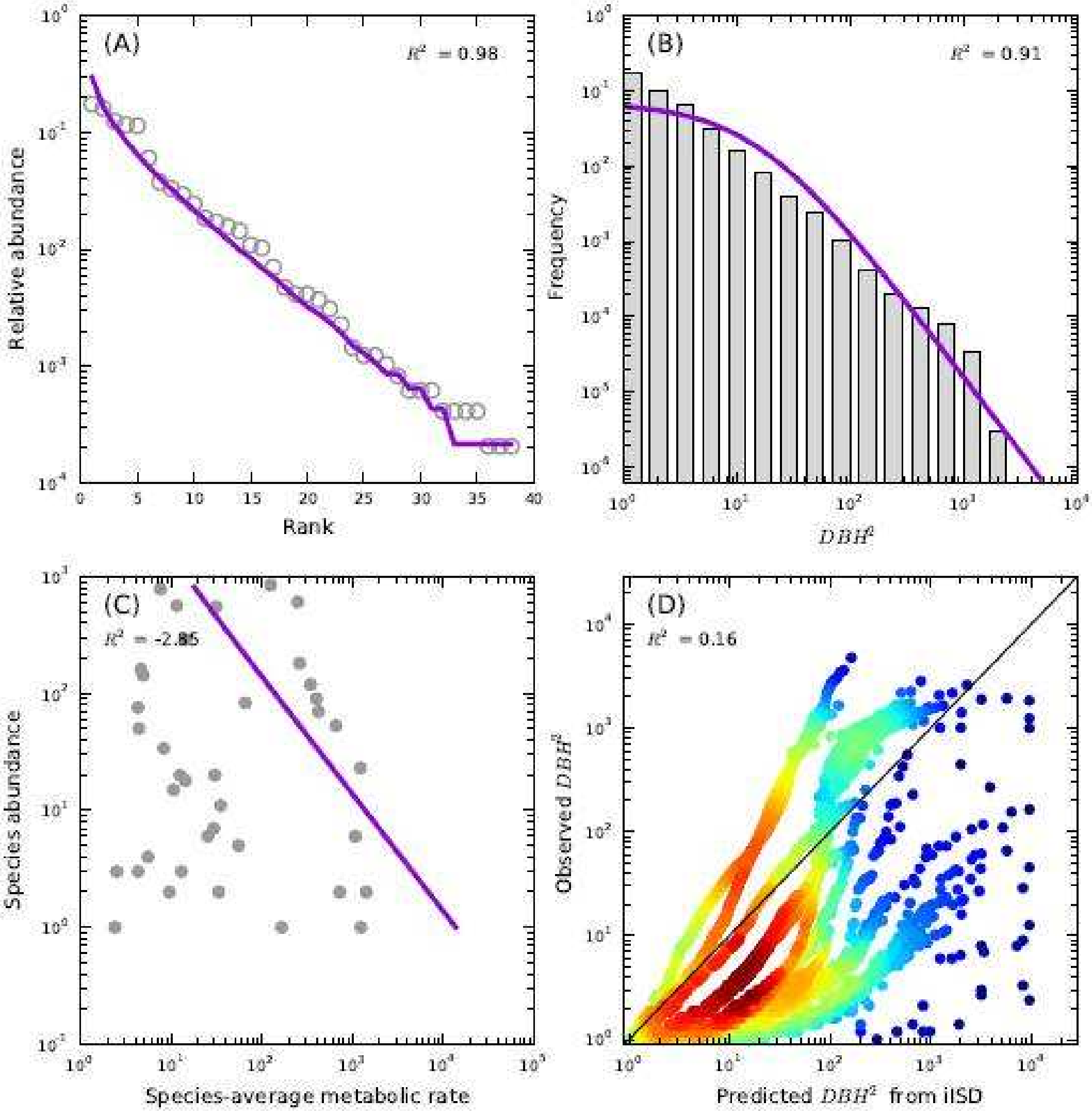

NC,13

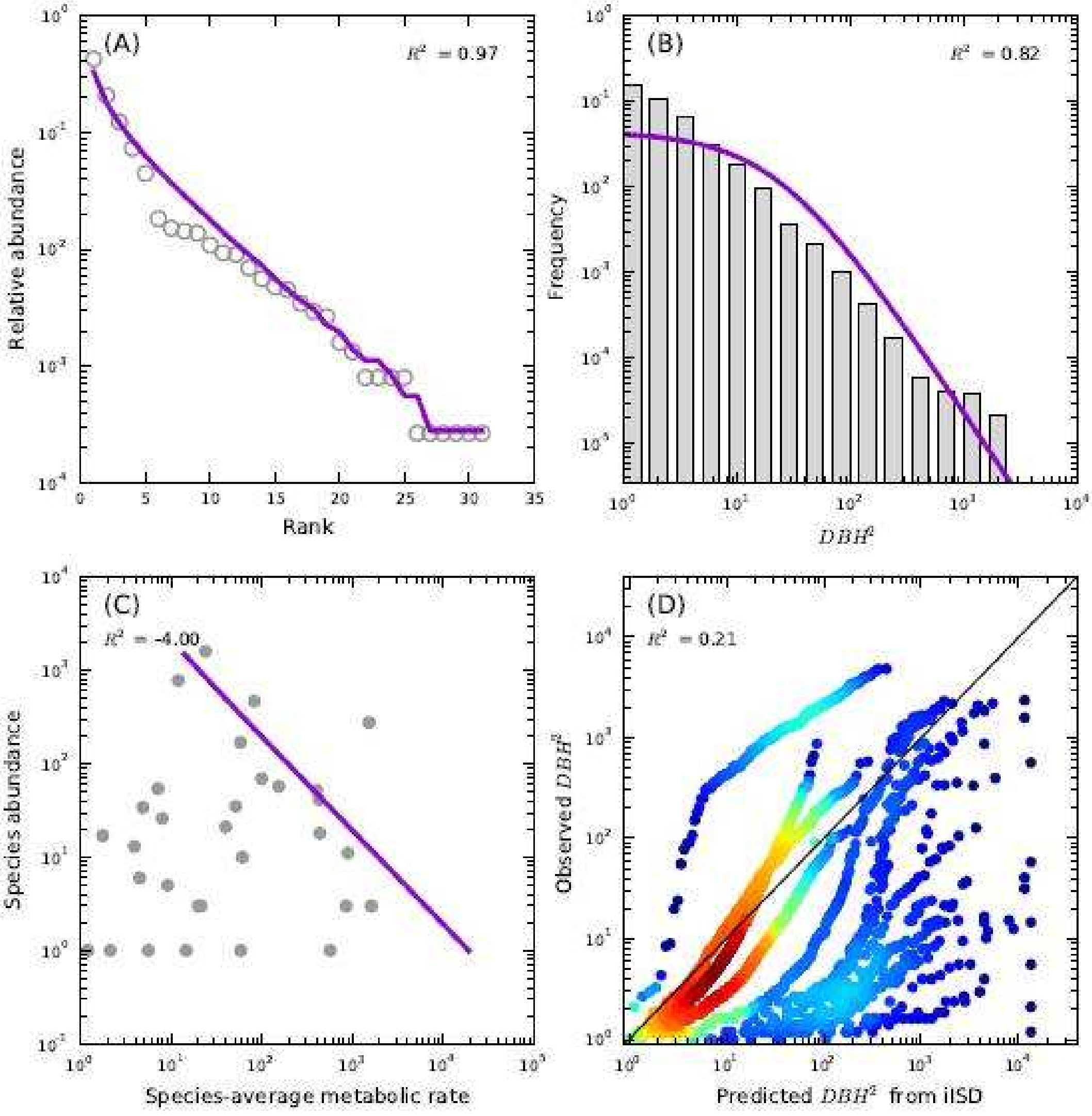

NC,14

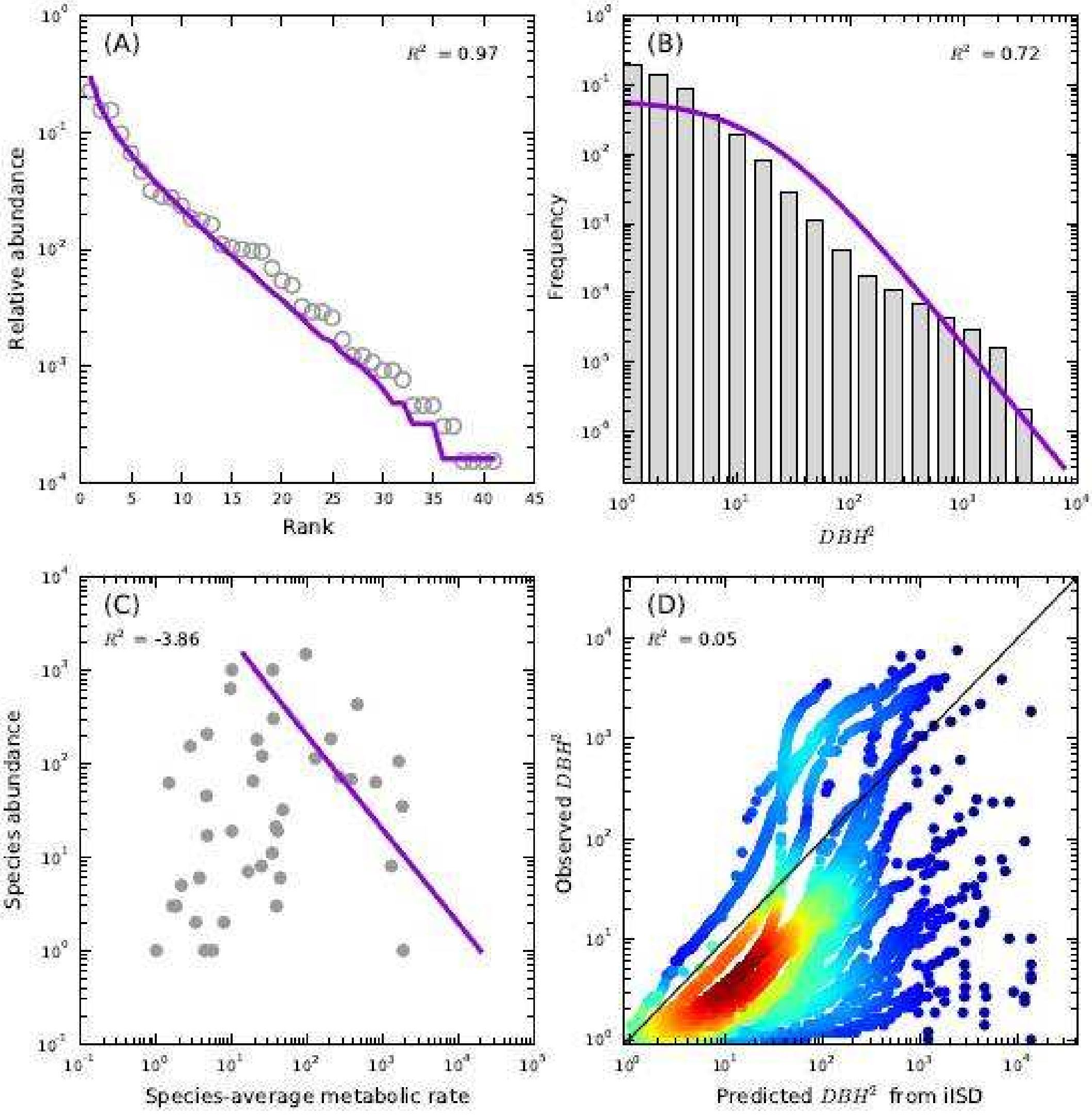

NC,4

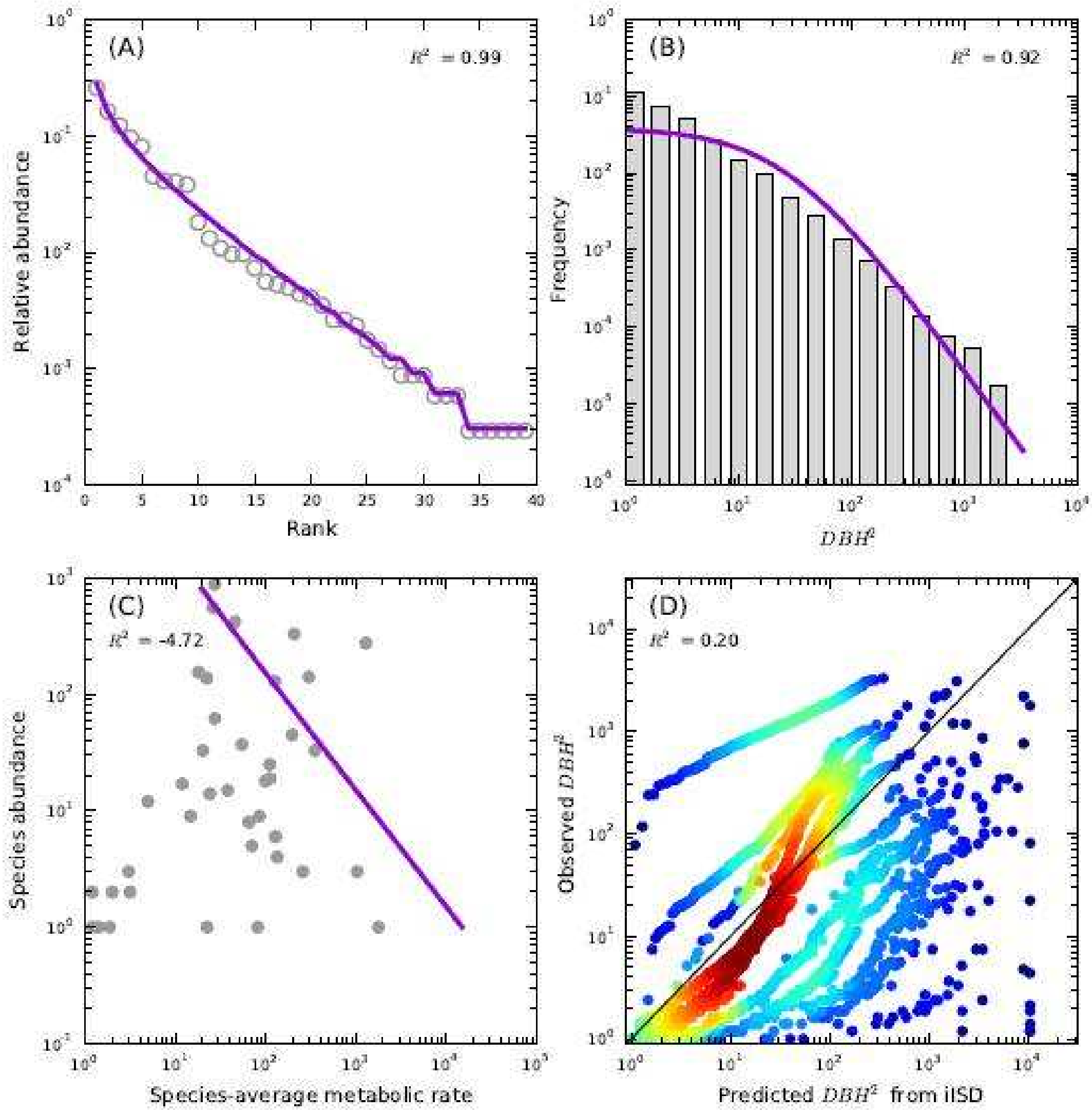

NC,93

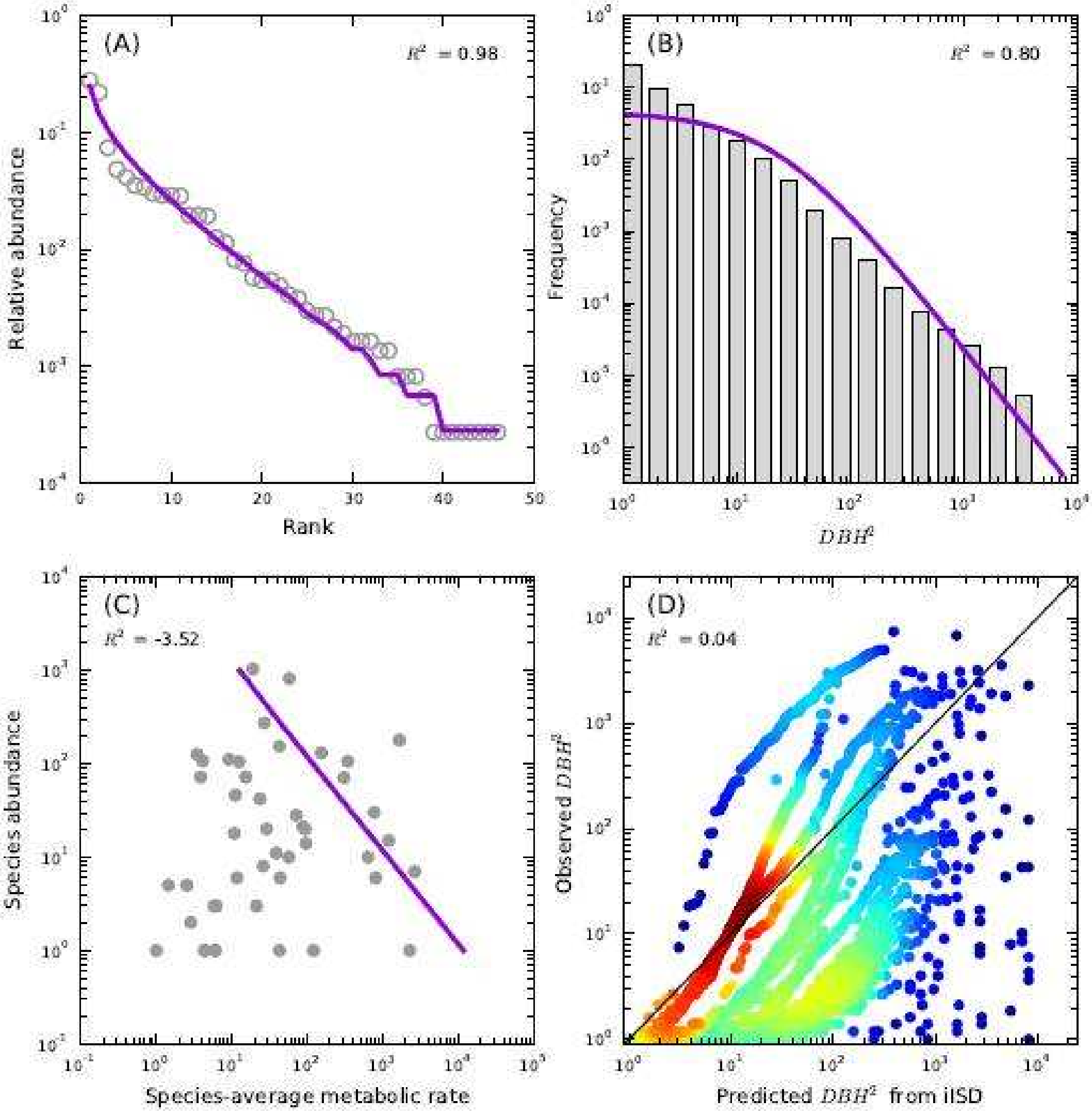

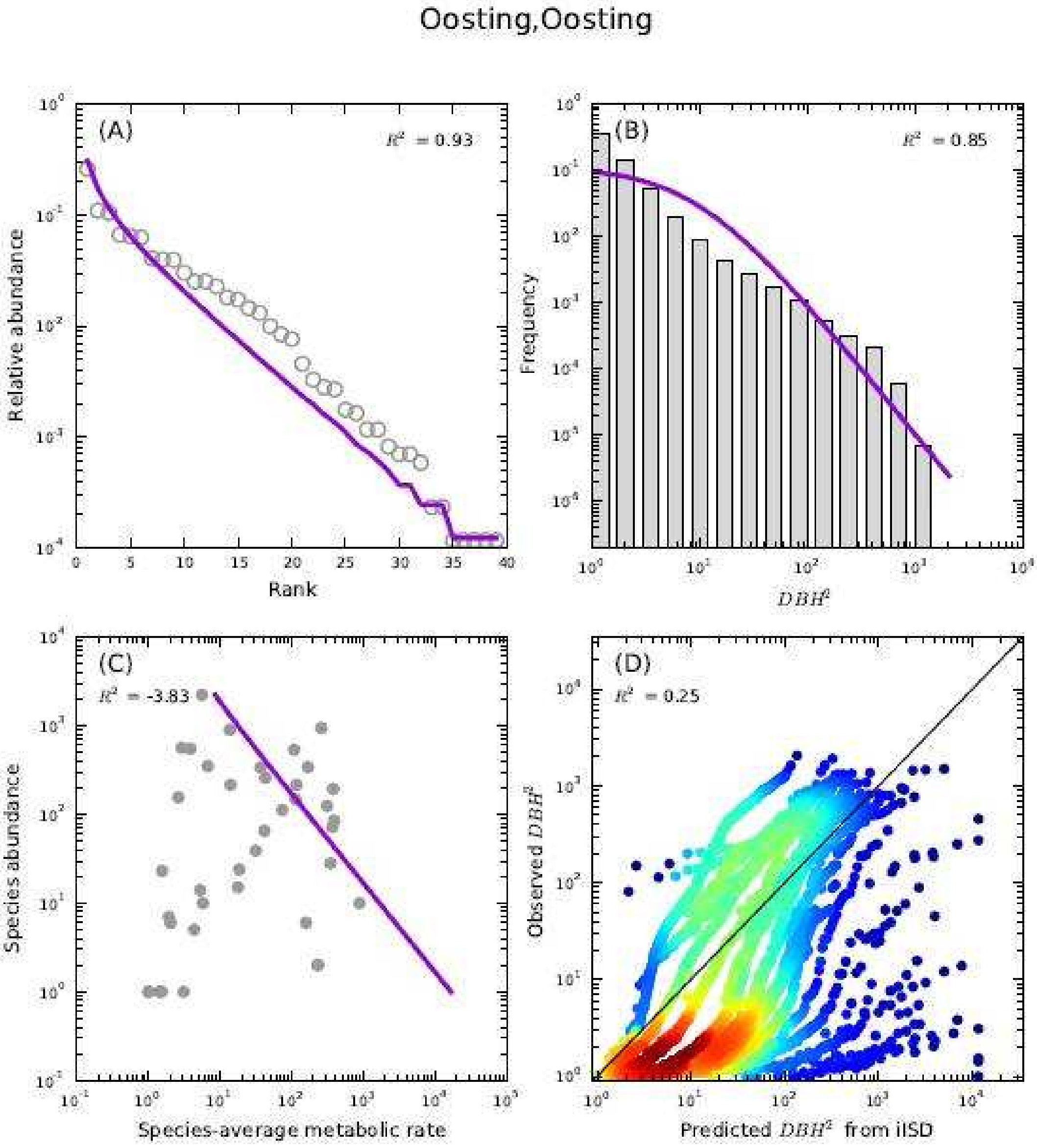

Serimbu,S-1

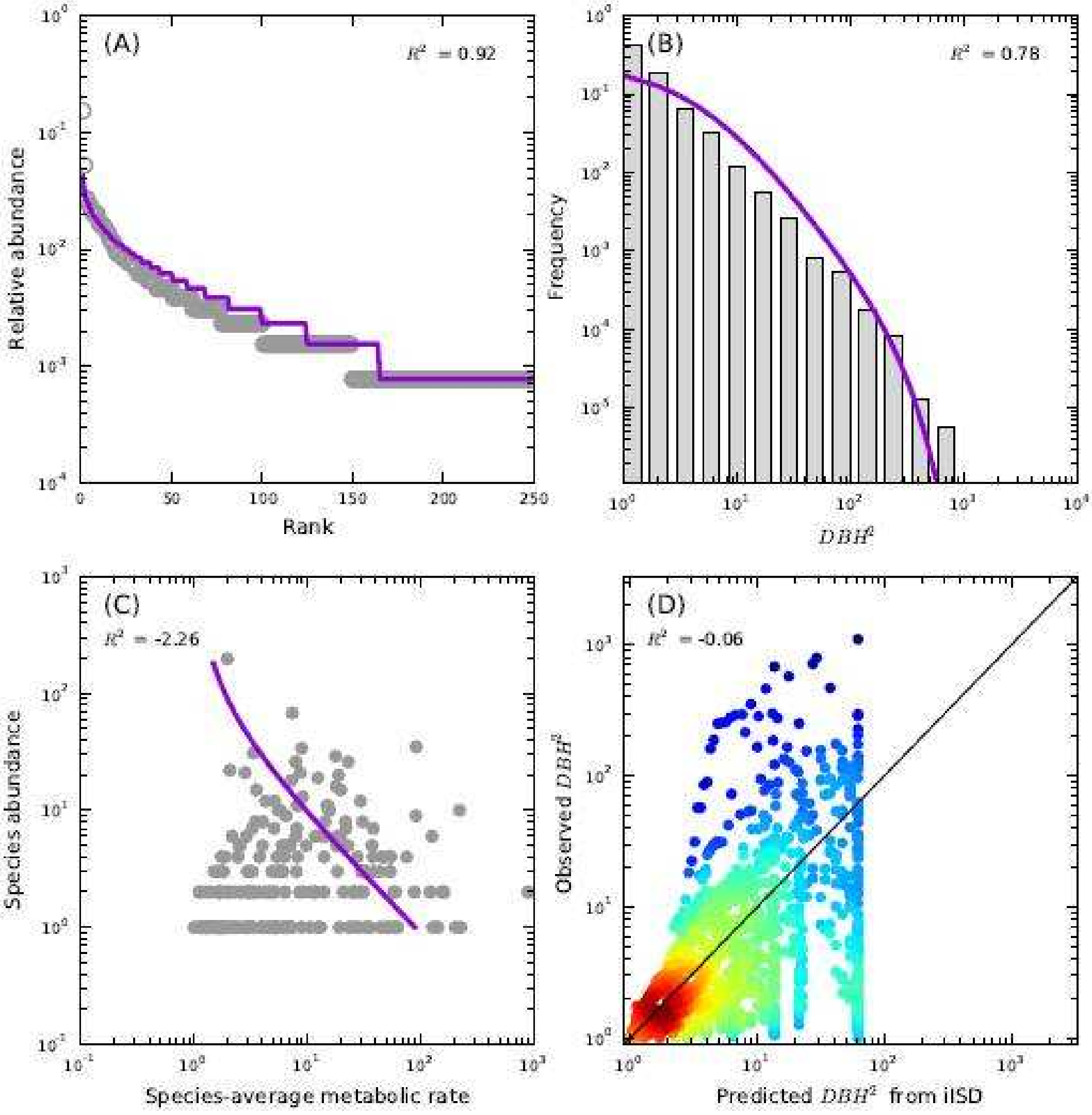

Serimbu,S-2

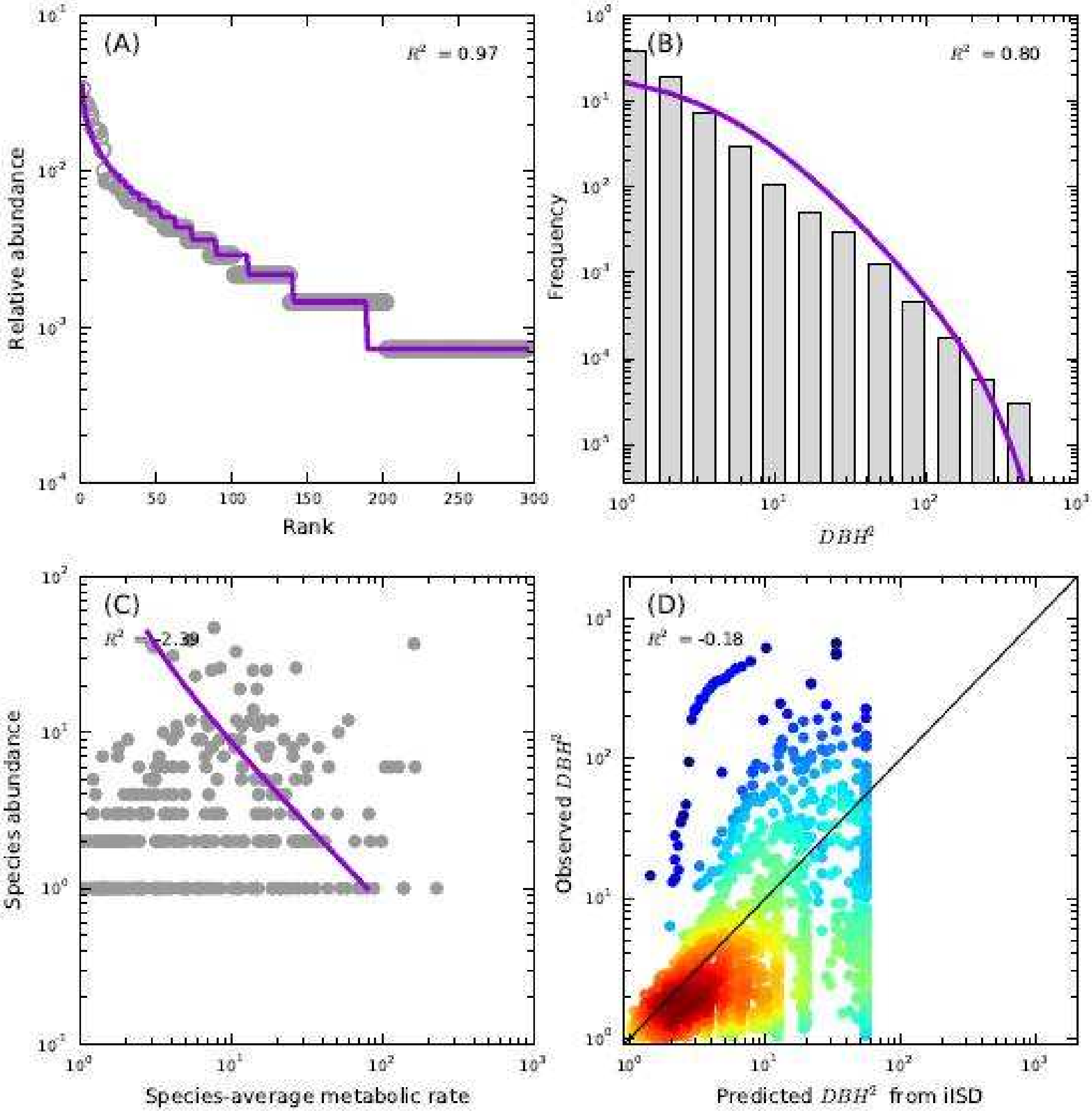

Shirakami,Akaishizawa

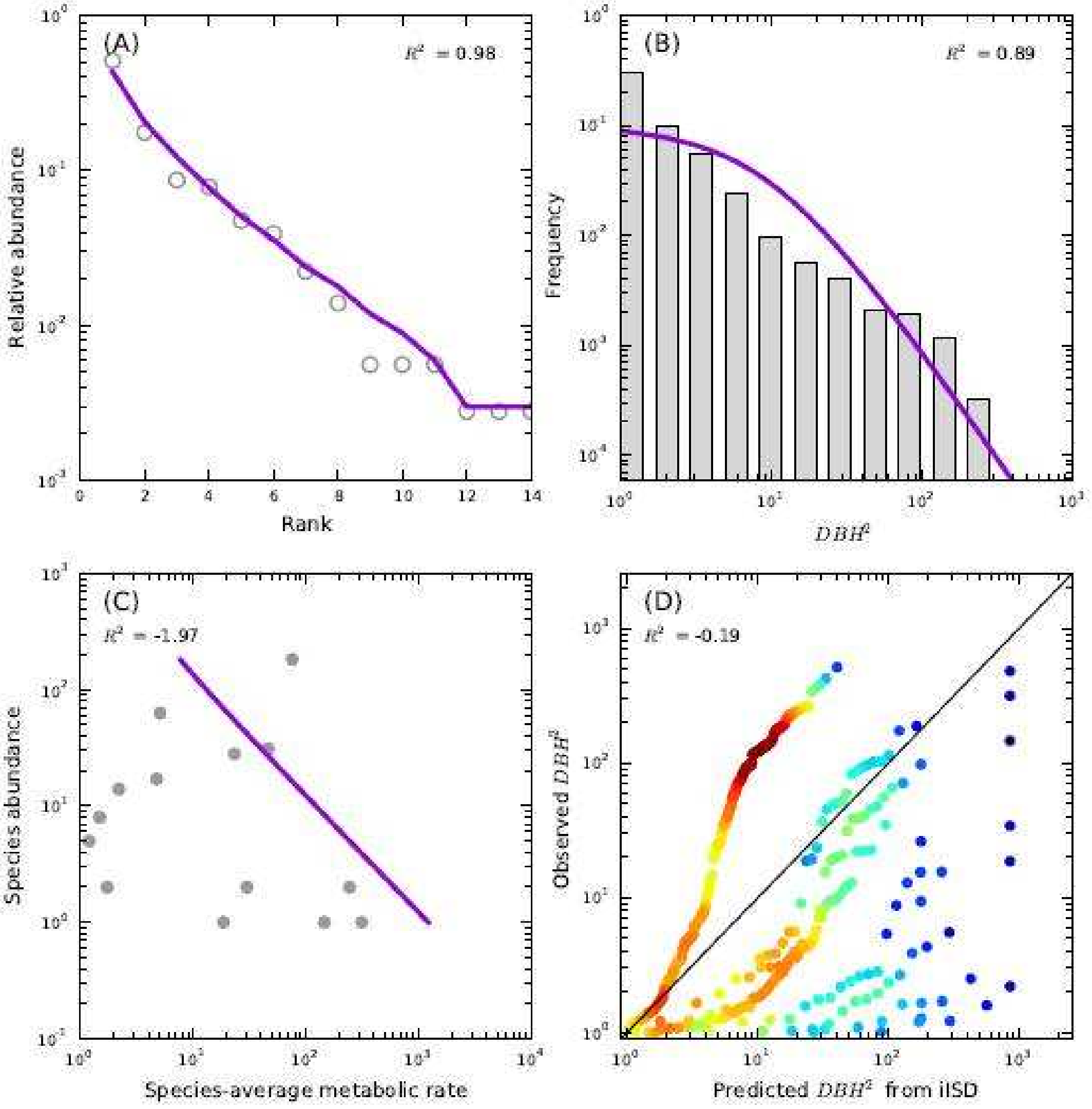

## Shirakami,Kumagera

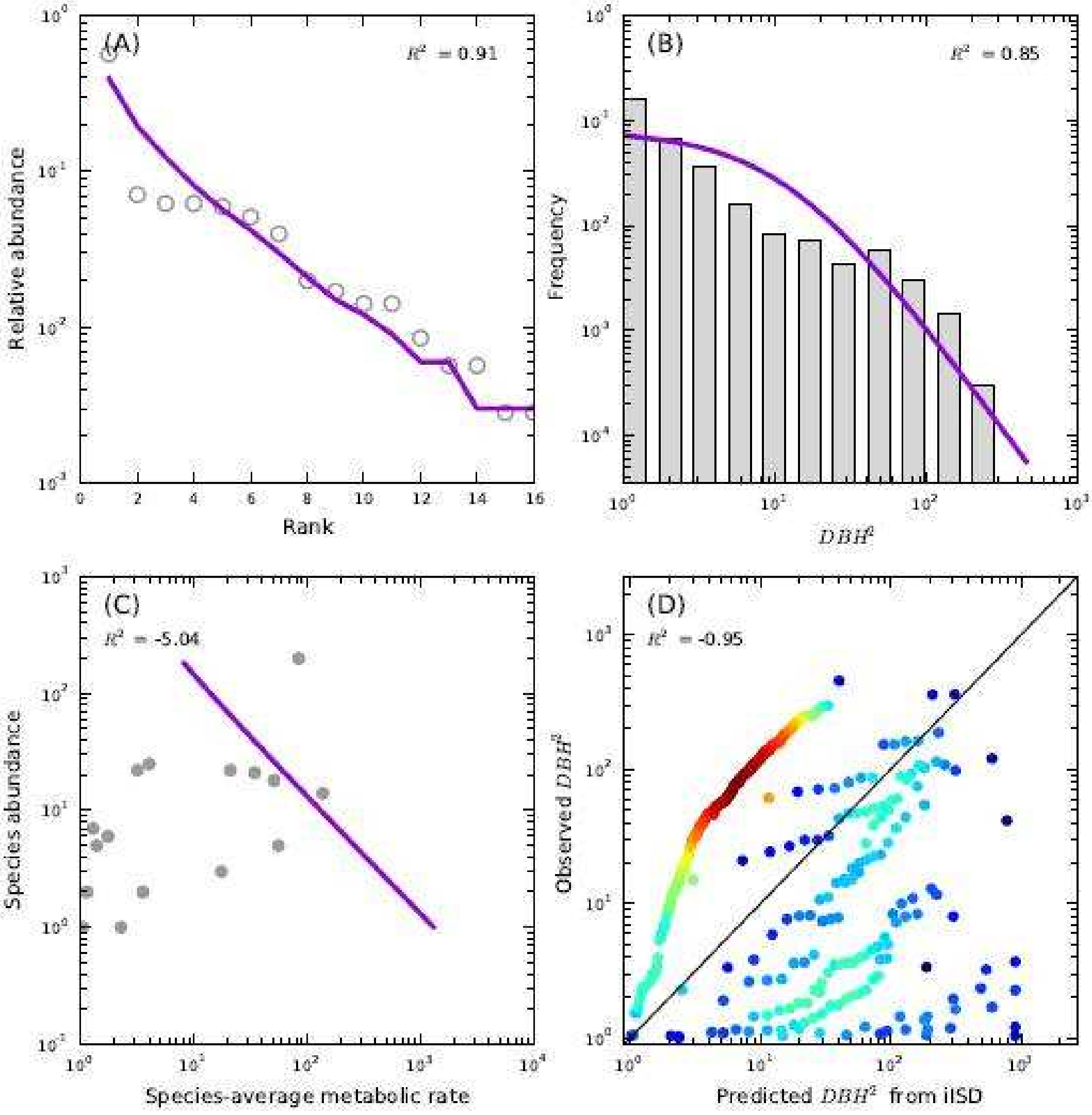

Sherman,sherman

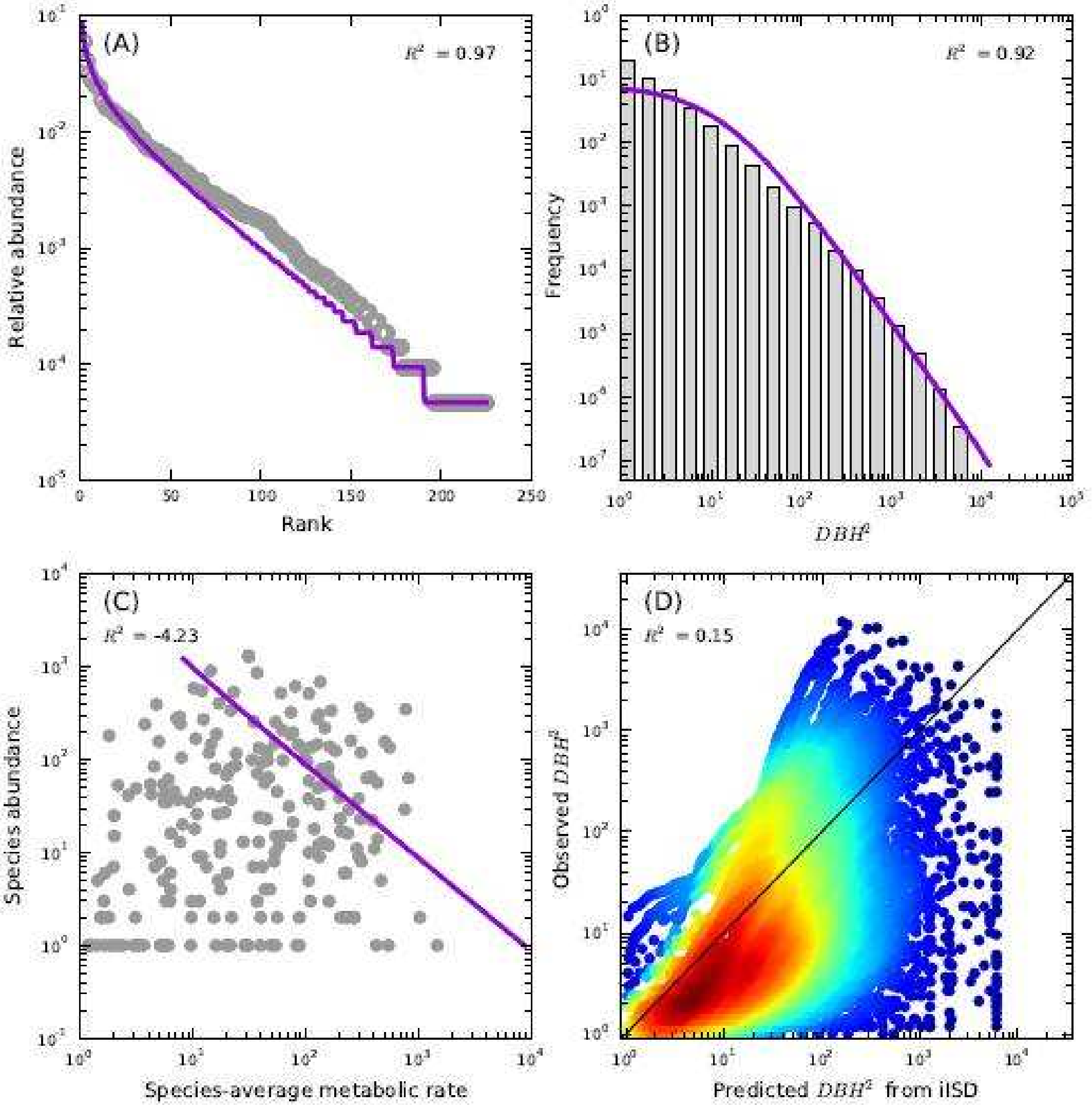